%% file: bare_jrnl_new_sample4.tex
\documentclass[lettersize,journal]{IEEEtran}
\usepackage{physics}
\usepackage{graphicx}
\usepackage[table,xcdraw]{xcolor}
\usepackage{caption}
\usepackage{float}
\usepackage{amsmath}
\usepackage{subfigure}
\usepackage{amsmath}
\usepackage{longtable}
\usepackage{booktabs}
\usepackage{float}
\usepackage{threeparttable}
\usepackage{xcolor}
\usepackage{diagbox}
\usepackage{cite}
\usepackage{hyperref}
\usepackage{bm}
\usepackage{enumitem}
\usepackage{multirow}
\newcommand{\comment}[1]{}

\begin{document}

\title{Video Quality Assessment:\\ A Comprehensive Survey}


\author{
Qi~Zheng,
Yibo~Fan$^{\star}$,
Leilei~Huang,
Tianyu~Zhu,
Jiaming Liu,
Zhijian Hao,
Shuo Xing,
Chia-Ju~Chen,
Xiongkuo~Min,
Alan~C.~Bovik$^{\dag}$,
Zhengzhong~Tu$^{\dag}$%
\thanks{Qi Zheng, Yibo Fan, Tianyu Zhu, Jiaming Liu, and Zhijian Hao are with Fudan University, Shanghai 200000, China (e-mail: qzheng21@m.fudan.edu.cn;
fanyibo@fudan.edu.cn;
zhuty22@m.fudan.edu.cn; liujm22@m.fudan.edu.cn; zjhao19@fudan.edu.cn; ).}
\thanks{Leilei Huang is with East China Normal University, Shanghai 200000, China (e-mail: llhuang@cee.ecnu.edu.cn).}
\thanks{Xiongkuo Min is with the Institute of Image Communication and Information Processing, Shanghai Jiao Tong University, Shanghai 200240, China (e-mail: minxiongkuo@sjtu.edu.cn).}
\thanks{Chia-Ju Chen and Alan C. Bovik is with the Department of Electrical and Computer Engineering, The University of Texas at Austin, Austin, TX 78712 USA (e-mail: ju40268@utexas.edu;bovik@utexas.edu).}
\thanks{Shuo Xing and Zhengzhong Tu are with the Department of Computer Science and Engineering, Texas A\&M University, College Station, TX 77840 (e-mail: shuoxing@tamu.edu;tzz@tamu.edu). This work was done prior to the employment of Zhengzhong Tu by Texas A\&M University, and he was not supported by any grant.}
\thanks{$^{\star}$: Corresponding author. $^{\dag}$: Equal advising.}
}

\markboth{Journal of \LaTeX\ Class Files,~Vol.~14, No.~8, August~2021}%
{Shell \MakeLowercase{\textit{et al.}}: A Sample Article Using IEEEtran.cls for IEEE Journals}


\graphicspath{{figures/}}

\maketitle

\input{sections/0_abstract}
\input{sections/1_introduction}

\input{sections/2_background}
\input{sections/3_subjective_study}
\input{sections/4_objective_assessment}
\input{sections/5_Performance_benchmark}

\input{sections/6_application}
\input{sections/8_conclusion}
\input{sections/9_acknowledgments}

\bibliographystyle{IEEEtran}
\bibliography{IEEEfull}

\vfill

\end{document}

%% file: sections/0_abstract.tex
\begin{abstract}
Video quality assessment (VQA) is an important processing task, aiming at predicting the quality of videos in a manner highly consistent with human judgments of perceived quality.
Traditional VQA models based on natural image and/or video statistics, which are inspired both by models of projected images of the real world and by dual models of the human visual system, deliver only limited prediction performances on real-world user-generated content (UGC), as exemplified in recent large-scale VQA databases containing large numbers of diverse video contents crawled from the web.
Fortunately, recent advances in deep neural networks and Large Multimodality Models (LMMs) have enabled significant progress in solving this problem, yielding better results than prior handcrafted models.
Numerous deep learning-based VQA models have been developed, with progress in this direction driven by the creation of content-diverse, large-scale human-labeled databases that supply ground truth psychometric video quality data.
Here, we present a comprehensive survey of recent progress in the development of VQA algorithms and the benchmarking studies and databases that make them possible.
We also analyze open research directions on study design and VQA algorithm architectures. Github link: https://github.com/taco-group/Video-Quality-Assessment-A-Comprehensive-Survey.

\end{abstract}


%% file: sections/1_introduction.tex
\section{Introduction}
\label{introduction}

\IEEEPARstart{R}{ecent} years have witnessed the rapid development of streaming media technologies and platforms, making video content the dominant form of Internet traffic. Streaming and social media videos play a central role in the daily lives of billions of people, in particular since the evolution of Web 2.0 has spurred an explosion of user-generated content (UGC). Moreover, recent advancements in generative AI technologies have fueled the rapid rise of AI-Generated Content (AIGC), enabling seamless content creation and transformation across various media platforms.
As such, it has become increasingly important for video service providers to improve the efficiency of the video transcoding techniques they deploy in the Cloud, while also delivering satisfying visual quality of experience (QoE) to customers. Balancing these goals conditioned on constantly changing network conditions remains a crucial and long-standing challenge of core interest.
Perceptually designed video quality assessment models, which act as `judges' of percieved quality, play an important role in monitoring and measuring visual content quality throughout global communication network and video processing chains, particularly in video coding, enhancement, and reconstruction algorithms.

Video quality studies can be broadly categorized into two groups: objective and subjective methods.
\textbf{Subjective VQA} involves humans in the loop, who view a number of video contents in a controlled environment, then render judgments of perceptual quality on each content.
These raw annotations are further normalized and averaged as mean opinion scores (MOS) or as difference mean opinion scores (DMOS).
While subjective studies are time and labor-consuming, they provide valuable psychometric datasets on which to benchmark, develop, compare, and calibrate video quality models.
\textbf{Objective VQA} methods rely on algorithmic models to predict the perceptual quality of video content.
According to the information available from reference signals that are input to an algorithm, objective VQA models can be categorized into full-reference (FR), reduced-reference (RR), or no-reference (NR) models.
FR VQA models compare visual differences between high-quality reference videos and distorted counterparts, hence way be viewed as perceptual video fidelity models.
RR VQA models are similar to FR methods, but only require significantly reduced amounts of signal information to be extracted from the reference videos to conduct quality prediction.
In many practical scenarios, however, the reference signals are inaccessible, \textit{e.g.,} videos uploaded by amateur photographers or videographers to social platforms such as YouTube and TikTok.
In such instances, NR VQA models are the only tools available to monitor and analyze such authentically distorted contents without any pristine reference.

\begin{figure}[!htp]
    \centering
    \includegraphics[width=8cm]{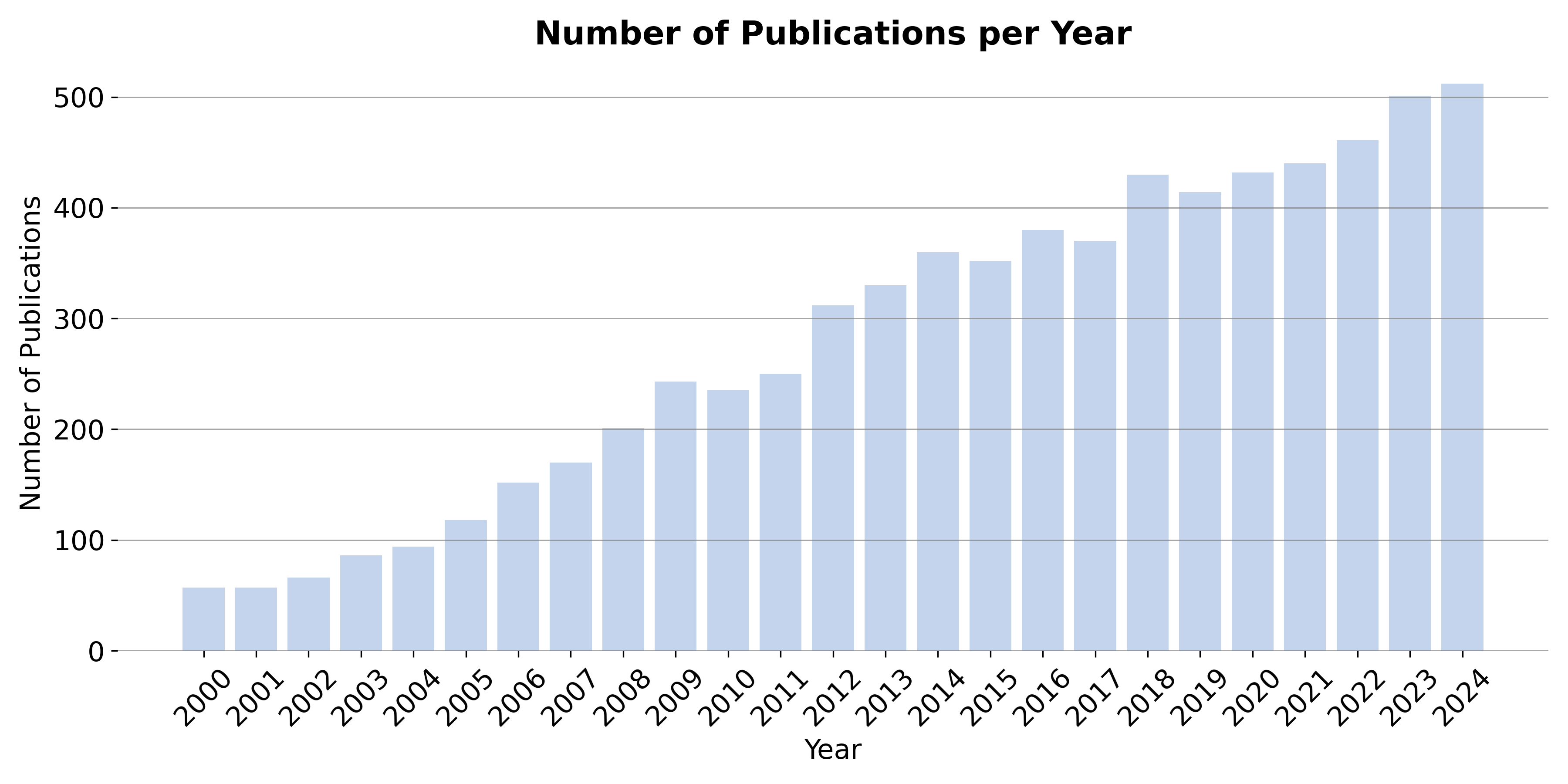}
    \caption{Number of publications on image and video quality assessment per year (from Google Scholar).}
    \vspace{-12pt}
    \label{fig: publications_per_year}
\end{figure}

Modeling the perception of video quality poses significant challenges.
Videos are degraded by a very wide variety of distortions, including noise, blur, ringing, banding, compression, and blockiness, which are very different to model, often occur together and interact, and impact perceptual video quality in ways that are also content-dependent.
The complexity of the problem is exacerbated by the commingling of multiple distortions, as well as the presence of complex object and camera motions.
Moreover, extracting quality-aware video features and predicting video quality over both space and time must account for temporal human memory effects, making the problem even more challenging.

Conventional VQA models typically rely on computing perceptually relevant visual differences and/or natural statistical regularities.
A simple mapping engine, such as a support vector regressor, is often used to learn how to map quality predictions from suitable distortion-aware features. Of course, the feature extraction process is integral to the success of these VQA models.
While these approaches have demonstrated significant success and are used in global industry deployments, there remains ample scope for further improvement.

Fortunately, deep learning has been revitalized by leveraging large datasets of human-labeled images~\cite{krizhevsky2017imagenet,deng2009imagenet}, enabling significant progress across high- and low-level computer vision tasks. Neural network architectures trained on vast amounts of data have demonstrated the ability to capture semantic features and universal representations, reducing reliance on manual feature selection and showcasing notable generalization capabilities.
Recent advancements in creating large-scale psychometric datasets for images and videos have further fueled the application of deep learning for quality prediction~\cite{8103112,ying2021patch,Youtube-UGC,min2020study}, as shown in Fig.~\ref{fig: publications_per_year}. Modern deep VQA models typically employ pre-training on ancillary tasks with abundant data, followed by fine-tuning on more focused video quality databases. While these databases are increasingly comprehensive, their size remains insufficient to train large models due to the high costs of conducting extensive subjective studies.
The adoption of deep networks has led to significant performance improvements in quality prediction by capturing diverse distortion phenomena and high-level semantic content while aligning with visual perception. However, the full potential of deep learning for video quality assessment remains unrealized, constrained by the scarcity of large, annotated psychometric datasets and limited understanding of human visual quality perception.

The recent advancement of large models~\cite{Kirillov_2023_ICCV,pmlr-v139-radford21a,NEURIPS2023_6dcf277e,Ye_2024_CVPR}, which are characterized by a significant number of parameters and trained on extensive data, has remarkably promoted the perceptual cognization of machines. Many researchers are exploring various manners of incorporating large language models (LLMs) and large multimodality models (LMMs) for IQA/VQA tasks, such as extending semantic-aware features in the prior knowledge embedded in large models~\cite{Kirillov_2023_ICCV,pmlr-v139-radford21a}, and enhancing the explainability of quality assessment with quality-aware prompts~\cite{Wu_2024_CVPR,wu2023qalignteachinglmmsvisual}, among others.

\begin{figure*}[!t]
\centering
\footnotesize
\def\xheight{0.85}
\setlength{\tabcolsep}{1pt}
\includegraphics[width=\xheight\linewidth]{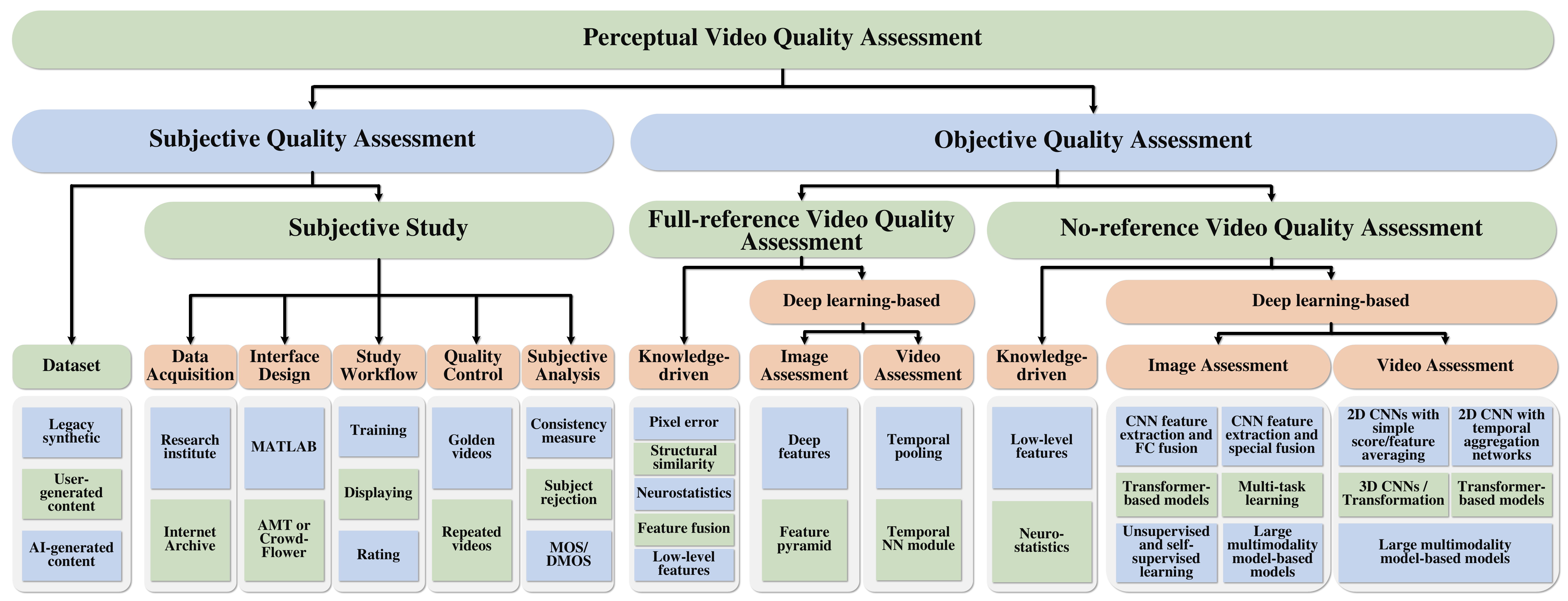}
\caption{Taxonomy of existing subjective and objective video quality assessment methods.}
\vspace{-12pt}
\label{fig:taxonomy}
\end{figure*}

Towards facilitating a better understanding current progress on modern video quality databases, and the development of deep learning-based models, we have sought to conduct a comprehensive overview of the past and recent advances and the state of the art of video quality modeling.
Our key contributions include:
\begin{itemize}[leftmargin=*]
    \item We review both subjective (database) and objective (algorithm) quality assessment models, providing not only a detailed taxonomy but also a nuanced analysis of their evolution and core methodologies (see Fig.~\ref{fig:taxonomy} and Fig.~\ref{fig:year_taxonomy}).
    \item We introduce readers to the elements of conducting subjective quality assessment studies, 
    and review many of the most popular and representative video quality datasets, which vary by types of video content and targeted use cases.
    \item We review traditional image and video quality assessment methods, as a set-up for a comprehensive discussion of recent deep learning-based VQA models.
    Commonly used (perceptual) loss functions are also discussed.
    \item We compare representative IQA/VQA models on databases of emerging content, providing insights on model design in terms of spatiotemporal information modeling and prior knowledge use in large models.
    \item We discuss practical applications, some of which are deployed at very large commercial scales, remaining core challenges, and future opportunities for improving and applying deep learning-based VQA models, towards facilitating and inspiring future research efforts and industry deployments in the video streaming and social media fields.
\end{itemize}

The remainder of the paper is organized as follows. Section~\ref{background} introduces neurostatistical video distortion models and key network layers in deep learning-based VQA models. Section~\ref{subjective-vqa} summarizes subjective video quality assessment methods and widely used datasets. Section~\ref{object_vqa} reviews full-reference and no-reference VQA models, emphasizing deep learning approaches and loss functions. Section~\ref{performance_bench} compares the performance of representative VQA models on emerging content databases. Finally, Section~\ref{sec:application_challenge} discusses practical applications and challenges, with Section~\ref{conclusion} concluding with forward-looking insights.

%% file: sections/2_background.tex
\section{Background}
\label{background}
To set the stage for the main topics, we will briefly review a few classical video quality assessment models, including structural similarity based methods and those that use neurostatistical distortion measurements. These established models, which are deep ingrained in principles of visual neuroscience, are quite relevant to processing in early network layers in deep perceptual video quality assessment models.

\subsection{Structural Similarity}
Natural photographic images are quite structured, and the human visual system (HVS) is able to efficiently extract structural information as part of the process of seeing.
The well-known full-reference image quality assessment (FR-IQA) method called SSIM~\cite{wang2003multiscale} measures the perceptual similarities of local path luminances, contrasts, and structures.
SSIM~\cite{wang2003multiscale} has been used in an extraordinary diversity of applications, including traditional FR-IQA~\cite{wang2003multiscale,sampat2009complex,wang2010information,6423791,xue2013gradient,liu2011image,zhang2011fsim,zhang2014vsi,shi2017perceptual}, FR-VQA~\cite{wang2004video,seshadrinathan2007structural,manasa2016optical}, and deep learning based FR-IQA~\cite{ding2020image,ding2021locally} and FR-VQA~\cite{sun2021deep}. 

It is instructive to further consider the construction of the most widely-deployed version of SSIM, called Multi-Scale SSIM or MS-SSIM~\cite{wang2003multiscale}. MS-SSIM contains three terms that are iterated over five scales, all combined in an exponentially-weighted product, with weights determined by a human study. The five contrast terms of MS-SSIM (just one in SSIM) may be interpreted as center-surround bandpass processes, computed over multiple scales, each rectified then divisively normalized by local bandpass energy. These terms may be viewed as a simple retinal model, with energy-normalized, non-directional center-surround receptive fields resembling the ganglionic responses. They are also similar to non-directional weighting functions learned in the earliest layers of CNNs, but with simple (\textit{e.g.}, ReLu) activations replaced by rectification and local energy normalization, which is a perceptual strategy effectively employed in deep autoencoder based image compression~\cite{balle2016end,chen2022learning}.
The structure terms of MS-SSIM, which involve saturation-biased, energy-normalized bandpass signal correlations resemble neural computations of correlations hypothesized to occur in extrastriate cortical area VZ~\cite{freeman2011metamers}, but which also resemble the feature correlations that occur in modern self-attention layers~\cite{vaswani2017attention}. While the correlations in Transformer architectures are learned, weighted autocorrelations, the correlations in MS-SSIM are only pre-weighted by perceptual optimization of the MS-SSIM exponents. The correlations are computed between bandpass signals and bandpass distorted signals (over scales), hence include autocorrelations of both signal and distortion, as well as cross-terms.

It is also worth noting that SSIM and MS-SSIM are differentiable (and also quasi-convex~\cite{channappayya2008design}), and have found extensive use not only for assessing the outcomes of deep image models~\cite{NEURIPS2022_94461854,NEURIPS2020_8a50bae2,NEURIPS2023_ccf6d8b4}, but also as network loss functions, enabling perceptual network optimization of any kind of deep image regressor~\cite{Lu_2019,Galteri_2017_ICCV,SHI2022108351}.

\subsection{Neurostatistical Quality Features}
High-quality photographic picture and videos reliably exhibit certain statistical regularities that are predictably altered by distortions~\cite{ruderman1994statistics,sheikh2006image}.
Specifically, pictures and videos that have bandpass filtered in space and/or time follow generalized Gaussian distributions~\cite{reininger1983distributions,mallat1989a}, and equivalent Gaussian-Scale mixture models~\cite{ruderman1994statistics,Wainwright1999scale,li2018vmaf}.
However, when distortions are introduced, the bandpass statistic of pictures and videos tend to predictably deviate from these models.
This has led to a plethora of powerful FR and NR picture and video quality prediction models that have been developed in wavelet~\cite{sheikh2006image,tu2021rapique,9746997}, spatial luminance~\cite{6353522,mittal2012no,xue2014blind,9921340}, chroma~\cite{kundu2017no,FRIQUEE}, space-time~\cite{9897565}, discrete cosine transform (DCT)~\cite{6705673,li2016spatiotemporal}, and other bandpass domains~\cite{tu2021rapique,9746997}.

The performances of MS-SSIM, VIF~\cite{sheikh2006image}, VMAF~\cite{Wainwright1999scale}), and other neuroscience-based models are so good that they are used to monitor and control the scaling and encode quality of most Internet traffic (which is about 80\% picture and video data). Indeed, deep models have a hard time beating these models on cinematic-grade streaming data, and are much more compute-intensive. However, on lower-grade, multi-distortion user-generated content (UGC), deep learning-based picture quality prediction models are much better able to map the internal statistical structures of distorted pictures and videos to human percepts of visual quality, especially in the blind image and video quality assessment (BIQA/BVQA) scenarios.

\subsection{Multi-layer Perceptron}
We also briefly review the primary types of network layers that constitute modern deep networks. The simplest and earliest is the multilayer perceptron (MLP)~\cite{hastie2009elements}, is a feedforward network consisting of multiple layers of interconnected nodes, each layer fully connected to the next. 
These are often referred to as ``fully connected networks" (FCNs).
When used in deep learning-based VQA models, MLPs usually serve as later regression layers, which serve to map deep features extracted by earlier layers to the final quality score predictions~\cite{kim2018deep,xu2020c3dvqa,feng2022deep,hou2020no,sun2022deep,you2019deep,li2015no,zhangyu2018blind,xu2021perceptual,wen2021strong}.
MLPs are also used to generate probability vectors that describe the distortions by applying softmax~\cite{li2015no,liu2018end}.

\subsection{Convolution Layer}
Convolutional layers extract features from input pictures and video frames by applying spatial convolution operations, which involve sliding a small kernel over the input data and computing the dot product between the kernel and each local input patch at each spatial position.
Convolutional layers are effective for extracting spatial and/or temporal features over multiple scales from visual data, making them well-suited for video quality assessment tasks.
A variety of video quality models apply convolutional neural networks (CNNs or ConvNets) to extract both low-level and high-level semantic features from video frames, to conduct video quality prediction~\cite{kim2018deep,varga2019noPool,wen2021strong,wu2021semantic,varga2022no,korhonen2020blind,tu2021rapique,Flickrvid-150k,varga2019noCNN,li2019quality,chen2020rirnet,xu2021perceptual,li2022blindly,li2022study,zhangyu2018blind}.
Other approaches employ 3D space-time CNNs to extract spatio-temporal features from blocks of video frames~\cite{feng2022deep,liu2018end,you2019deep}.
Some VQA models leverage both 2D and 3D CNNs to separately analyze spatial and temporal quality degradation~\cite{xu2020c3dvqa,chen2021deep,hou2020no,ying2021patch,wang2021rich,sun2022deep}.
These sucecesses are supported by studies that have shown that perceptual distances are well modeled by deep activations within convolutional networks~\cite{zhang2018unreasonable,tariq2020deep,sun2021deep,danier2022flolpips}.

\subsection{Recurrent Layer}
Videos are spatio-temporal, as are many distortions, hence capturing short- and long-term temporal correspondance is essential for accurate perceptual video quality prediction. 
Recurrent layers, like long short-term memory (LSTM) blocks~\cite{hochreiter1997long} and gated recurrent units (GRU)~\cite{dey2017gate} have been widely used to model temporal video dependencies.
LSTMs and GRUs may either process multi-frame clips~\cite{you2019deep} or single frames~\cite{Flickrvid-150k,varga2019noCNN,xu2021perceptual,li2022blindly,telili20222bivqa}, when predicting video quality scores.
%

\subsection{Attention Layers}
Attention layers~\cite{vaswani2017attention} have become important components of modern deep learning models. They are able to selectively weight local space-time regions or channels that may be more important for feature representation learning.
Typical attention models include convolutional spatial attention mechanism~\cite{woo2018cbam} and channel squeeze-and-excitation attention models~\cite{hu2018squeeze}.
Various methods have been proposed for calculating attention weights.
For example, some models use Gaussian functions~\cite{20219506420} to compute the attention weights, while others use average and maximum pooling~\cite{xu2021perceptual}, standard variance~\cite{20229452150}, or graph convolutions~\cite{xu2021perceptual} to obtain the weights.
The choice of attention mechanism depends on the nature of the data and the specific requirements of the VQA task at hand.

\subsection{Transformer}
Originally proposed for natural language processing tasks~\cite{vaswani2017attention}, Transformers have since been successfully applied to various computer vision tasks~\cite{vaswani2017attention,dosovitskiy2020image,liu2021swin}.
The core components of Transformers are multi-head self-attention mechanisms which can effectively capture global spatial relationships by learning content-dependent adaptive weights.
Since large-scale spatial and temporal associations can play an important role in making video quality predictions, Transformer models have also been studied in this context.
However, the quadratic computational complexity of self-attention mechanisms presents challenges when processing large amounts of video data.
To address this issue, several studies have explored the use of sparsely sampled clip-~\cite{you2021long} and/or patch-level~\cite{xing2021starvqa} features to reduce the number of Transformer tokens.
Other have devised less expensive window-based attention mechanisms~\cite{liu2021swin} having only linear complexity with respect to image size~\cite{wu2022discovqa,wu2022fast}.
Overall, the Transformer architecture is a promising direction for modeling long-range (memory) dependencies that affect perceived video quality, although further research is needed to explore this potential.

\subsection{Large Models}
Large models have shown impressive capabilities on visual understanding tasks, thus it is intuitive to leverage them as perceptual quality assessment tools. Several recent efforts have demonstrated their effectiveness on the quality assessment task. Among them, the large segmentation model SAM~\cite{Kirillov_2023_ICCV} extracts highly generic semantic structural information while exhibiting exceptional generalization ability, and has been adopted as a semantic feature extraction module by several IQA models~\cite{li2023samscore,li2023sam}. By training on extensive image-text pair data, the multi-modal model CLIP~\cite{pmlr-v139-radford21a} can build semantic-level relationships between texts and visual information. CLIP can be used as either a semantic-aware feature extraction module~\cite{He_2024_CVPR,Yuan_2024_CVPR} or a quality index customized by perception-aware prompts~\cite{Wang_Chan_Loy_2023,wu2023exploringopinionunawarevideoquality,wu2023towards}. Large language models can be extended to multi-modal tasks by incorporating visual inputs via multi-modal models like CLIP, resulting in large multimodal models. A series of recently developed IQA/VQA models successfully leverage LMMs to perform prompt-driven~\cite{wu2023q,Wu_2024_CVPR} or embedding-based~\cite{ge2024lmm} quality evaluation.

%% file: sections/3_subjective_study.tex
\section{Subjective Video Quality Assessment}
\label{subjective-vqa}
\subsection{Subjective Study Introduction}
\label{ssec:subjective-study intuoduction}
Subjective VQA by a sufficiently large enough sample of human subjects is the most reliable method to assess perceptual video quality. 
Subjective VQA studies provide valuable data by obtaining human scores on corpuses of distorted videos. 
These may be very specific, \textit{e.g.}, directed towards distortions arising from frame rate variations, or they may be extremely diverse, as with UGC content. 
In any case, they provide "golden benchmarks" against which to design, assess, and compare the performances of objective VQA models.

Subjective VQA includes the In-lab study and the studies may be broadly divided into two types: those that are conducted under controlled conditions in a laboratory, and those conducted online via crowd-sourcing. 
The former allows for the assessment of quality under constrained conditions, usually by reliable human raters, while the latter is an efficient and successful way of gathering much larger numbers of human annotations of video quality, but with much less control over viewing devices and environment, and participant network conditions and overall reliability.

\begin{figure}[!t]
    \centering
    \includegraphics[width=6.5cm]{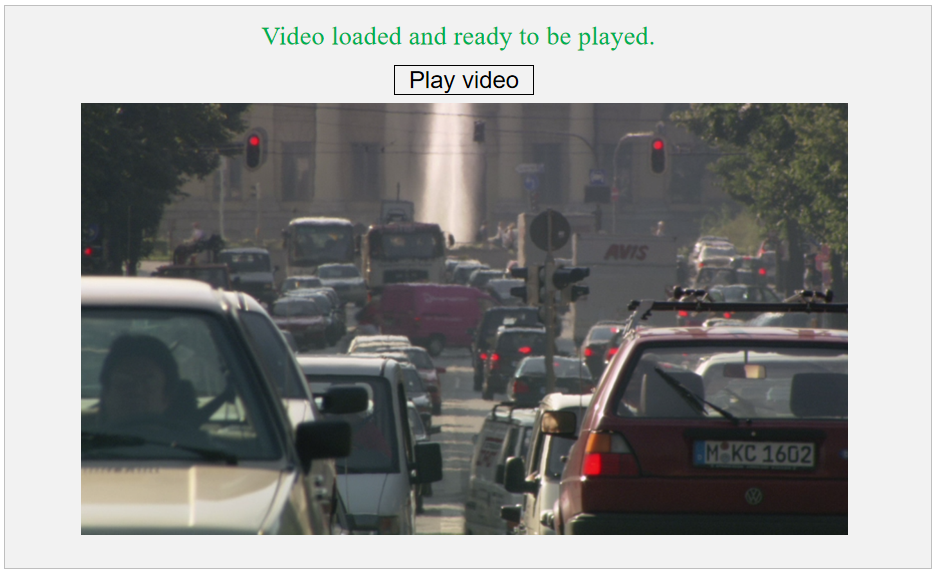}
    \caption{Example of a visual interface used when playing video.}
    \vspace{-8pt}
    \label{fig: playing interface}
\end{figure}

\begin{figure}[!t]
    \centering
    \includegraphics[width=6.5cm]{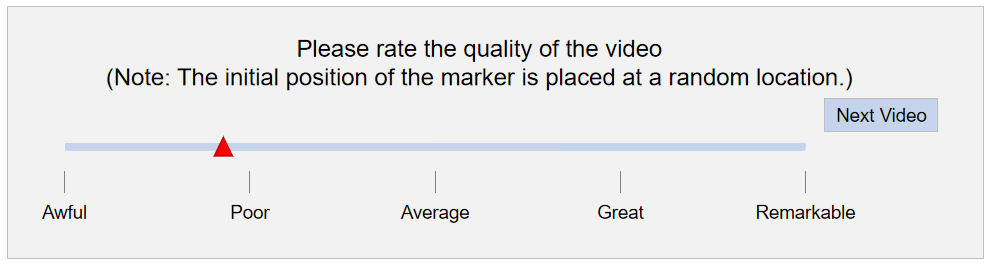}
    \caption{Examplar discrete rating scale.}
    \vspace{-8pt}
    \label{fig: rating interface}
\end{figure}




\subsubsection{\textbf{Data Acquisition}}
\label{data aquisition}
The creators of early VQA databases obtained open-source videos from public websites, willing academic or industry sources, or by capturing their own content. 
Because of copyright of issues, obtaining videos that could be openly shared with others was difficult, and early VQA datasets such as LIVE VQA~\cite{LIVE-VQA} only included about a dozen unique contents. 
Since the main application in those days was the emergent digital television, these contents were uncompressed, cinematic quality videos captured with high-end professional cameras, then converted to digital format with the utmost care, to guarantee that the reference videos would be as distortion free as possible. 
Once a suitable set of high-quality videos has been collected, then a set of representative distortions are applied, taking care to create a wide range of perceived seventies. 
Ideally, the range of distortion exceeds that encountered in the target application, since, without the benefit of very large numbers of human labels (typically only in the tens of thousands in this kind of study), model-building is enhanced by a wider range of less clustered samples. 
The applied distortions include standard-compliant compression~\cite{LIVE-VQA}, wireless transmission artifacts~\cite{moorthy2012video}, frame-rate variations~\cite{9497087}, flicker~\cite{choi2018video}, scaling distortions~\cite{lee2021sub}, bit depth (dynamic range variations)~\cite{ebenezer2024hdr}, and more.

The other major category of VQA datasets are those containing real, unchanged video content that was generated by consumer-level users and then typically shared on social media platforms. 
To gather data that is representative of this UGC data, VQA researchers typically acquire much larger numbers of open-source videos from public platforms like the Internet Archive, and YouTube,~\cite{ying2021patch,LIVE-VQC,ying2020patches}.
If a very large number of videos are collected, these may be statistically sampled based on low-level attributes (such as brightness, colorfulness, temporal activity, etc.) to better match typical social media videos~\cite{ying2021patch},~\cite{ying2020patches}.
In the end, the goal is to acquire a large set of videos, typically numbering in the tens of thousands, that have not been altered in any way (including resizing) that were taken by typical real-world casual users. 
These UGC videos contain very diverse mixtures of distortions.

\begin{figure}[!t]
    \centering
    \includegraphics[width=6.5cm]{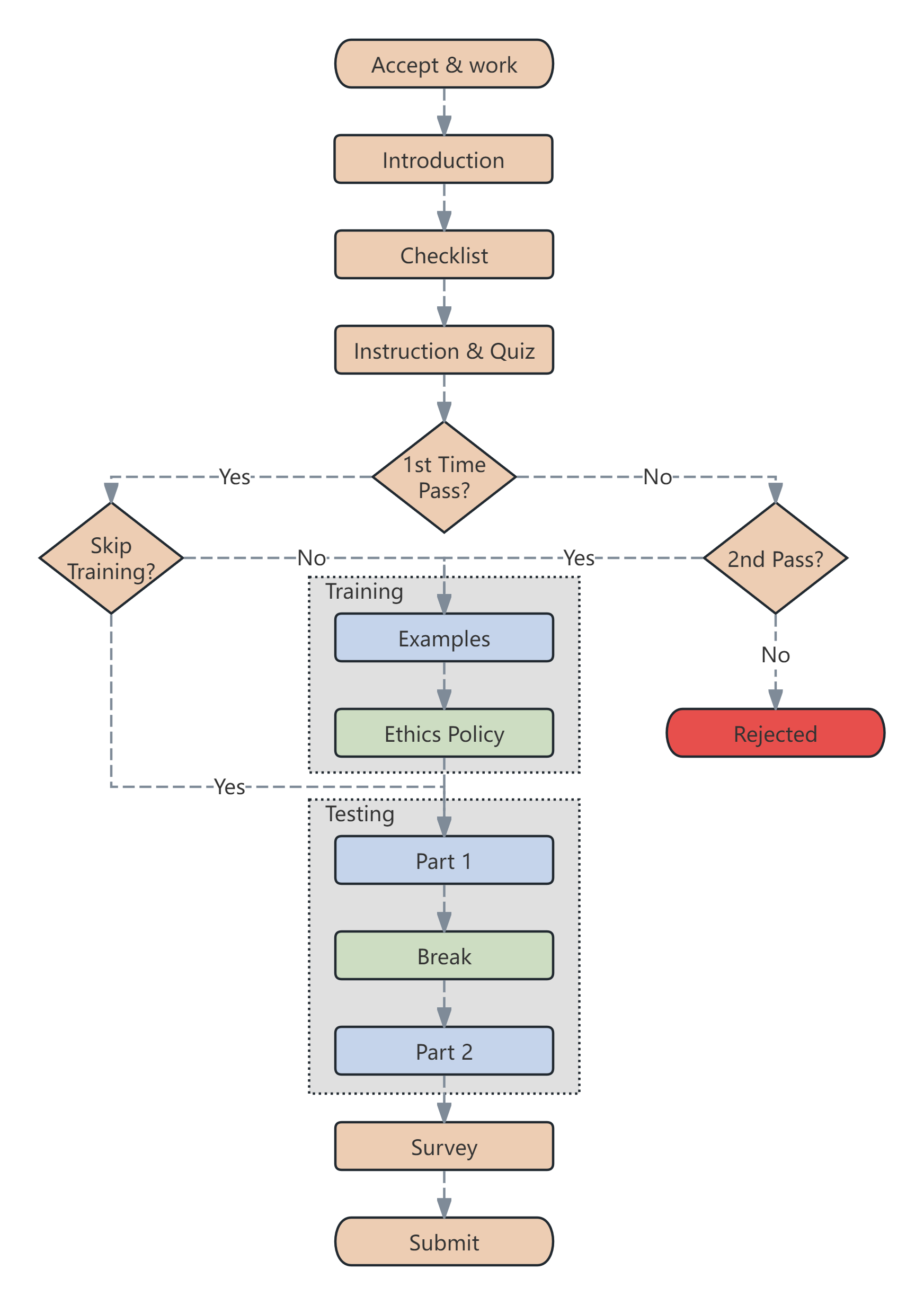}
    \vspace{-5pt}
    \caption{Flowchart of a online crowd-sourcing study~\cite{Telepresence}.}
    \vspace{-12pt}
    \label{fig: study workflow}
\end{figure}

\subsubsection{\textbf{Study Interface Design}}
\label{interface design}
Crowd-sourcing studies using platforms like Amazon Mechanical Turk (https://www.mturk.com/) and CrowdFlower (https://www.crowdflower.com/) are an effective way to gather much larger numbers of human annotations, thereby allowing for the development of data-driven approaches to more general VQA problem scenarios~\cite{LIVE-VQC,Telepresence}.
As shown in Fig.~\ref{fig: playing interface} and Fig.~\ref{fig: rating interface}, each video is played, without display scaling, on a white or gray background.
Once viewed, a rating interface appears that prompts subjects to record their quality scores using a sliding bar. 
While this is happening, next video to be viewed is cached so that it is entirely downloaded before it is played back, to avoid any re-buffering events or stalling that might change a subject's quality judgments. 
Plugins to speed up the videos are also detected and prohibited.

Several survey questions are asked at the beginning of each session, to collect subject demographics and to record any video and audio device specifications that have not already been measured.
Participants whose AMT acceptance rates (known as reliability scores) or equipment specifications do not meet requirements are not allowed to proceed. 
Before their participation starts, each subject is asked to complete a series of steps to eliminate distractions, ensure audio devices are on (if relevant), adjust seating, and wear corrective lenses.

For studies done in the laboratory, MATLAB has been a popular platform for developing user interfaces for human subjective quality studies. 
To further ensure uninterrupted playback, all the displayed sequences are pre-loaded into memory before they begin to play, to avoid any latencies due to slow hard disk access of large video files. 
The videos are viewed by the subjects on a monitor both contemporary to and appropriate for the task at hand, i.e. able to faithfully present the videos at modern resolutions and bit depths reflecting current commercial products, and configurable, if needed, for specific scenarios, such as changing frame rates.
In most instances, the display device should present the videos at their native resolutions, and at higher refresh rates than the native frame rates.
For subjects taking part in these studies, they are generally naive volunteers, and 
usually are asked to participate in a short series of vision tests, including a Snellen test of visual acuity, an Ishihara test of color vision, and if appropriate, a Randot test of 3D vision.

\subsubsection{\textbf{Study Workflow}}
\label{study workflow}
 The general workflows of in-lab and crowd-sourced studies are similar, as depicted in Fig.~\ref{fig: study workflow}. At the beginning, each subject is given a brief set of instructions regarding the task they will participate in. They are then ordinarily shown a few sample videos exhibiting a range of distortions and qualities that they might encounter in the study. 

Methods of video display are generally classified as single, double, or multiple stimulus depending on whether one, two, or several videos are displayed simultaneously or in sequence for each subject to view and quality-rate. 
The single stimulus method is well suited to a large number of emerging multimedia applications, such as quality monitoring for Video on Demand, IPTV, Internet streaming, etc. 
One simple reason is that presenting a single stimulus better reflects real viewing. 
Moreover, it is difficult to compare two videos displayed in space (\textit{e.g.} side by side) or in time (one followed by the other). 
It also significantly reduces the amount of time needed to conduct the study (given a fixed number of human subjects) as compared to a double stimulus study\cite{LIVE-VQA}. 
Nevertheless, there are situations where a double stimulus study may be called for, \textit{e.g.} if the video distortions are very subtle.

After a short training session to allow the participants to become familiar with the interface, each subject is asked to rate a set of videos. 
AVQA studies normally use discrete scales such as the Absolute Category Rating (ACR), Degradation Category Rating (DCR), and Comparison Category Rating (CCR)\cite{MCL-V} scales to record quality judgments, while others use a continuous rating scale to allow the subjects to record with greater freedom and increased sensitivity. 
To avoid fatigue, the experiment is usually divided into several sessions of short durations (typically 30-45 minutes) with breaks.
Subjects who give very similar quality scores to all videos are disqualified. 
A wide variety of protocols are used to detect and eliminate the scores of subjects who are distracted, inattentive, or afterwise poorly participating.

\begin{table*}[]
\centering
\setlength{\tabcolsep}{3pt}
\scriptsize
\caption{Taxonomy of subjective VQA databases: legacy databases with synthetic distortions against UGC databases with large-scale in-the-wild videos. Legacy databases and UGC databases are indicated by \textit{italic} and orthographic font, respectively.}
\label{tab:pgc_ugc_databases}
\resizebox{\textwidth}{!}{%
\begin{tabular}{lllllllllllll}
\toprule
Year & Name & \begin{tabular}[c]{@{}c@{}}Unique\\ contents\end{tabular} & \begin{tabular}[c]{@{}c@{}}Total\\ videos\end{tabular} & Resolution & \begin{tabular}[c]{@{}c@{}}Frame\\ Rate\end{tabular} & \begin{tabular}[c]{@{}c@{}}Video\\ Length(s)\end{tabular} & Format & \begin{tabular}[c]{@{}c@{}}Distortion\\ Type\end{tabular} & Subjects & \begin{tabular}[c]{@{}c@{}}Rates per \\video\end{tabular} & Data & Env \\ \midrule
2008 & \textit{LIVE-VQA}~\cite{LIVE-VQA} & 10 & 150 & 786×432 & 25/50 & 10 & YUV+264 & Compression, Transmission & 38 & 29 & DMOS+$\sigma$ & In-lab \\
2014 & \textit{CVD2014}~\cite{CVD2014} & 5 & 234 & 720p, 480p & 9-30 & 10-25 & AVI & Camera, Capture & 210 & 30 & MOS & In-lab \\
2015 & \textit{MCL-V}~\cite{MCL-V} & 12 & 108 & 1080p & 24-30 & 6 & YUV+264 & Compression, Scaling & 45 & 32 & MOS & In-lab \\ 
2015 & \textit{BVI-HFR}~\cite{mackin2018study} & 22 & 88 & 1080p & 15-120 & 10 & YUV420& Temporal downsampling & 29 & 29 & MOS & In-lab \\
2021 & \textit{LIVE-YT-HFR}~\cite{madhusudana2021subjective} & 16 & 480 & 2160p, 1080p & 24-120 & 6-8 & YUV420& \begin{tabular}[c]{@{}l@{}}Compression, \\ Temporal downsampling\end{tabular}& 85 & 223 & MOS, DMOS & In-lab \\
2022 &\textit{BVI-VFI}~\cite{10304617} & 36 & 540 & 540p & 30-120 & 5 & MP4 & Frame interpolation & 189 &  57& DMOS & In-lab \\\midrule
2017 & KoNViD-1k~\cite{Konvid-1k} & 1200 & 1200 & 540p & 24-30 & 8 & MP4 & In-the-wild & 642 & 114 & MOS+$\sigma$ & Crowd \\
2018 & LIVE-VQC~\cite{LIVE-VQC} & 585 & 585 & 1080p-240p & 19-30 & 10 & MP4 & In-the-wild & 4776 & 240 & MOS & Crowd \\
2019 & YouTube-UGC~\cite{Youtube-UGC} & 1380 & 1380 & \begin{tabular}[c]{@{}l@{}}4k(HDR)-360p\end{tabular} & 15-60 & 20 & MKV & In-the-wild & \textgreater{}8k & 123 & MOS+$\sigma$ & Crowd \\
2019 & FlickrVid-150k~\cite{Flickrvid-150k} & 153841 & 153841 & 540p & 24-120 & 5 & MPEG-4 & In-the-wild & - & 5 (89) & MOS & Crowd \\
2021 & LSVQ~\cite{ying2021patch} & 38811 & 38811 & Diverse & Diverse & 5-12 & - & In-the-wild & 6300 & 35 & MOS & Crowd \\
2021 & Youku-V1K~\cite{Youku-v1k} & 1072 & 1072 & 1080p & - & 10 & - & UGC, PGC & - & 15+ & MOS+$\sigma$ & Crowd \\
2021 & PUGCQ~\cite{PUGCQ} & 10k & 10k & $\leq$1080p & - & 5 & MP4 & Professional UGC & 50 & 50 & MOS & Crowd \\
2021 & YT-UGC+~\cite{YT-UGC+} & 1380 & 1380 & \begin{tabular}[c]{@{}l@{}}4k(HDR)-360p\end{tabular} & 15-60 & 20 & MKV &UGC, Compression & \textgreater{}8k & 10(labels) & DMOS, MOS+$\sigma$ & Crowd \\
2021 & LIVE-YT-Gaming~\cite{YT-Gaming} & 600 & 600 & 1080p-360p & 30/60 & 8-9 & MP4 & UGC gaming & 61 & 3 & MOS & Crowd \\
2022 & Tele-VQA~\cite{Telepresence} & 2320 & 2320 & - & - & 7 & - & In-the-wild & 526 & 34 & MOS & Crowd \\
2023 & Maxwell~\cite{10.1145/3581783.3611737} & 4543 & 4543 & 240p-1080p & - & 9 & MP4 & In-the-wild & 35 & 440 & \begin{tabular}[c]{@{}l@{}}MOS, \\ Ternary choices \end{tabular} & In-lab\\ 
2023 & DIVIDE-3k~\cite{Wu_2023_ICCV} & 3590 & 3590 & 240p-1080p & 24-30 & 2-13 & MP4 & In-the-wild & 35 & 125 & MOS+$\sigma$ & In-lab\\
2023 & TaoLive~\cite{Zhang_2023_CVPR} & 418 & 3762 & 720p, 1080p & - & 8 & MP4 & UGC, Compression  & 44 & 44 & MOS & In-lab\\
2024 & KVQ~\cite{Lu_2024_CVPR} & 600 & 4200 & - & - & - & MP4 & \begin{tabular}[c]{@{}l@{}}UGC, Enhancement,\\Pre-processig, Transcoding \end{tabular} & 15 & 15 & \begin{tabular}[c]{@{}l@{}}MOS, \\ Ranking score \end{tabular} & In-lab \\
\bottomrule
\end{tabular}%
}
\vspace{-12pt}
\end{table*}

\subsubsection{\textbf{Crowd-sourcing Quality Control}}
\label{crowd-sourcing quality control}
When conducting online crowd-sourcing studies, quality control is essential to obtaining reliable labels. 
While there is a great advantage to access to many more subjects (often orders of magnitude more), the data is much noisier unless precautions are taken to detect and remove poorly connected or dishonest ``workers".

One common practice is to include “golden” videos selected from other databases featuring similar content, on which highly reliable subjective scores were previously obtained. 
These scores are compared with the worker’s inputs to determine if there are wide enough divergences. 
In this way, many unreliable subjects may be prevented from further participation. 
For more details on these and other methods of improving online crowdsourced data reliability in this context, see~\cite{ying2021patch,Telepresence,ying2020patches}.
Since some subjects are less serious, distracted, or otherwise poorly focused on their tasks, it is important to apply subjection rejection protocols such as ITU recommendations BT.500~\cite{Telepresence}, BT.500.11~\cite{BT.500.11}, BT.500.15, or the more accurate and comprehensive SUREAL~\cite{li2017recover}.

Another practice is to select a small random set of videos which are repeated elsewhere in a subject's session as a control. 
Again, failure of a subject to produce consistent scores on these may lead to their elimination. 
Some users may have inadequate hardware to receive and display the videos, or poor/low bandwidth connectivity. 
While the latter may be ameliorated by requiring each video to be completely downloaded before display, the session may be terminated with a polite explanation, since the fault is not the workers.
Since Amazon Mechanical Turk includes tools to measure technical aspects of a user's device or connectivity, these are important and practical precautions.

\subsubsection{\textbf{Subjective Analysis}}
\label{subjective recovery analysis}
To establish the internal integrity of the final set of collected subjective scores, it is important to examine the consistency of the recorded Mean Opinion Scores (MOS).
This is commonly done by randomly dividing the subjects pool into two equal and disjoint sets, then computing the Spearman Rank Correlation Coefficient (SRCC) between the two corresponding sets of MOS.

Repeating this random calculation a reasonable number of times, then taking the average (and/or median) SRCC as a consistency measure, is useful for assessing the difficulty of the task and/or the degree of agreement of the subjects. Values above SRCC=0.85 are generally desirable, while SRCC=0.95 is only occasionally reached (on easier VQA tasks). 
Additional subject rejection methods, especially useful for large online studies where the subjects can be less reliable, include differences between each subject's MOS and those obtained from a known, reliable set of subjects (perhaps recorded in the laboratory), on a set of ``golden'' videos~\cite{ying2021patch, LIVE-VQC}.
Another method is to present a small number of videos (typically 5) twice within a subject's session, but spaced in time, then determine whether they give sufficiently similar responses on repeated videos as on the first presentations of those contents.

In addition to MOS, difference mean opinion scores (DMOS)~\cite{5404314,9665209} are also commonly recorded in single stimulus experiments using hidden reference removal~\cite{10.1117/12.509908}. In each session, a difference score on each video rated by each subject is computed by subtracting the quality score from the corresponding reference quality score. The difference scores of reference videos are zero and are removed. Z-scores~\cite{10.1117/12.201231} are then computed to normalize the difference scores from each session, then are collected over all sessions to form a matrix indexed by subjects and videos. Finally, DMOS is obtained by linearly rescaling Z-scores to the range of $[0,100]$.

\subsection{Profession Content Datasets}
\label{legacy datasets}
Most early VQA databases were directed towards television and streaming, and contain original, professionally generated studio content which is subsequently distorted under laboratory conditions. 
These datasets generally comprise 10-20 unique high-quality source videos, processed in a specific manner to simulate one of a few synthetic impairments (\textit{e.g.} MPEG encoding, or packet loss), wherein MOS (or DMOS) are obtained from a relatively small (20-40) group of subjects in a controlled laboratory environment. 
These datasets are limited in representation of the complex, highly variable characteristics of videos.

\begin{figure*}[!t]
    \centering
    \subfigure[LIVE-VQA]{
    \label{fig.live-vqa}
    \includegraphics[height=3cm]{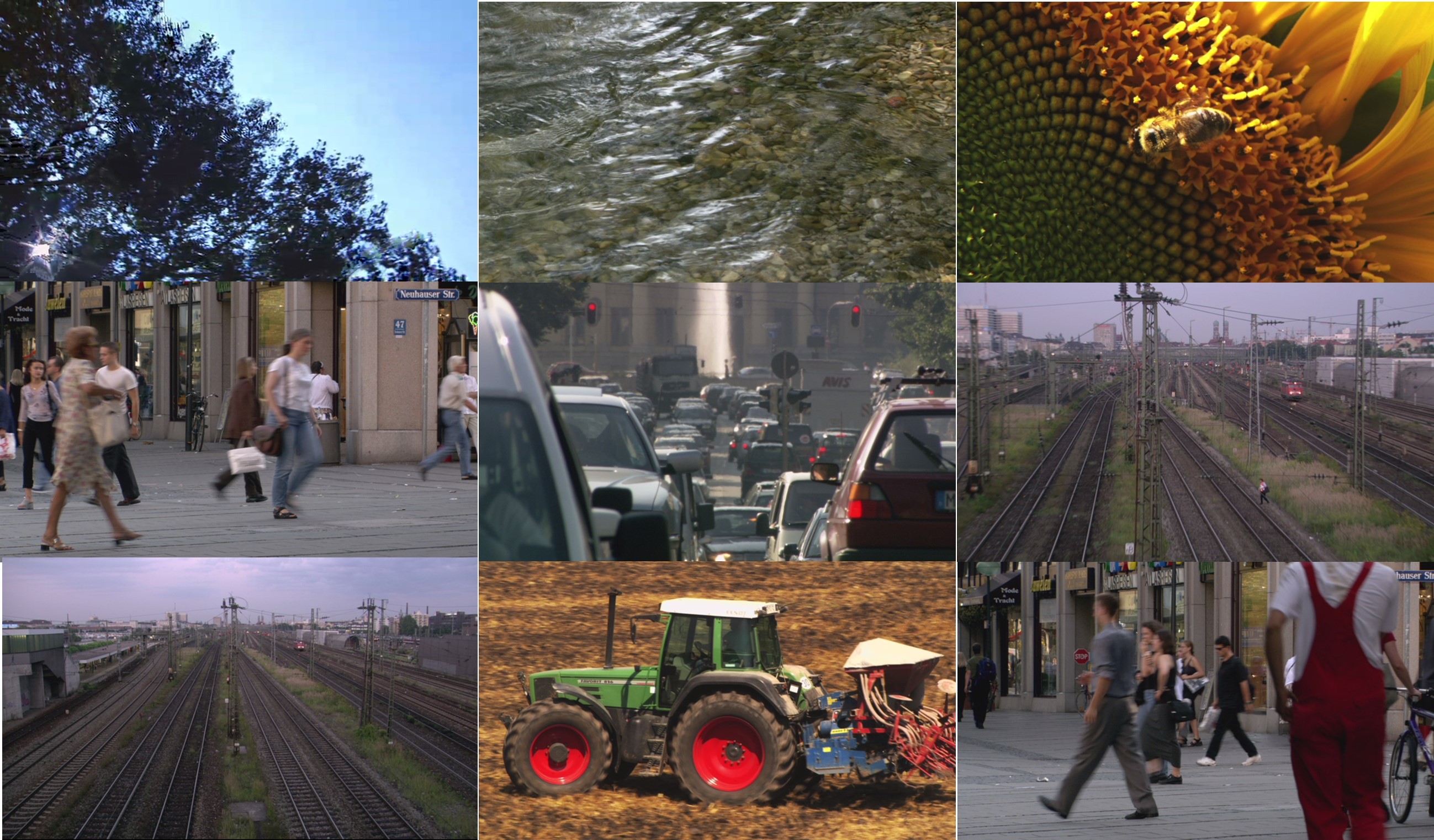}}
    \subfigure[CVD2014]{
    \label{fig.cvd2014}
    \includegraphics[height=3cm]{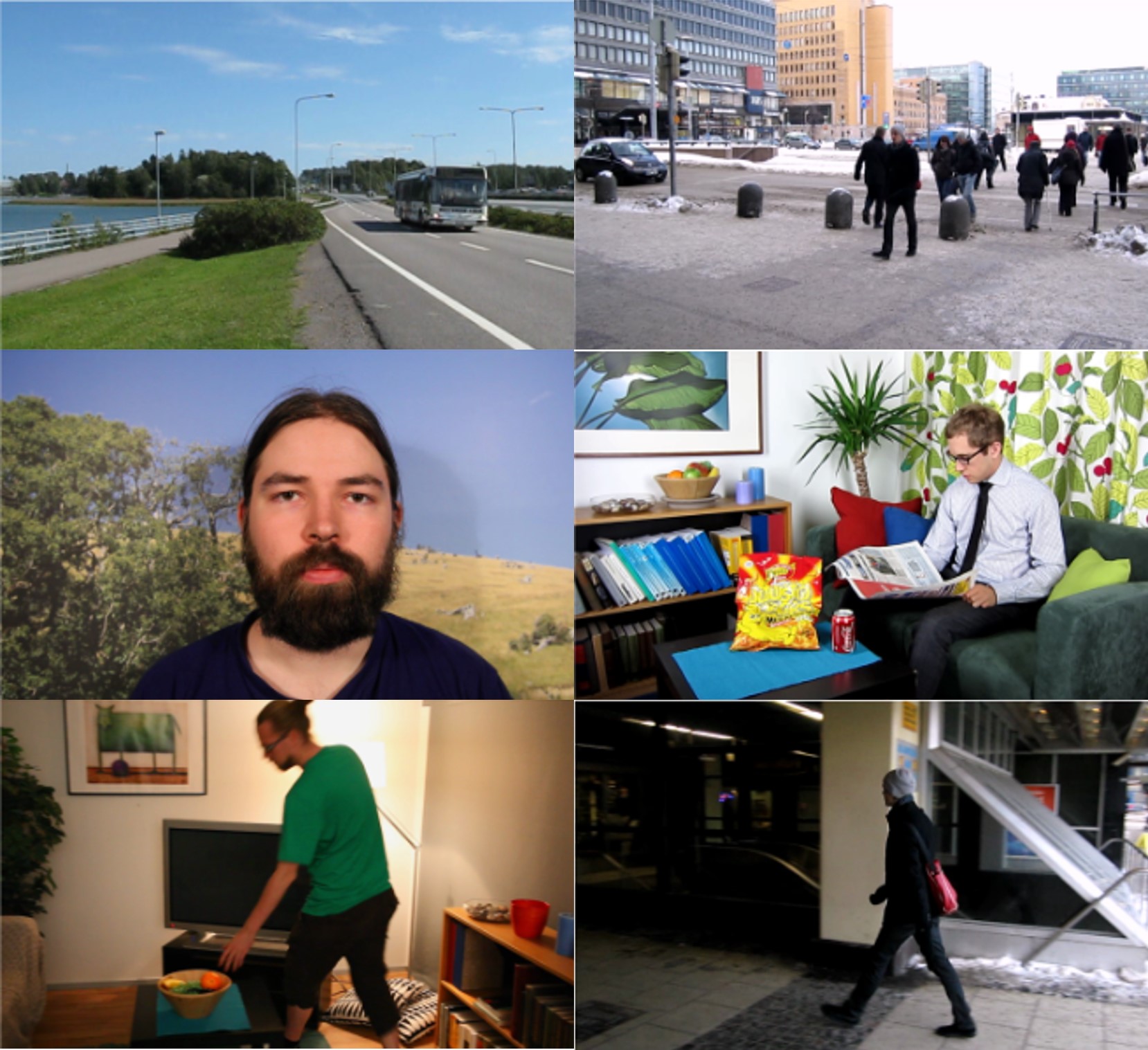}}
    \subfigure[MCL-V]{
    \label{fig.mcl-v}
    \includegraphics[height=3cm]{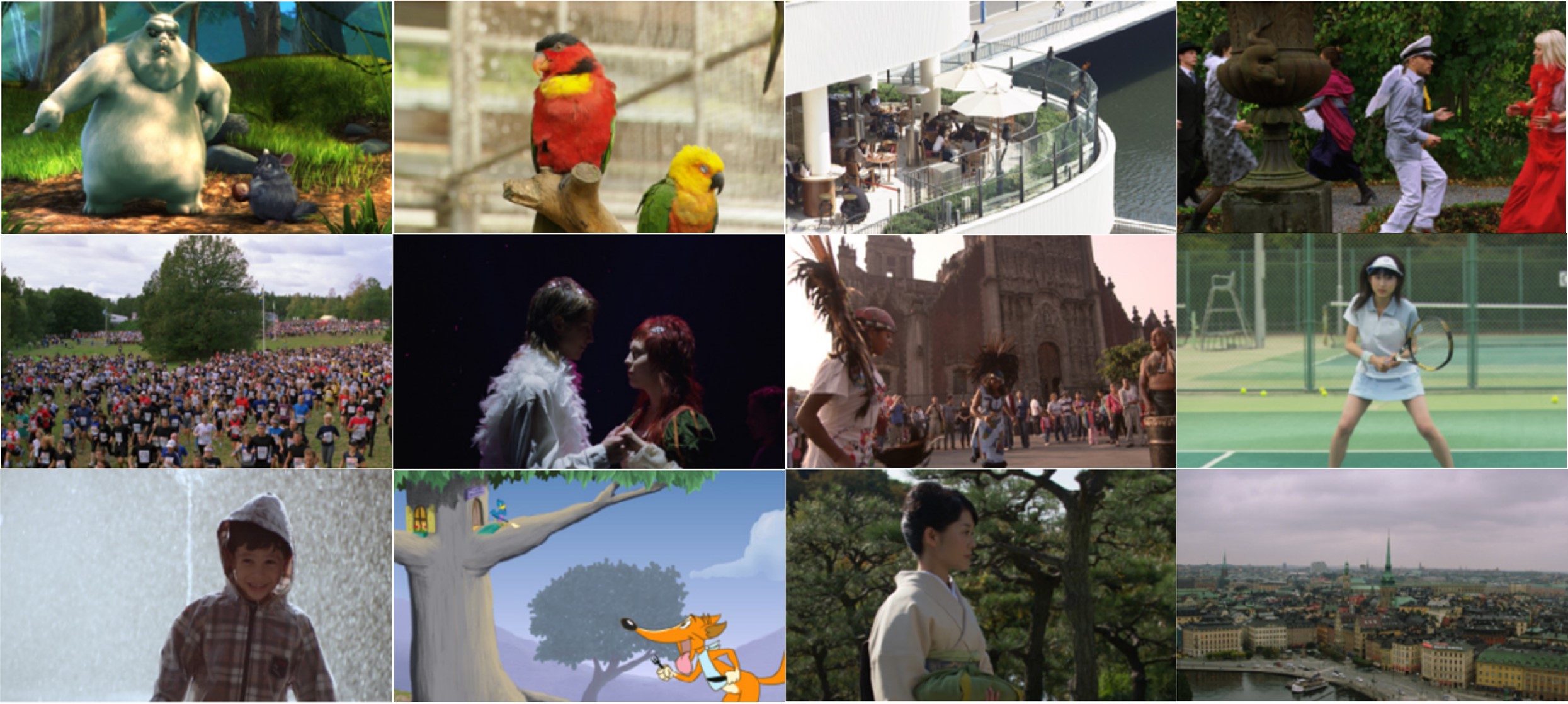}}\\
    \subfigure[LIVE-YT-HFR]{
    \label{fig.live-hfr}
    \includegraphics[height=3cm]{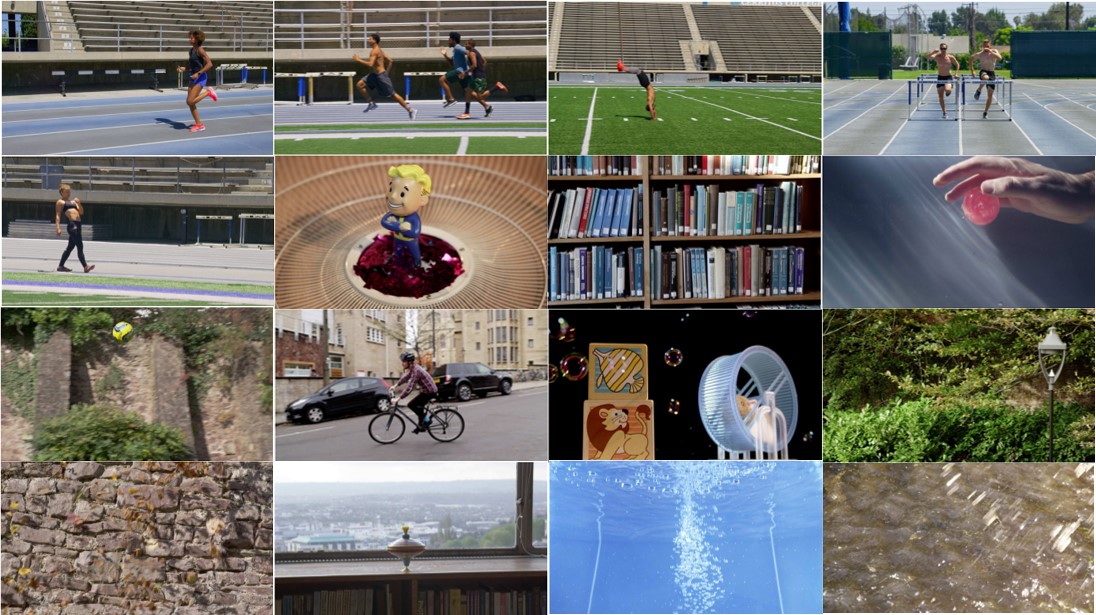}}
    \subfigure[BVI-HFR]{
    \label{fig.bvi-hfr}
    \includegraphics[height=3cm]{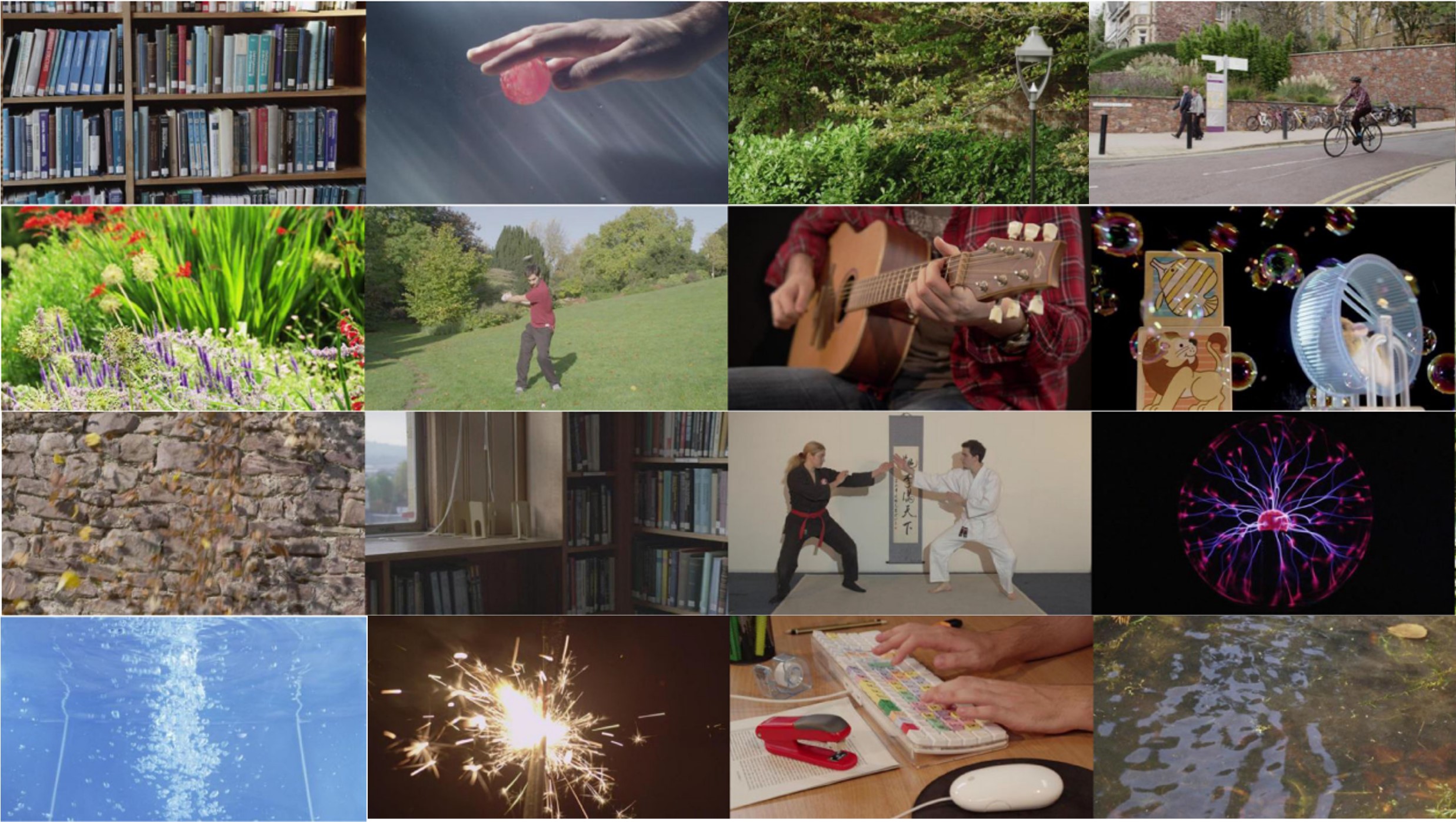}}
    \subfigure[BVI-VFI]{
    \label{fig.bvi-vfi}
    \includegraphics[height=3cm]{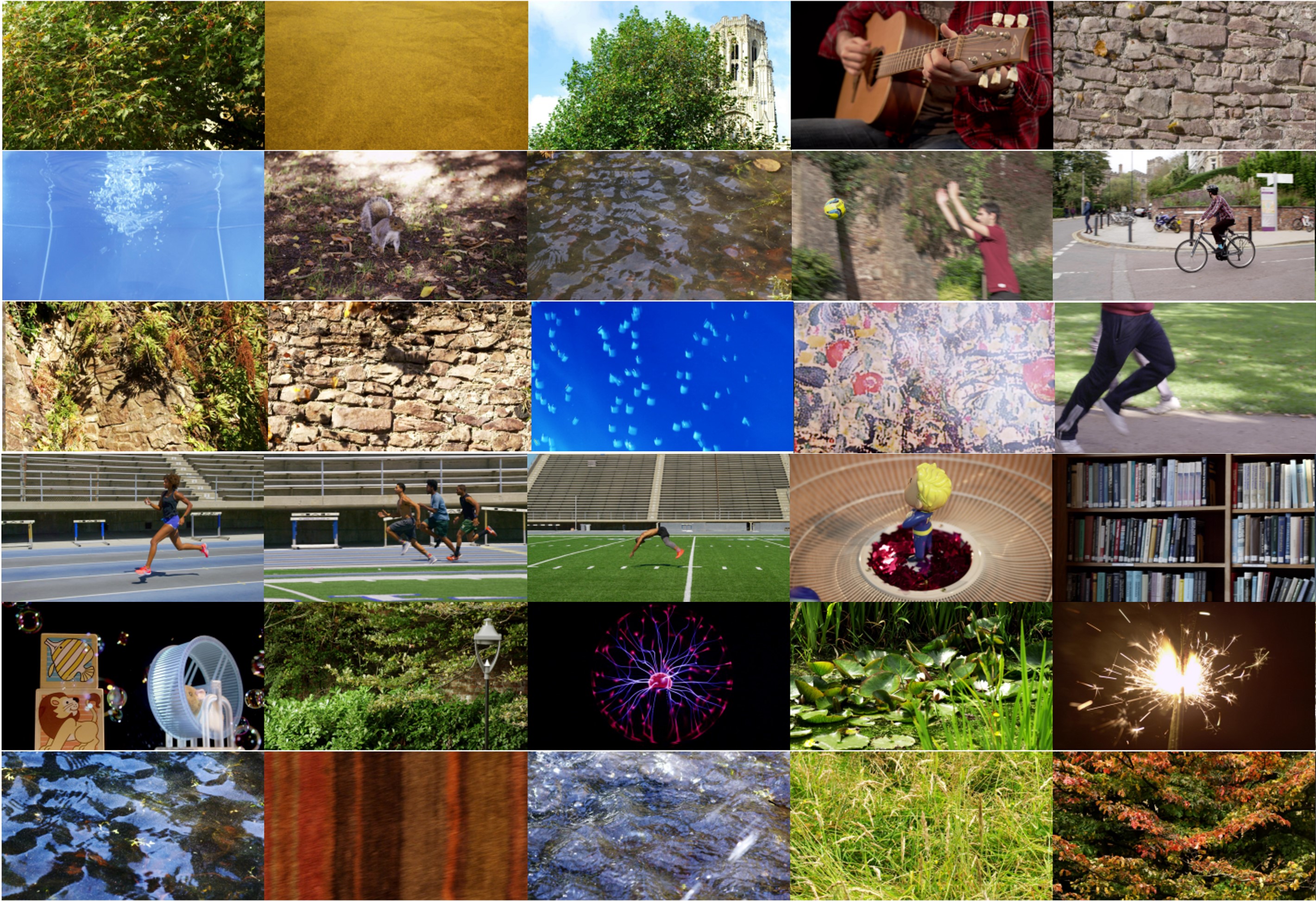}}
    \\
    \subfigure[KoNViD-1k]{
    \label{fig.konvid-1k}
    \includegraphics[height=3cm]{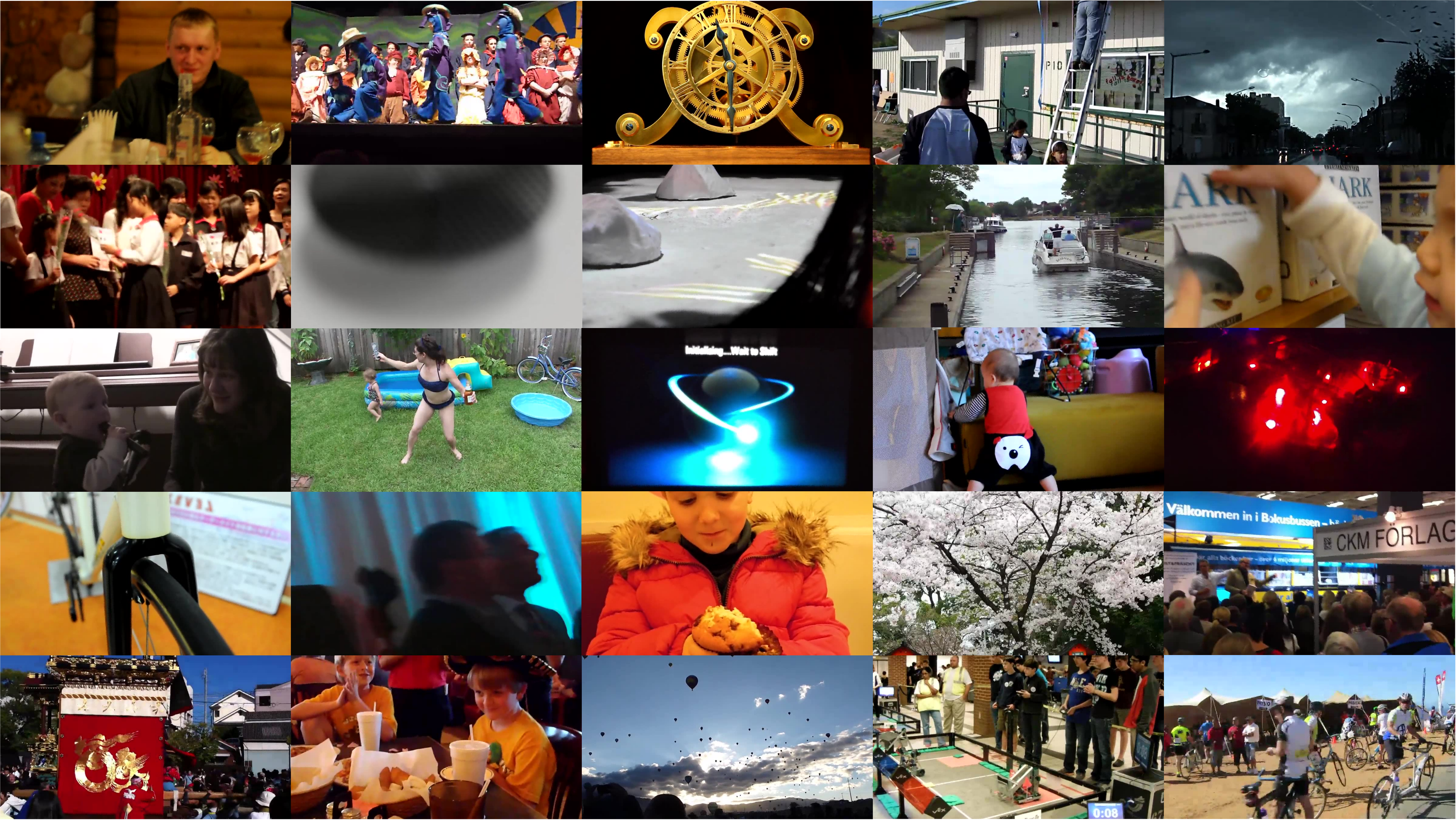}}
    \subfigure[LIVE-VQC]{
    \label{fig.live-vqc}
    \includegraphics[height=3cm]{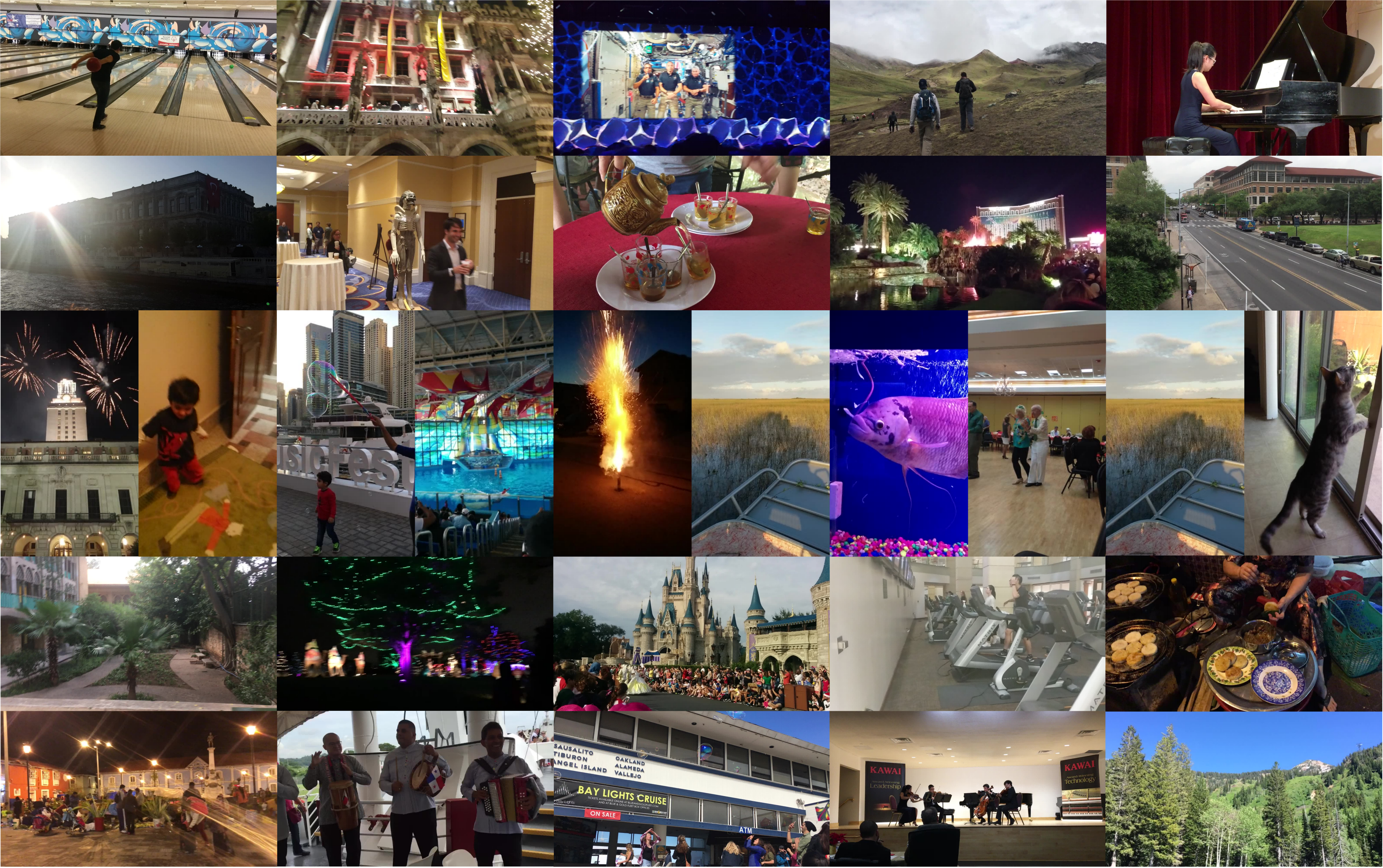}}
    \subfigure[YouTube-UGC]{
    \label{fig.youtube-ugc}
    \includegraphics[height=3cm]{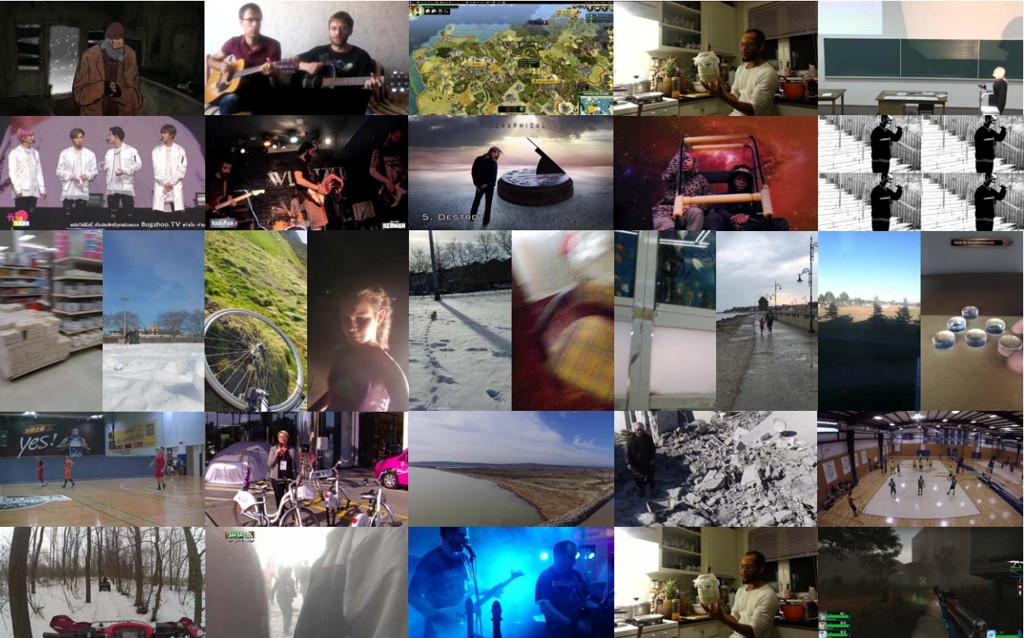}}
     \\
    \subfigure[LIVE-YT-Gaming]{
    \label{fig.yt-gaming}
    \includegraphics[height=3cm]{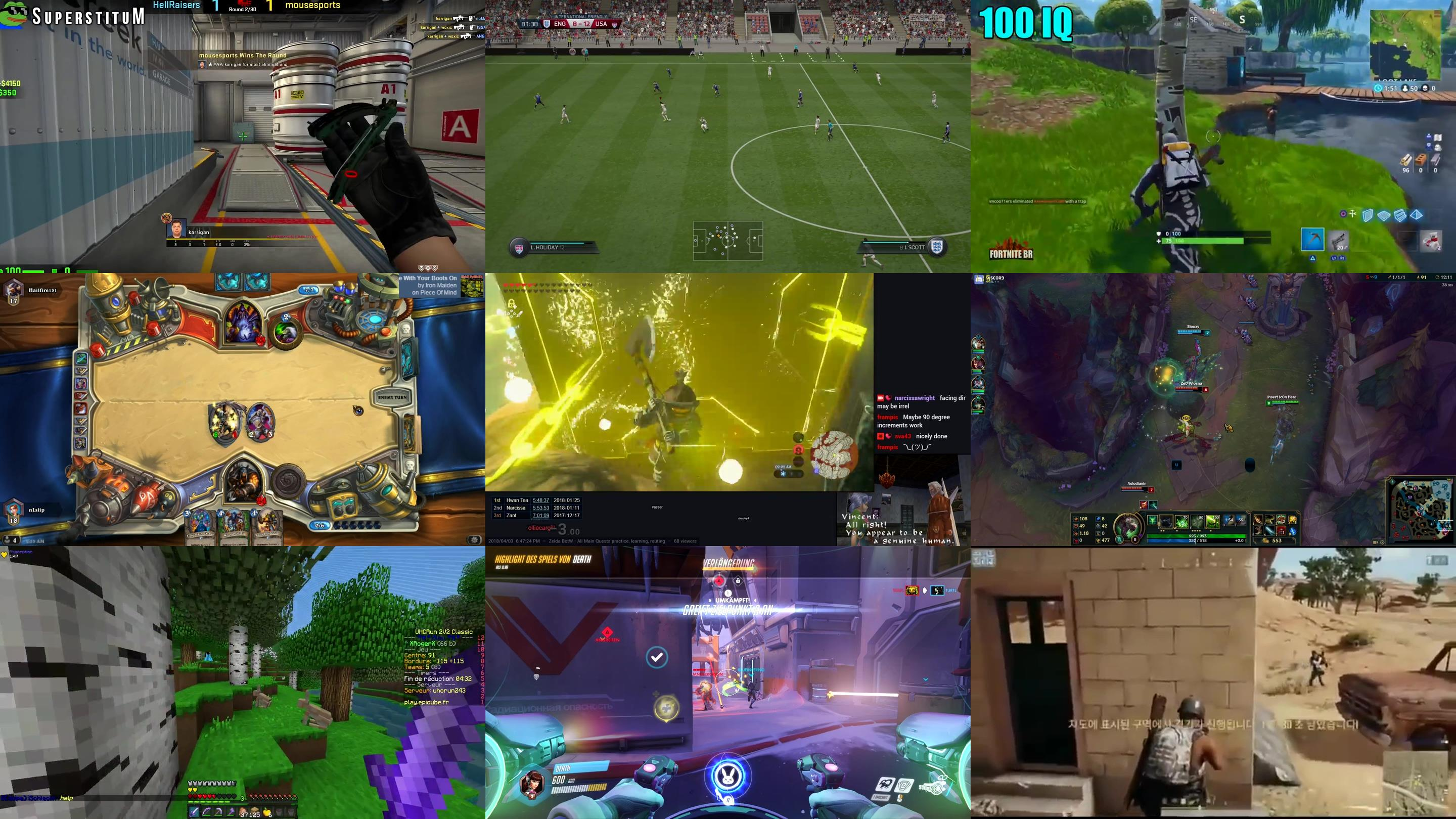}}
    \subfigure[TaoLive]{
    \label{fig.taolive}
    \includegraphics[height=3cm]{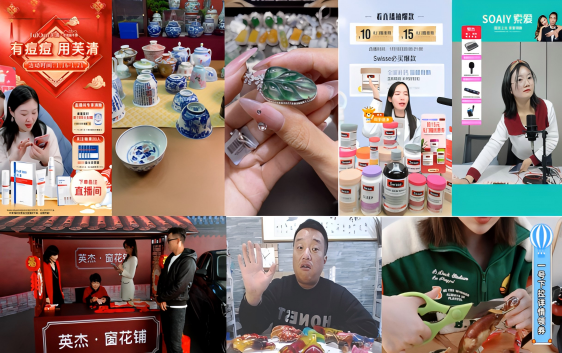}}
    \subfigure[KVQ]{
    \label{fig.kvq}
    \includegraphics[height=3cm]{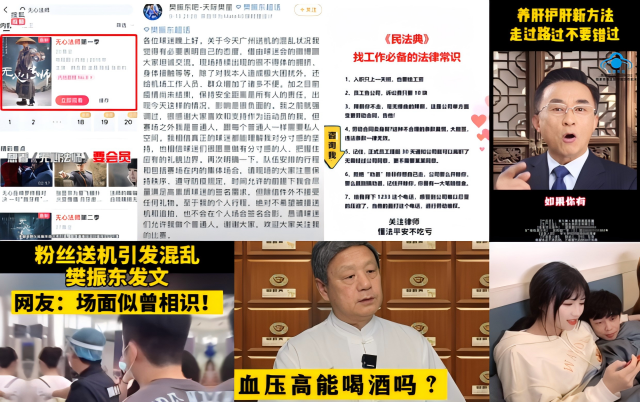}}
    \\
    \subfigure[T2VQA-DB]{
    \label{fig.t2vqa-db}
    \includegraphics[height=3cm]{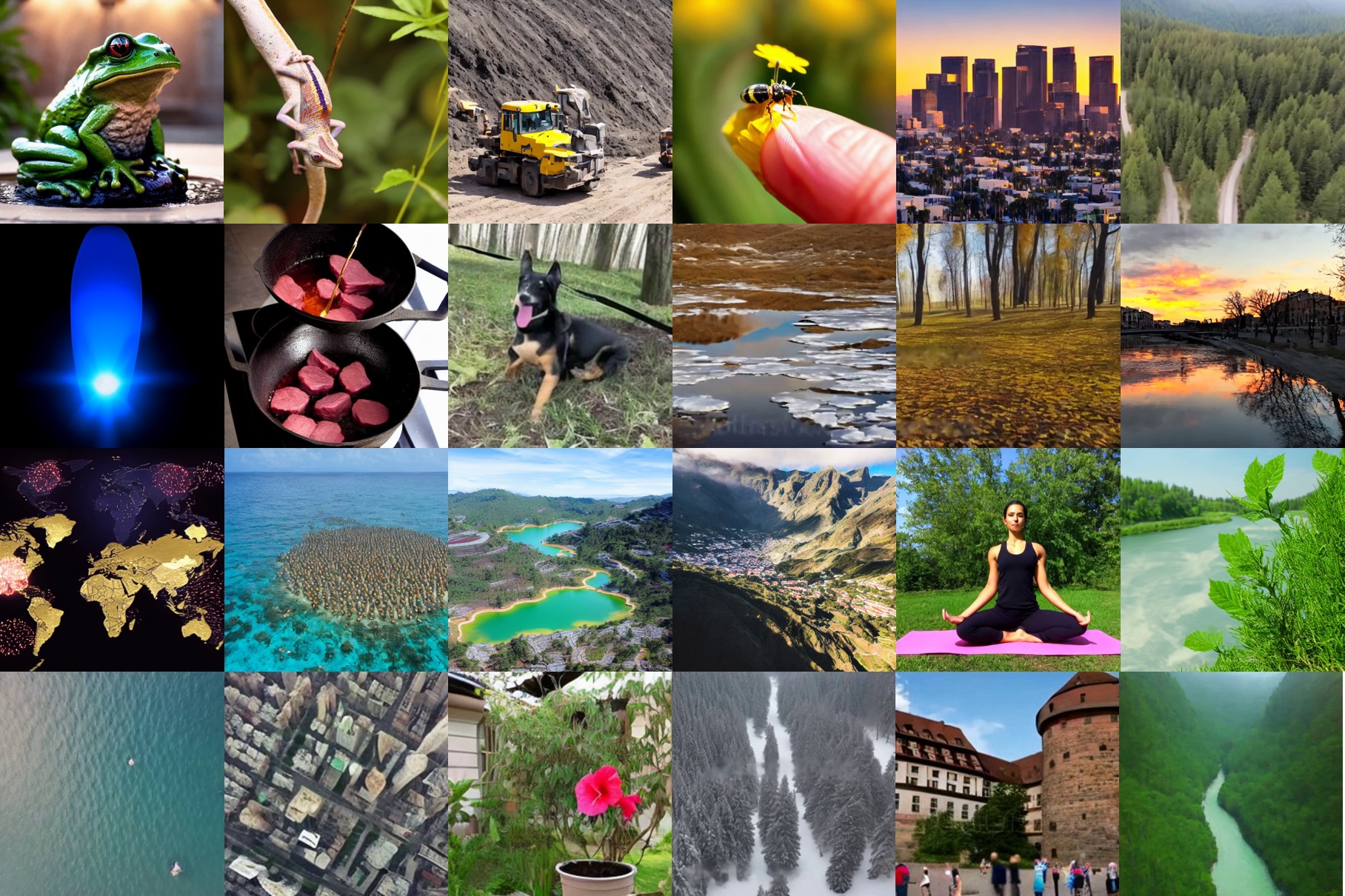}}
    \subfigure[GAIA]{
    \label{fig.gaia}
    \includegraphics[height=3cm]{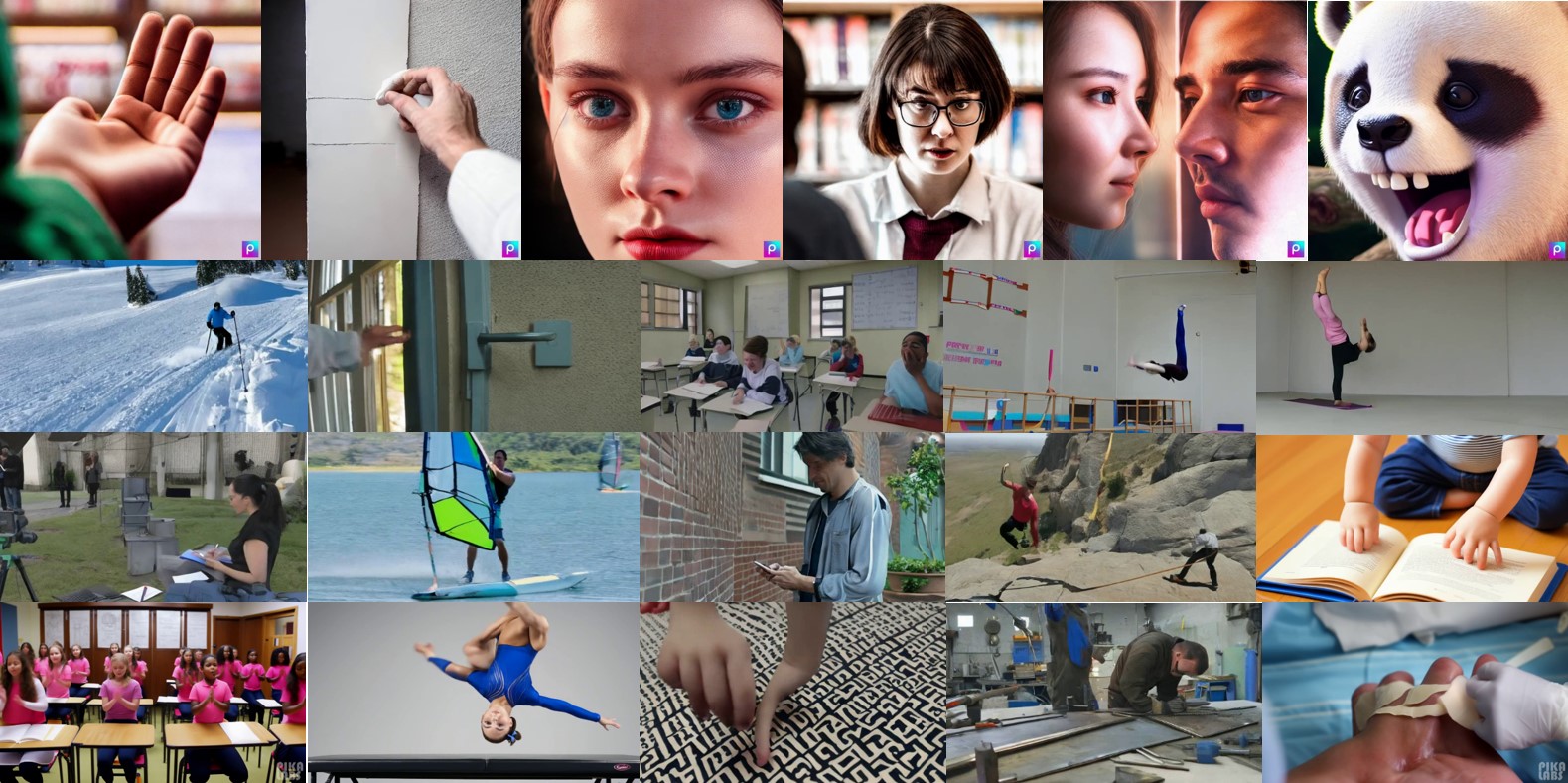}}
    \subfigure[FETV]{
    \label{fig.fetv}
    \includegraphics[height=3cm]{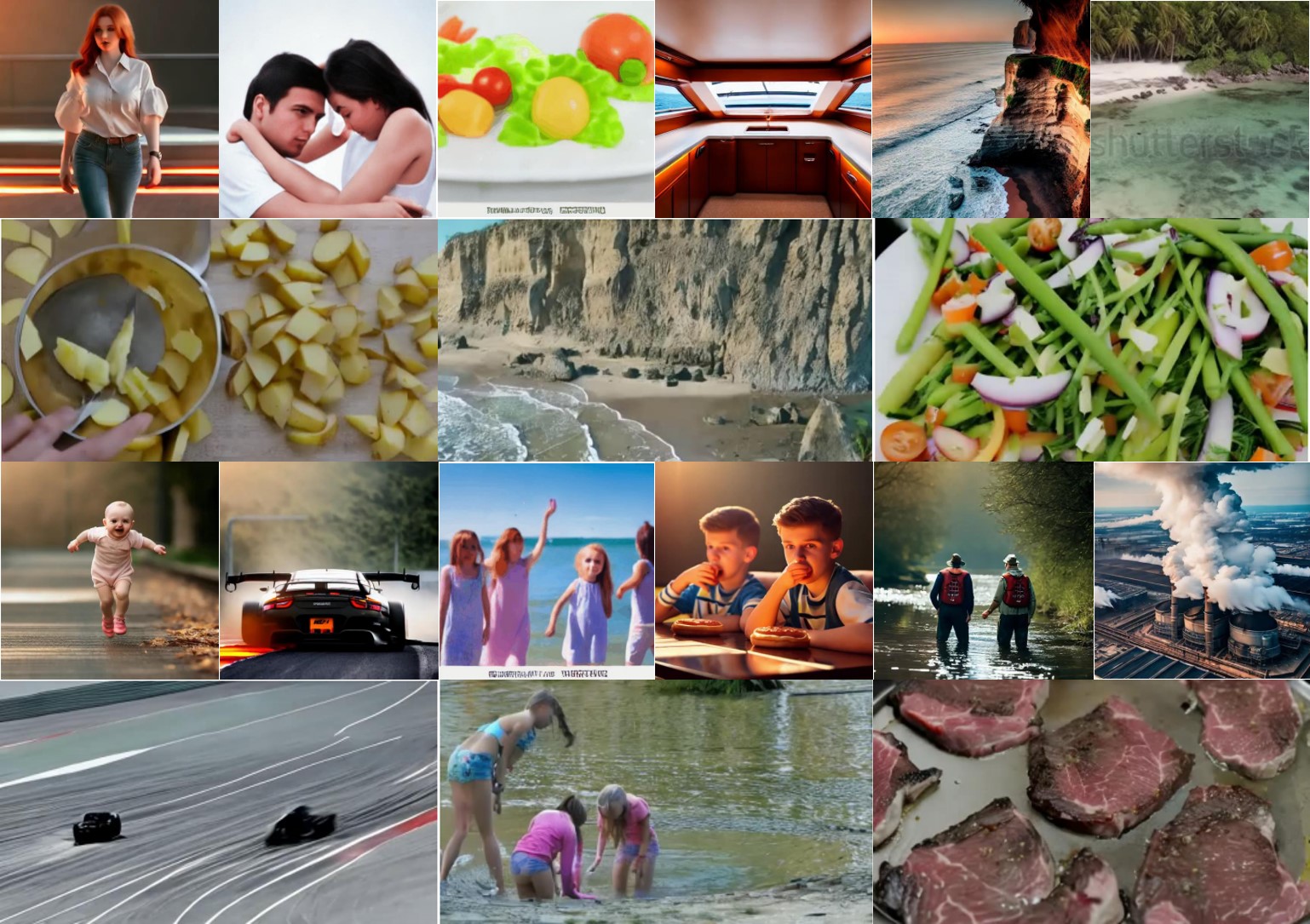}} \\
    \subfigure[EvalCrafter]{
    \label{fig.evalcrafter}
    \includegraphics[height=3cm]{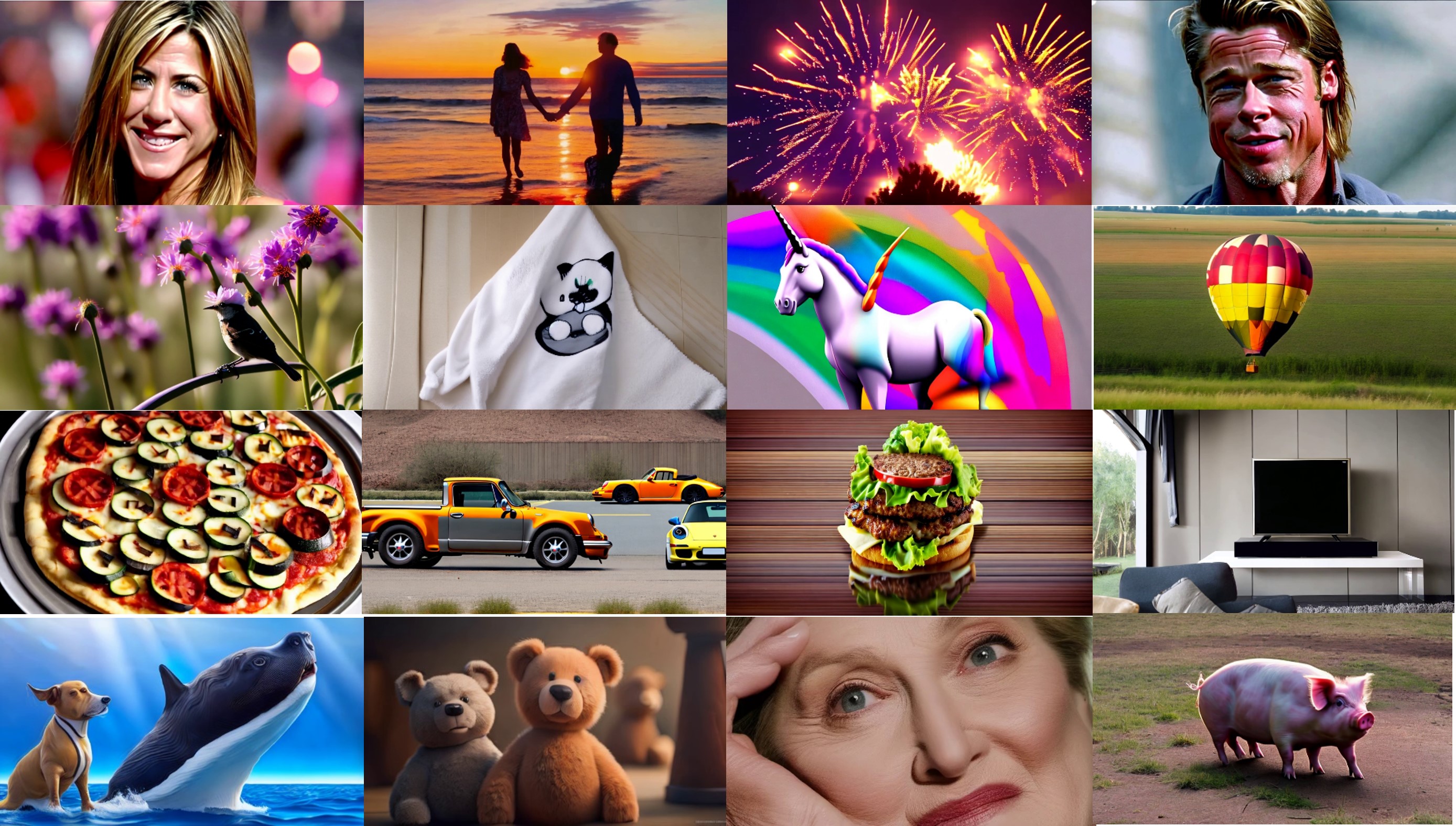}}
    \subfigure[VBench]{
    \label{fig.vbench}
    \includegraphics[height=3cm]{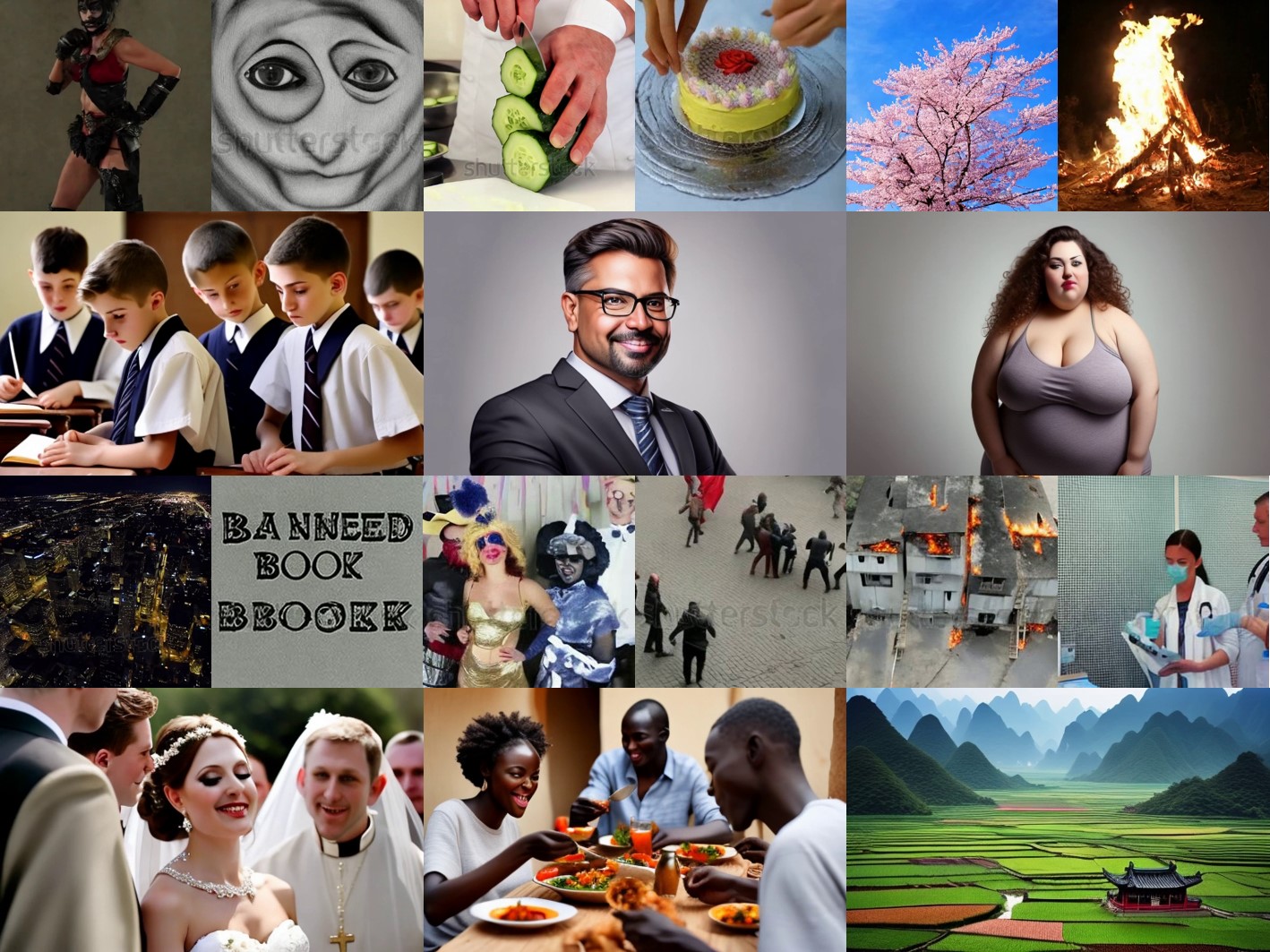}}
    \subfigure[Chivileva et al.]{
    \label{fig.fetv}
    \includegraphics[height=3cm]{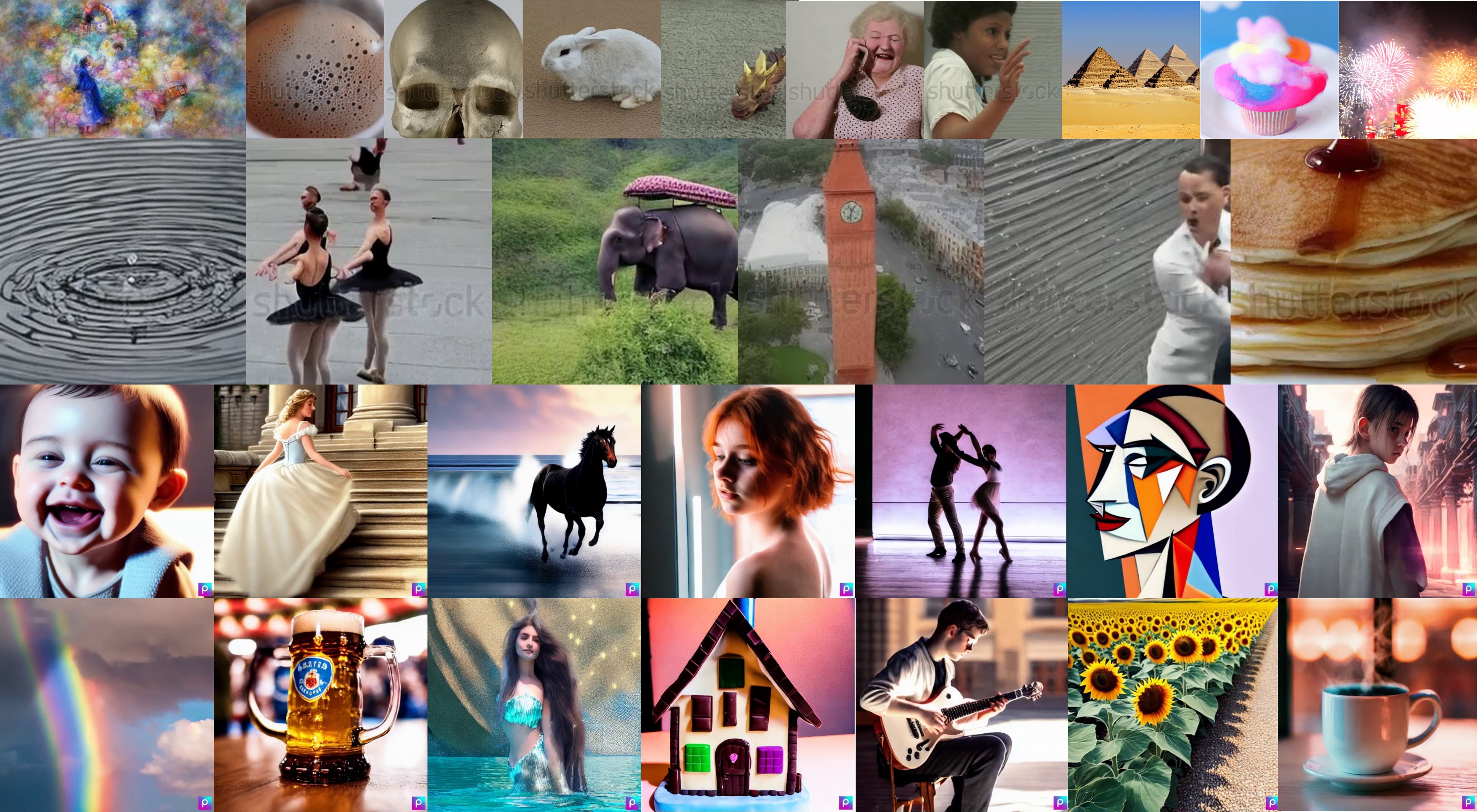}}
    \caption{Samples of video contents selected from legacy datasets ((a)-(f)), UGC datasets ((g)-(l)), and AIGC datasets ((m)-(r)).}
    \label{fig:sample-frames-legacy}
\end{figure*}

Table~\ref{tab:pgc_ugc_databases} summarizes and classifies several of the most widely-used and successful public VQA databases created during the past 15 years. 
Fig.~\ref{fig:sample-frames-legacy} (a)-(f) shows sample frames taken from six popular legacy databases.
The first useful VQA database was the LIVE Video Quality Database\cite{LIVE-VQA} published in 2010, which includes 10 reference videos and 150 videos simulating compression and transmission distortions. 
Other databases containing original and artificially-distorted videos include such as CVD2014\cite{CVD2014}, MCL-V\cite{MCL-V}. 
As video dimensions have increased in other ways than spatial resolution, other video quality databases have become available that contain subjectively labeled high motion~\cite{shang2021study}, high frame rates (HFR)~\cite{9497087,mackin2018study}, high dynamic range (HDR)~\cite{10325414}, and frame interpolation~\cite{10304617} videos.

\subsubsection{\textbf{LIVE-VQA}}
\label{live-vqa}
The LIVE Video Quality Database\cite{LIVE-VQA} consists of 10 uncompressed reference videos and 150 distorted videos created using four common streaming degradations, such as compression and packet loss.
The videos in LIVE-VQA were all in 720P formats and span a wide range of visual qualities.
Each video was rated by 38 human subjects in a single stimulus presentation.
The subjects scored the video quality on a continuous quality scale.
\subsubsection{\textbf{CVD2014}}
\label{cvd2014}
This database\cite{CVD2014} contains 234 videos from 78 cameras without post-processing distortions. Instead of only MOS, it includes quality descriptions from participants, covering dimensions like sharpness, graininess, color balance, darkness, and jerkiness.
\subsubsection{\textbf{MCL-V}}
\label{mcl-v}
The MCL-V dataset\cite{MCL-V} includes 12 uncompressed HD video clips with diverse content (\textit{e.g.}, cartoons, faces, action), in YUV420p format, 1080p resolution, and frame rates of 24fps or 30fps. These 6-second clips were distorted using H.264/AVC compression and scaling, resulting in 96 videos. Ratings from 45 subjects using a pairwise comparison method were converted into absolute scores using the Bradley-Terry model.

\subsubsection{\textbf{BVI-HFR}} 
This database~\cite{mackin2018study} examines the impact of high frame rates on video quality. It includes 22 4K, 120fps source sequences, downsampled to 1080p and frame rates of 60fps, 30fps, and 15fps, resulting in 88 videos. Quality ratings were collected from 29 participants in a controlled in-lab study.

\subsubsection{\textbf{LIVE-YT-HFR}} 
LIVE-YT-HFR~\cite{madhusudana2021subjective} studies the combined effects of compression artifacts and frame rate variation, and contains 480 videos from 16 unique contents distorted by combinations of 6 frame rates (24, 30, 60, 82, 98, 120fps) and 5 compression levels (CRF 0 - 63). An in-lab human study was conducted to yield 19,000 human quality ratings obtained from a pool of 85 human subjects.

\subsubsection{\textbf{BVI-VFI}} BVI-VFI~\cite{10304617} is proposed to understand how the quality of temporally interpolated content is perceived. It employs five commonly used video frame interpolation algorithms on 36 source videos to generate 540 interpolated videos. More than 10,800 ratings were collected by a large-scale human study involving 189 human participants.

Since increased motion, as occurs in live sports streaming~\cite{shang2021study}, increased frame rates~\cite{9497087}, and increased bit depths~\cite{10325414}, all operating in conjunction with compression, present special challenges of different distortions, higher bandwidth requirements, and a previous lack of models able to predict these kinds of distortions, new databases are required to allow for the development of VQA targets able to target these expanded classes of distortion scenarios~\cite{ebenezer2021chipqa,madhusudana2021st,ebenezer2023making}.


\subsection{UGC Datasets}
\label{ugc-datasets}
Most of the databases just described are characterized by a small number of unique contents (10-15) rated by a small number of human subjects ($<100$), implying limited (but often sufficient) amounts of data.
However, in the context of (generally) non-professional, a vast variety of distortions may occur during video acquisition or rendering, resizing, compression, sharing, re-compression, processing, transmission, and reception. 
Several of this multitude of possible space-time distortions may afflict any video, commingling and interacting to create new, unnamable distortions.
To model this plethora of possible distortions, much larger datasets of real-world videos are required to span the space of possible required representations.
Likewise, much larger amounts of human quality annotations are required to offer the possibility of mapping video measurements to accurate subjective quality predictions.
This has led to the creation of large-scale crowd-sourced video quality studies, inspired by the success of the LIVE Challenge picture quality database~\cite{ghadiyaram2015massive}.
Table~\ref{tab:pgc_ugc_databases} provides a summary of some of these, and descriptions of several popular and recent UGC VQA databases are introduced as follows.
Fig.~\ref{fig:sample-frames-legacy} (g)-(l) shows sample frames taken from six popular UGC databases.

\subsubsection{\textbf{KoNViD-1k}}
\label{konvid-1k}
KoNViD-1k~\cite{Konvid-1k} contains 1,200 public-domain videos showcasing diverse ``in the wild'' distortions sampled from YFCC100m~\cite{thomee2016yfcc100m}. A crowdsourced study gathered over 73,000 ratings using a five-category quality scale.

%
\subsubsection{\textbf{FlickrVid-150k}}
\label{flickrvid-150k}
FlickrVid-150k~\cite{Flickrvid-150k} includes 153,841 videos with coarse quality ratings and another 1,596 videos with finer labels. Videos were re-encoded for uniformity but lost some diversity and authenticity.

%
\subsubsection{\textbf{LSVQ}}
\label{ying2021patch}
LSVQ~\cite{ying2021patch} is the largest VQA dataset with 38,811 diverse videos and 116,433 cropped spatial, temporal, and spatiotemporal patches. It includes about 5.5 million quality ratings from 6,284 subjects using a single-stimulus, continuous rating scale protocol.

%
\subsubsection{\textbf{LIVE-VQC}}
\label{live-vqc}
LIVE-VQC~\cite{LIVE-VQC} consists of 585 authentic videos captured by 80 inexpert videographers using 43 different models of 101 personal camera devices, thereby obtaining wide ranges of complex, authentic distortions. Over 205,000 opinion scores were collected from 4,776 participants.

%
\subsubsection{\textbf{YouTube-UGC}}
\label{youtube-ugc}
YouTube UGC~\cite{Youtube-UGC} features 1,500 videos across different resolutions and categories, including features like High Dynamic Range (HDR). Each video was rated by over 100 participants.

%

\begin{table*}[]
\centering
\setlength{\tabcolsep}{1.5pt}
\scriptsize
\caption{Taxonomy of subjective AIGC VQA databases.}
\label{tab:aigc_video_databases}
\resizebox{\textwidth}{!}{%
\begin{tabular}{llllllllllll}
\toprule
Year & Name & \begin{tabular}[c]{@{}c@{}}Total\\ videos\end{tabular} & Resolution & \begin{tabular}[c]{@{}c@{}}Frame\\ Rate\end{tabular} & \begin{tabular}[c]{@{}c@{}}Video\\ Length(s)\end{tabular} & Models & Prompts & Subjects & Ratings & Env & Quality Aspects \\ \midrule
 2023 & Chivileva et al.~\cite{chivileva2023measuringqualitytexttovideomodel} & 1,005 & - & - & -& 5 & 201& 24 & 48,240& Crowd & Perception, alignment  \\
 2023 & EvalCrafter~\cite{Liu_2024_CVPR} & 2,500 & $256\times256$-$1280\times720$ & 8, 24 & 2-4& 5 & 700& 7 & 8,647& - &  \begin{tabular}[c]{@{}l@{}}Video quality, Text alignment, Motion quality,\\ Temporal Consistency, Subjective likeness\end{tabular}   \\
 2023 & FETV~\cite{NEURIPS2023_c481049f} & 2,476 & \begin{tabular}[c]{@{}c@{}}$256\times256$,$480\times480$\\ $512\times512$, $576\times320$ \end{tabular} &  8 & 2-4& 4 & 619& 3&28,116 & - & \begin{tabular}[c]{@{}l@{}}Static quality, Temporal quality, \\ Overall alignment, Fine-grained alignment \end{tabular}  \\
 2023 & VBench~\cite{Huang_2024_CVPR} & 6,984 & $256\times256$,$512\times512$ & 8, 10 &2-3& 4 & 1,746& -& -& - & Video quality, Video-Condition Consistency  \\
 2024 & T2VQA-DB~\cite{kou2024subjectivealigneddatasetmetrictexttovideo} & 10,000 & $512\times512$ & 4 & 4& 9 & 1,000& 27& 270,000& - & Video fidelity, Text alignment\\
 2024 & GAIA~\cite{chen2024gaiarethinkingactionquality} & 9,180 & $256\times256$-$2048\times1536$ & 4-50 & 2.8 & 18 & 510& 54& 971,244 & In-lab & Video action quality\\
2024 & LGVQ~\cite{zhang2024benchmarkingaigcvideoquality} & 2,808 & $256\times256$-$1408\times768$ & 4-24 & 8-96 & 6 & 468 & 54 & 454,896 & - & \begin{tabular}[c]{@{}l@{}}Spatial quality, Temporal quality,\\ Text-to-video alignment  \end{tabular} \\
 
\bottomrule
\end{tabular}%
}
\vspace{-12pt}
\end{table*}

\subsubsection{\textbf{Youku-V1K}}
\label{youku-1k}
Youku-V1K~\cite{Youku-v1k} includes 1,072 videos from youku.com, annotated with over 22,000 ratings using a discrete scale to study content with intentional editing effects like blur of unimportant or background objects.

%
\subsubsection{\textbf{PUGCQ}}
\label{pugcq}
PUGCQ~\cite{PUGCQ} contains 10,000 videos of professionally-generated content labeled with attributes like `face', `noise', and `blur'. The dataset aims to maximize content diversity using features from pre-trained models.

%
\subsubsection{\textbf{YT-UGC+}}
\label{yt-ugc+}
YT-UGC+~\cite{YT-UGC+} refines YouTube-UGC categories with 610 detailed labels and includes three transcoded variants: Video-On-Demand (VOD), Video-On-Demand with Lower Bit-rate (VODLB), and Constant Bit Rate (CBR). Videos are categorized into three DMOS levels.

%

\subsubsection{\textbf{LIVE-YT-Gaming}}
\label{yt-gaming}
Gaming videos differ from other UGC content due to distinct visuals and specific distortions like improper graphics settings, frame drops, temporal lags, and low-quality recording software. 
LIVE-YT-Gaming~\cite{YT-Gaming} is the first database to focus on real UGC gaming content, with 600 clips from 59 games across diverse resolutions and frame rates. It includes 18,600 ratings from 61 participants.

%
\subsubsection{\textbf{Tele-VQA}}
\label{telepresence}
Tele-VQA~\cite{Telepresence} comprises 2,320 videos, including 1,129 virtual meeting recordings and 1191 YouTube videos, with 78,880 subjective ratings from 526 participants, evaluating both video and audio quality.

%
\subsubsection{\textbf{Maxwell}}
\label{Maxwell}
Maxwell~\cite{10.1145/3581783.3611737} contains 4,543 videos with over two million opinions annotated by 35 human participants. It collects multidimensional quality ratings covering technical and aesthetic factors to guide model design.
Different from previous datasets, these explanation-level human opinions in Maxwell highlight correlations between various quality factors of the human vision system.

%
\subsubsection{\textbf{DIVIDE-3k}}
\label{DIVIDE-3k}
DIVIDE-3k~\cite{Wu_2023_ICCV} features 3,590 videos sourced from YFCC-100M~\cite{10.1145/2812802}, Kinetics-400~\cite{DBLP:journals/corr/KayCSZHVVGBNSZ17}, and LSVQ~\cite{ying2021patch}, annotated with 450,000 opinions, including scores for aesthetic, technical, and overall quality.

%
\subsubsection{\textbf{TaoLive}}
\label{TaoLive}
TaoLive~\cite{Zhang_2023_CVPR} explores both in-capture distortions and live-streaming distortions using 418 UGC videos from Taobao~\cite{taoliveonline} live streaming platform, compressed into 3,762 variants with 44 participants rating the videos.

%

\subsubsection{\textbf{KVQ}}
\label{KVQ}
KVQ~\cite{Lu_2024_CVPR} studies short-form videos on Kwai~\cite{kwai}, featuring 4,200 videos processed via enhancement, pre-processing, and transcoding. It uses a mixed scoring manner of MOS and ranking labels from 15 participants.

\subsection{AIGC Datasets}
\label{aigc-datasets}
The emergence of large text-to-image (T2I) and text-to-video (T2V) models has introduced challenges in assessing the perceptual quality of AI-generated content (AIGC), including text alignment, naturalness, rendering, and temporal consistency. Addressing these issues necessitates subjective studies to benchmark and calibrate objective AIGC VQA models.
%
%
Recent years have seen the emergence of multiple datasets that model the subjective quality of AI-generated images, focusing on human preferences~\cite{Wu_2023_ICCV,NEURIPS2023_33646ef0,NEURIPS2023_73aacd8b}, perceptual quality~\cite{10222021,10.1007/978-981-99-9119-8_5,chen2024exploringnaturalnessaigeneratedimages}, text alignment~\cite{10262331,AIGIQA-20K_CVPRW}, and compression-aware quality~\cite{li2024cmcbenchnewparadigmvisual}. However, there are still few AIGC VQA datasets available, and more work is needed in this direction. Existing AIGC VQA databases are listed in Table~\ref{tab:aigc_video_databases}, and sample frames are shown in Fig.~\ref{fig:sample-frames-legacy} (m)-(r).

\subsubsection{\textbf{Chivileva et al.}} 
Chivileva et al.~\cite{chivileva2023measuringqualitytexttovideomodel} developed a dataset of 1,005 videos generated by five T2V models from 201 carefully selected prompts covering diverse scenarios. 48,240 ratings were collected from 24 subjects on perception and alignment quality.


\subsubsection{\textbf{EvalCrafter}} EvalCrafter~\cite{Liu_2024_CVPR} 
contains 700 prompts grouped into four metaclasses, used to generate 2,500 videos from five T2V models. It features 16 objective quality assessment tools for four quality aspects and 8,647 subjective ratings from seven subjects on five quality dimensions.

\subsubsection{\textbf{FETV}} 
FETV~\cite{NEURIPS2023_c481049f} includes 619 prompts categorized by content, attribute control, and complexity. It evaluates four T2V models across four quality aspects, including spatiotemporal quality and alignment, using assessments from three human subjects.

\subsubsection{\textbf{VBench}} 
VBench~\cite{Huang_2024_CVPR} benchmarks T2V quality with 16 evaluation dimensions linked to 100 prompts each. It includes a subjective study to validate VBench's alignment with human opinions, whereby human preference annotations were collected on videos generated by four T2V models.

\subsubsection{\textbf{T2VQA-DB}} 
T2VQA-DB~\cite{kou2024subjectivealigneddatasetmetrictexttovideo} features 10,000 videos from nine T2V models based on 1,000 graph-selected prompts. Opinions on text-video alignment and fidelity were collected from 27 subjects.


\subsubsection{\textbf{GAIA}} 
GAIA~\cite{chen2024gaiarethinkingactionquality} focuses on human action quality in AI-generated videos. It contains 9,180 videos from 18 T2V models generated using 510 GPT-4~\cite{openai2023gpt4} prompts, and is rated on three quality aspects by 54 subjects.

\subsubsection{\textbf{LGVQ}} 
LGVQ~\cite{zhang2024benchmarkingaigcvideoquality} explores motion of various scenarios by decomposing prompts into foreground, background, and motion categories. 2,808 videos were generated from six T2V models based on 468 prompts, and then rated by 54 subjects for spatial quality, temporal quality, and text alignment.


\begin{figure*}[!t]
\centering
\footnotesize
\setlength{\tabcolsep}{1pt}
\includegraphics[width=\linewidth]{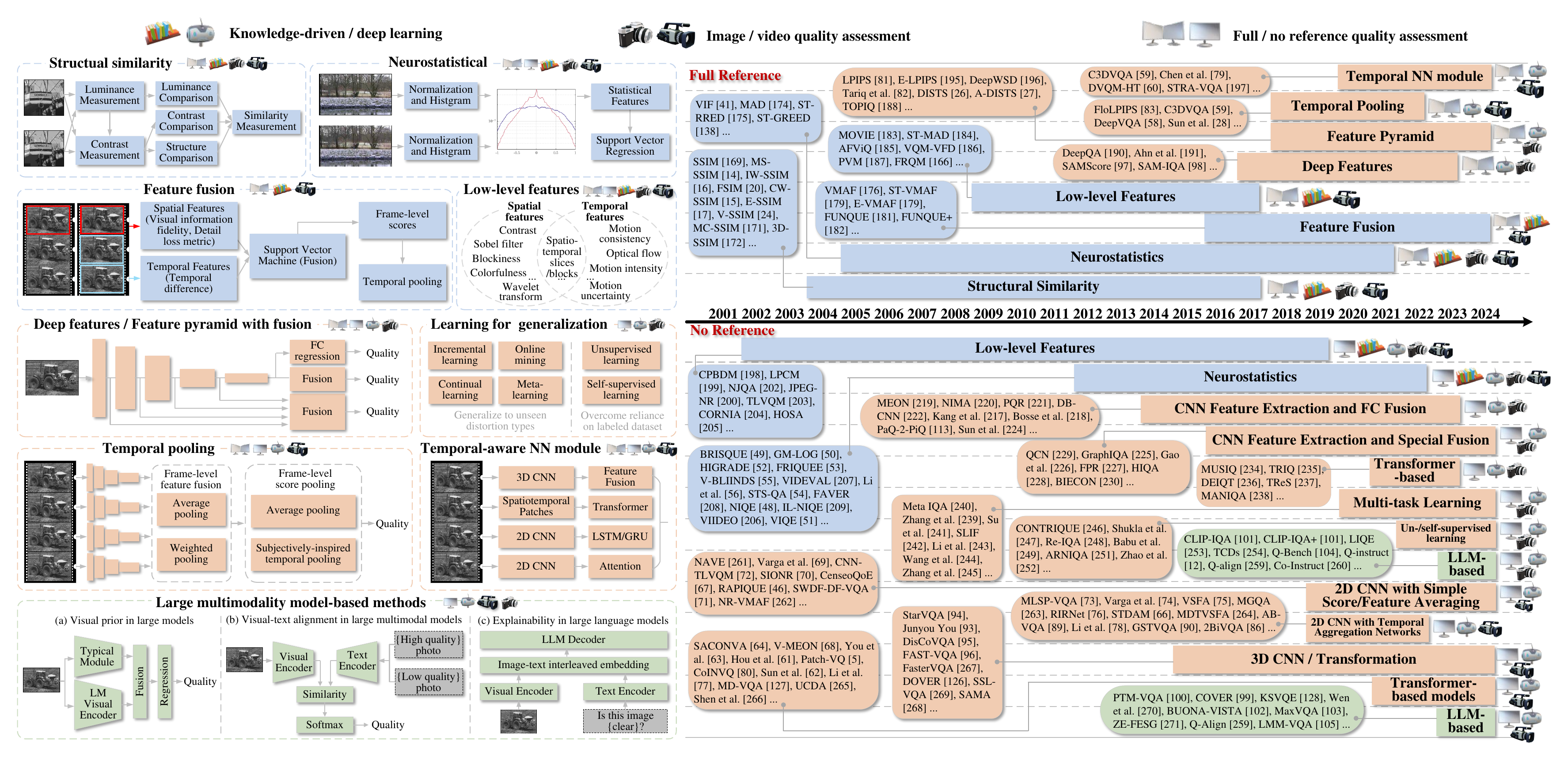}
\vspace{-5pt}
\caption{
Classification and evolution of objective quality assessment models. The left figure presents framework diagrams for each category, while the right figure highlights their temporal progression and representative models. Categories are color-coded as follows: knowledge-driven (blue), deep learning-based (orange), and large model-based (green).
}
\vspace{-12pt}
\label{fig:year_taxonomy}
\end{figure*}

%% file: sections/4_objective_assessment.tex
\section{Objective Video Quality Assessment}
\label{object_vqa}

In this section, we provide an in-depth review of video quality assessment methods covering both full-reference and no-reference models, with a focus on deep learning-based approaches and commonly used loss functions. We analyze, categorize, and summarize the current mainstream quality assessment methods, highlighting model performance and efficiency in quality prediction. Fig.~\ref{fig:year_taxonomy} illustrates an overarching summary of different algorithm categories and their development over time, offering a structured overview of each type of model and their evolution.

\subsection{Full-reference Video Quality Assessment}
\label{fr_vqa}
Full-reference video quality assessment (FR-VQA) models seek to quantify (perceptual) differences between reference videos and their distorted counterparts, and map them to predictions of visual quality.
Since FR models have available undistorted content information, they generally demonstrate higher correlations against human judgments than non-reference models. 
FR models can be classified into knowledge-driven and deep learning-based algorithms: 1) \textbf{Knowledge-driven} algorithms are inspired by models of the human visual system, and calculate perceptual distance between pixels, structures, or other relevant properties such as gradient difference.
2) \textbf{Deep learning-based} methods accept reference and distorted pairs of videos or video frames as inputs, learning both perceptual quality/distortion representations as well as deep semantic abstractions, and how distorted and semantic data relate when forming quality predictions.
In this section, we discuss both kinds of methods by providing a concise overview of IQA models before delving into a comprehensive examination of VQA models. 

\subsubsection{\textbf{Knowledge-driven FR VQA Methods}}
The last two decades have seen the development of a remarkable variety of knowledge-driven FR-IQA models. These models can be extended to FR-VQA tasks by simply using them to estimate frame-wise quality, then temporally pooling the frame-level measurements~\cite{zhang2017frame,tu2020comparative} to obtain overall video quality scores.
However, this approach does not capture temporal distortions or their impacts on visual perception. 
Significant effort has been applied to develop true FR-VQA models by knowledge-based temporal distortion modeling.
In the following, existing FR IQA/VQA models are further classified into five sub-categories, and discussed in turn.

\noindent\underline{\textbf{Type i: Pixel-error-based VQA.}}
The earliest full-reference image quality assessment (FR-IQA) algorithms measured pixel-level errors, such as the mean squared error (MSE) or peak signal-to-noise ratio (PSNR).
However, due to the lack of consideration of how people view and process visual distortions, these simple numerical measures often exhibited poor correlation with human perceptions~\cite{sheikh2006statistical}.

\noindent\underline{\textbf{Type ii: Structural Similarity-based VQA.}}
Things began to change with the introduction of a perceptually motivated IQA model that is accurate and efficient, the Primetime Emmy Award winning Structural Similarity or \textbf{SSIM}~\cite{wang2004image} Index.
Further advancements of SSIM soon followed, including multi-scale structural similarity (\textbf{MS-SSIM})~\cite{wang2003multiscale}, wavelet-domain spectral similarity (\textbf{CW-SSIM})~\cite{sampat2009complex}, information content weighted (\textbf{IW-SSIM})~\cite{wang2010information}, gradient and phase congruency similarity (\textbf{FSIM})~\cite{zhang2011fsim}, edge strength similarity (\textbf{ESSIM})~\cite{6423791}, gradient magnitude similarity~\cite{xue2013gradient,liu2011image,zhang2011fsim}, and visual saliency (\textbf{VSI})~\cite{zhang2014vsi,shi2017perceptual}.
Owing to the simplicity and effectiveness of SSIM~\cite{wang2004image}, significant efforts have been dedicated to extending it to the video domain.
%
%
An early \textbf{``video SSIM''}~\cite{wang2004video} operates by space-time sampling estimates of perceptual video quality at three levels: spatially local, frame, and sequence, employing two weighting pooling methods based on luminance estimation and frame motion.
%
\textbf{Wang and Li}~\cite{wang2007video} proposed an alternative weighting scheme of SSIM based on human perception of motion information.
However, weighted pooling of spatial SSIM scores does not necessarily account for temporal distortions. Succeeding works explored modeling temporal information in quality assessment.
\textbf{V-SSIM}~\cite{seshadrinathan2007structural} is a motion compensated variant of SSIM~\cite{wang2004image} that models motion information using optical flow.
\textbf{MC-SSIM}~\cite{5604290} is another motion-compensated variant of SSIM, which evaluates structural retention between motion-compensated regions in a frame.
\textbf{Manasa and Channappayya}~\cite{manasa2016optical} compute temporal quality estimates using local optical flow statistics and spatial quality estimates using MS-SSIM~\cite{wang2003multiscale}, subsequently pooling both estimates into single video quality scores.
Instead of separately analyzing spatial and temporal distortion, \textbf{3D-SSIM}~\cite{6466936} generates a 3D quality map by applying SSIM measurements within local 3D blocks, which are then pooled into an overall video quality score using a weighted scheme based on local information content.

\noindent\underline{\textbf{Type iii: Neurostatistics-based VQA.}}
Another major advance was the discovery of neurostatistical models of the responses of visual neurons to visual distortions. This approach was best exemplified by the full reference Visual Information Fidelity (\textbf{VIF})~\cite{sheikh2005information,sheikh2006image} model, which quantifies statistical deviations of visual information arising from distortions, as measured under a modified natural scene statistics model.
As discussed later, a variety of popular and powerful no-reference (NR) visual quality prediction models have been devised using similar neurostatistical perceptual distortion models. 
Another strategy, called most apparent distortion (\textbf{MAD})~\cite{larson2010most} explicitly separates distortion strength into either apparent or invisible categories, subsequently modeling two distinct measurement strategies on high and low-quality images, respectively.
The \textbf{ST-RRED} model\cite{rs2012video} expands on the single-picture VIF\cite{sheikh2006image} IQA model by modeling temporal distortions using neurostatistical models of spatially-bandpass frame differences.
ST-RRED remains highly competitive on most VQA databases and has the added efficiency of allowing for reduced-reference versions via a wavelet-domain subsampling strategy.
\textbf{ST-GREED}~\cite{madhusudana2021st} goes further by addressing distortions arising from variable frame rates (VFR) combined with compression, using neurostatistical spatial and temporal band-pass video models.
ST-GREED performs remarkably better than any prior existing models on a large subjective database of compressed VFR videos ranging from 30 to 120 frames/sec~\cite{9497087}.
However, the same authors show that previous VQA models like SSIM, MS-SSIM, ST-RRED, VMAF~\cite{arjovsky2017wasserstein}, and even PSNR can be greatly improved in a simple way by using concepts from ST-GREED~\cite{pc2022making}.

\noindent\underline{\textbf{Type iv: Feature fusion-based VQA.}}
VIF and ST-RRED form the foundation of the Netflix Emmy-winning Video Multi-Method Assessment Fusion (\textbf{VMAF})~\cite{li2016toward} models, which simplifies them by using only four spatial channels of VIF, average absolute frame differences, and a detail-loss measurement feature\cite{sli2011image}. 
These features are used to train VMAF on an internal VQA database using a support vector regression (SVR) model.
VMAF has been used to control the quality of all video encodes streamed globally for years, and has been adopted by many other streaming video providers, making it competitive with SSIM/MS-SSIM, which controls the quality of a large percentage of all television content, and much social media, for example picture/video content uploaded and subsequently streamed by Facebook.
\textbf{ST-VMAF} and \textbf{E-VMAF}~\cite{bampis2018spatiotemporal} enhance the performance of VMAF~\cite{li2016toward} by enriching its temporal features and by employing multiple regression models, respectively.
To alleviate the computational burden presented by multiple models in fusion-based VQA methods, \textbf{FUNQUE}~\cite{9897312} shares computation by decomposing the features of different models into a common transform domain. The succeeding framework \textbf{FUNQUE+}~\cite{10375336} includes low-complexity fused-feature models to boost quality prediction accuracy.

\begin{figure*}[!t]
\centering
\includegraphics[width=0.9\linewidth]{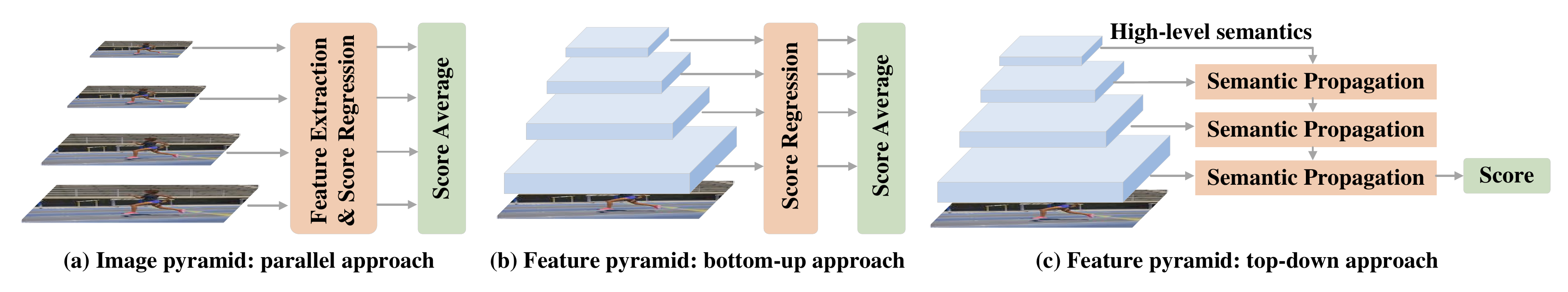}
\caption{Three types of IQA framework based on how they extract and employ multi-scale features: the parallel, bottom-up and top-down methods~\cite{10478301}.}
\label{fig:IQA_pyramid}
\vspace{-8pt}
\end{figure*}


\noindent\underline{\textbf{Type v: Low-level motion feature-based VQA.}}
FR VQA models of this type often model temporal distortions within well-designed motion feature extraction modules, based on motion trajectories, wavelet transform, etc.
The concept of V-SSIM is expanded upon by the \textbf{MOVIE} index~\cite{seshadrinathan2009motion}, which evaluates space-time video quality along motion trajectories, using a deep brain model of extracortical area MT.
\textbf{ST-MAD}~\cite{vu2011spatiotemporal} applies the picture quality model MAD~\cite{larson2010most} along spatiotemporal slices (STS) or cuts of a space-time video.
\textbf{AFViQ}~\cite{you2013attention} conducts perceptual quality assessment of foveated videos by evaluating perceptible contrast-sensitive differences in the wavelet domain.
\textbf{VQM-VFD}~\cite{pinson2014temporal} extracts a variety of features from spatiotemporal blocks to identify quality degradations.
\textbf{PVM}~\cite{zhang2015perception} video quality perception by adaptively modeling texture masking and blur detection using a dual-tree complex wavelet transform.
Additionally, several studies have focused on predicting video quality at varying frame rates.
This is important since many distortions, such as stutter and motion bur, manifest over longer time intervals than existing models like VMAF and SSIM are able to measure.
This is of particular importance when there is high motion, as in sports videos.
In addition to the aforementioned ST-GREED model, \textbf{FRQM}~\cite{zhang2017frame} analyzes variable frame rate (VFR) video quality using a temporal wavelet decomposition followed by a spatial subband combination.

\subsubsection{\textbf{Deep Learning-Based FR IQA Methods}}
\label{sssec:dl_fr_iqa}
While deep learning-based IQA methods can be extended to VQA by applying frame-level scores followed by temporal pooling, deep neural networks and learning strategies offer significantly more scope for tailored design for VQA. 
The temporal dynamics of videos, including motion and frame dependencies, necessitate sophisticated NN architectures, resulting in a broader range of approaches within the VQA domain.
Due to the complexity and diversity of current deep learning-based VQA methods, it is important to review IQA and VQA separately.
This separation allows for a more thorough examination of the unique challenges and innovations within each area, providing a clearer understanding of their individual advancements.
We summarize deep learning-based FR IQA and VQA models in Table~\ref{table:fr_models}, which begins with an overview of the core attributes and methodologies of these techniques.

Deep learning-based FR IQA models usually rely on a typical workflow whereby deep features are first extracted from reference images, distorted images, and/or error images, and then fused and regressed into image quality scores. 
Depending on the level of features used for quality evaluation, these models can be further divided into deep feature-based and feature pyramid-based models, the former of which use only final-layer deep features, while the latter constructs a pyramid of features from multiple layers.

\noindent\underline{\textbf{Type i: Deep feature-based IQA.}}
\textbf{Bosse et al.}~\cite{bosse2017deep} propose a unified deep network for FR- and NR-IQA, employing a Siamese network for patch-level feature extraction, followed by feature fusion, patch-wise quality regression and pooling. For NR-IQA, the reference branch and feature fusion are omitted. 
\textbf{DeepQA}~\cite{kim2017deepsen} uses a CNN to model a visual sensitivity function, computes an error map as normalized log differences between reference and distorted images, and applies sensitivity-based weighting to derive a perceptual error map. A nonlinear regressor maps the pooled error map to subjective scores.
Building on DeepQA, \textbf{Ahn et al.}\cite{ahn2021deep} improve distortion sensitivity prediction by using UNet~\cite{ronneberger2015u} for spatial preservation, incorporating the reference image into input signals, and adding convolutional layers to expand the receptive field. Quality scores are calculated as the mean perceptual error map values.

Recent advancements in large models have significantly impacted fields like computer vision and natural language processing.
The Segment Anything Model (SAM) has extended its utility from semantic segmentation to image quality assessment.
%
\textbf{SAMScore}~\cite{li2023samscore} evaluates semantic structural similarity in image translation tasks by computing the spatial-wise cosine similarity of SAM-derived semantic embeddings, averaged into a final score.
\textbf{SAM-IQA}~\cite{li2023sam} utilizes SAM's encoder to extract features in both frequency and spatial domains. For full-reference tasks, feature distances in both domains are calculated before quality regression, while for no-reference tasks, the features are directly input into the regression module.


\noindent\underline{\textbf{Type ii: Feature pyramid-based IQA.}}
The concept of multi-scale feature extraction in MS-SSIM can be used in deep learning-based models as well, as seen in feature pyramid-based models shown in Fig.~\ref{fig:IQA_pyramid}. These can be further divided into bottom-up approaches~\cite{zhang2018unreasonable,ding2020image} and top-down approaches~\cite{10478301}, according to the mechanisms used to fuse features from different layers~\cite{10478301}. We first introduce representative bottom-up approaches as follows.

\textbf{LPIPS}~\cite{zhang2018unreasonable} measures deep feature distances using networks like SqueezeNet~\cite{iandola2016squeezenet}, AlexNet~\cite{krizhevsky2014one}, and VGG~\cite{simonyan2014very}. Reference and distorted patches are processed, channel-wise normalized, and compared using cosine distance, averaged spatially and across layers. Finally, a small network is trained to predict subjective quality using ranking loss, excelling in handling geometric distortions, which makes LPIPS popular for tasks like super-resolution. 
\begin{figure}[!t]
    \centering
    \includegraphics[width=8.5cm]{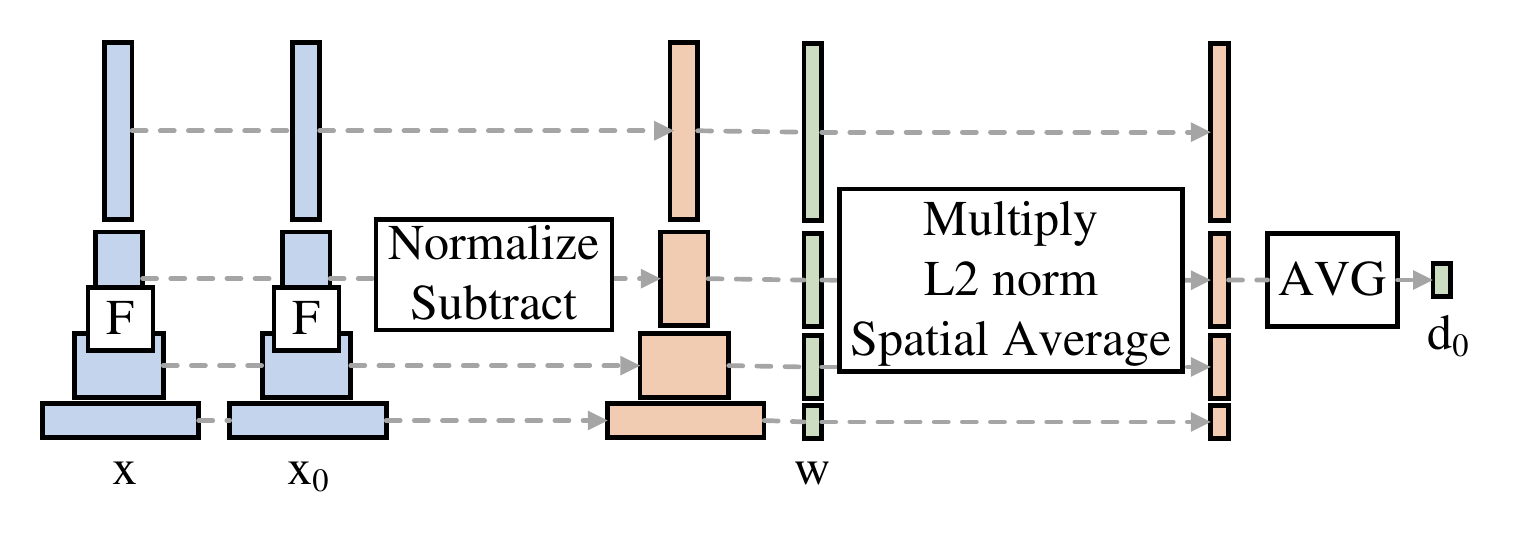}
    \vspace{-10pt}
    \caption{The framework of LPIPS~\cite{zhang2018unreasonable}.}
    \vspace{-10pt}
    \label{fig:LPIPS}
\end{figure}
Kettunen et al.\cite{kettunen2019lpips} highlight LPIPS's vulnerability to adversarial attacks and introduce \textbf{E-LPIPS}, improving robustness with geometric/color transformations, multi-layer distance computation, dropout, and average pooling.
\textbf{DeepWSD}~\cite{liao2022deepwsd} extends LPIPS by using Wasserstein distance to assess image quality without labeled data. Deep features from VGG-16 are reshaped and compared across five stages, and additional metrics like an ``EUL'' index, combining Euclidean norm with adaptive weights, enhance perceptual fidelity assessment.
\textbf{Tariq et al.}~\cite{tariq2020deep} analyze pre-trained CNN feature maps to identify channels sensitive to visual distortions. Spatial frequency and orientation measures are combined into a single Perceptual Efficacy (PE) score, reflecting channel effectiveness in predicting perceptual quality.
\textbf{DISTS}~\cite{ding2020image} integrates sensitivity to structural distortions and tolerance to texture resampling. Using VGG-16, it replaces max pooling with weighted L2 pooling, computes texture and structure distances via global means and correlations, and combines these into a weighted quality score. 
\textbf{A-DISTS}~\cite{ding2021locally} enhances DISTS by incorporating locally adaptive structure and texture similarity using variance-to-mean ratios of features to weight similarities across space and channels.

Unlike the above bottom-up paradigm, \textbf{TOPIQ}~\cite{10478301} adopts a top-down approach for both FR and NR image quality assessment. Multi-scale features from the first five layers of ResNet-50~\cite{he2016deep} are unified via a gated local pooling module and enhanced with self-attention blocks. Cross-scale attention is then applied progressively from high-level to low-level features, with the final features pooled and regressed into image-level quality scores.


\begin{table*}[!t]
\renewcommand{\arraystretch}{0.95}
\centering
\fontsize{6.2pt}{8pt}\selectfont
\caption{Overview of deep full-reference IQA/VQA methods, indicated by \textit{italic}/orthographic font, respectively.
}
\label{table:fr_models}
\renewcommand\arraystretch{1.2}
\resizebox{\textwidth}{!}{%
\begin{tabular}{c|c|c|c|c}
\toprule 
\makebox[0.05\textwidth][c]{Type} & \makebox[0.05\textwidth][c]{Method} & \makebox[0.1\textwidth][c]{\begin{tabular}[c]{@{}c@{}} Architecture \\ (Feature extraction + \\ Quality fusion)\end{tabular}} &
\makebox[0.15\textwidth][c]{\begin{tabular}[c]{@{}c@{}} Core block \\ (Pretrained models or \\ crafted modules)\end{tabular}} &
\makebox[0.3\textwidth][c]{Key idea} \\
\hline\\[-1.em]
\multirow{10}{*}{\begin{tabular}[c]{@{}c@{}} Deep feature \\ based IQA \end{tabular}} &
\begin{tabular}[c]{@{}c@{}}\emph{Sebastian Bosse}  
 \\ \emph{et al.~\cite{bosse2017deep}}\end{tabular} & \begin{tabular}[c]{@{}c@{}} CNN + MLP \& \\ Weighted pooling \end{tabular} & \begin{tabular}[c]{@{}c@{}}Siamese Network, \\Spatial weighted Pooling\end{tabular} & \begin{tabular}[c]{@{}c@{}}Both average pooling and weighted average regression are evaluated \\in a unified framework for FR and NR image quality assessment.\end{tabular}\\
\cline{2-5} 
& \emph{DeepQA~\cite{kim2017deepsen}} & \begin{tabular}[c]{@{}c@{}} CNN + MLP \& \\ Weighted pooling \end{tabular}  & \begin{tabular}[c]{@{}c@{}}Crafted CNN,\\ Perceptual map generation\end{tabular} & \begin{tabular}[c]{@{}c@{}}Visual sensitivity map weighted pixels on an objective error\\ map produce  a perceptual error map.\end{tabular}\\
\cline{2-5}  
& \begin{tabular}[c]{@{}c@{}}\emph{Sewoong Ahn}  
 \\ \emph{et al.~\cite{ahn2021deep}}\end{tabular} & \begin{tabular}[c]{@{}c@{}} CNN + MLP \& \\ Weighted pooling \end{tabular}  & \begin{tabular}[c]{@{}c@{}}DeepQA, \\Skip-Connection\end{tabular} & \begin{tabular}[c]{@{}c@{}}Modifies DeepQA to enrich spatial information, using\\a Unet structure, adding layers, and directly predicting quality scores.\end{tabular}\\ 
\cline{2-5} 
& \emph{SAMScore~\cite{li2023samscore}}   & \begin{tabular}[c]{@{}c@{}}SAM + \\ Similarity pooling\end{tabular}  & \begin{tabular}[c]{@{}c@{}}Image encoder of SAM, \\ Cosine similarity\end{tabular} & \begin{tabular}[c]{@{}c@{}}Leverage SAM encoder to extract semantic embeddings, based on which \\ spatial-wise cosine similarity is calculated and averaged into the overall quality. \end{tabular}\\ \cline{2-5} 
& \emph{SAM-IQA~\cite{li2023sam}}   & SAM + MLP& \begin{tabular}[c]{@{}c@{}}Image encoder of SAM, \\ Fourier/classic convolution\end{tabular} & \begin{tabular}[c]{@{}c@{}}Embeddings of SAM encoder are gone through spatial-frequency feature  \\ extraction, and the obtained features are regressed into the overall quality.\end{tabular} \\ \cline{1-5} 
\multirow{14}{*}{\begin{tabular}[c]{@{}c@{}} Feature pyramid \\ based IQA \end{tabular}} 
& \emph{LPIPS~\cite{zhang2018unreasonable}} & \begin{tabular}[c]{@{}c@{}}CNN \\+ Distance pooling\end{tabular}  & \begin{tabular}[c]{@{}c@{}}
AlexNet/VGG,\\ Distance Calculation \end{tabular}& \begin{tabular}[c]{@{}c@{}}Calculates the distance between network activations of pairs of image patches,\\ averaged across spatial dimensions and layers of the network.\end{tabular}\\
\cline{2-5} 
&\begin{tabular}[c]{@{}c@{}}\emph{Markus Kettunen}  
 \\ \emph{et al.~\cite{kettunen2019lpips}}\end{tabular} & \begin{tabular}[c]{@{}c@{}}CNN \\+ Distance pooling\end{tabular}  & \begin{tabular}[c]{@{}c@{}}Adversarial Attack,\\ LPIPS\end{tabular} & \begin{tabular}[c]{@{}c@{}}Boosts LPIPS by transforming the input images, computing over \\all layers, applying dropout, and using averaging pooling.\end{tabular}\\
 \cline{2-5} 
& \emph{DeepWSD~\cite{liao2022deepwsd}} & \begin{tabular}[c]{@{}c@{}}CNN \\+ Distance pooling\end{tabular}  & \begin{tabular}[c]{@{}c@{}}VGG-16, \\Wasserstein Distance\end{tabular} & \begin{tabular}[c]{@{}c@{}}The Wasserstein distance and Euclidean distance at each stage of CNN \\are measured and averaged to predict the final quality scores. \end{tabular}\\
\cline{2-5} 
& \begin{tabular}[c]{@{}c@{}}\emph{Taimoor Tariq}  
 \\ \emph{et al.~\cite{tariq2020deep}}\end{tabular} & \begin{tabular}[c]{@{}c@{}}CNN \\+ Distance pooling\end{tabular}  & \begin{tabular}[c]{@{}c@{}} Pretrained models, \\Channel selection\end{tabular}& \begin{tabular}[c]{@{}c@{}}Analyzes the capabilities of deep features for estimating quality\\ degradations by measuring frequency and orientation selectivity.\end{tabular}\\
\cline{2-5} 
& \emph{DISTS~\cite{ding2020image}} & \begin{tabular}[c]{@{}c@{}}CNN \\+ Similarity pooling\end{tabular}  & \begin{tabular}[c]{@{}c@{}}VGG-16, \\Similarity Calculation\end{tabular} & \begin{tabular}[c]{@{}c@{}}The first FR IQA model insensitive to texture resampling, \\measures perceptual similarity within a deep representation.\end{tabular}\\
\cline{2-5} 
& \emph{A-DISTS~\cite{ding2021locally}} & \begin{tabular}[c]{@{}c@{}}CNN \\+ Similarity pooling\end{tabular}  & \begin{tabular}[c]{@{}c@{}}Separate Structure and\\ Texture\end{tabular} & Adaptively weights local structure and texture separated by a dispersion index.\\
\cline{2-5} 
& \emph{TOPIQ~\cite{10478301}} & \begin{tabular}[c]{@{}c@{}}CNN + Feature \\ fusion \& MLP\end{tabular} & \begin{tabular}[c]{@{}c@{}}ResNet50, Self-attention, \\ Cross-attention\end{tabular} & \begin{tabular}[c]{@{}c@{}} Exploits multi-scale features and incrementally transfers high-level \\ semantic information to low-level representations in a top-down fashion. \end{tabular}\\
 \midrule
\multirow{8}{*}{\begin{tabular}[c]{@{}c@{}} Temporal pooling \\ based VQA \end{tabular}} 
& FloLPIPS~\cite{danier2022flolpips} & \begin{tabular}[c]{@{}c@{}}CNN \\+ Distance pooling\end{tabular}  & \begin{tabular}[c]{@{}c@{}}LPIPS, Optical flow, \\ Weighted spaital average \end{tabular}& \begin{tabular}[c]{@{}c@{}}Re-designs the spatial averaging step of LPIPS by weighting by\\ the differences of estimated optical flow of adjacent frames.\end{tabular}\\
\cline{2-5} 
& C3DVQA~\cite{xu2020c3dvqa} & \begin{tabular}[c]{@{}c@{}} 2D CNN + 3D CNN \\ \& Mask pooling \end{tabular} & 2D CNN, 3D CNN & \begin{tabular}[c]{@{}c@{}}A CNN with 2D and 3D kernels learns distortion visibility thresholds\\ which mask residual frames to exaggerate noticeable distortions.\end{tabular}\\
\cline{2-5} 
& DeepVQA~\cite{kim2018deep} & \begin{tabular}[c]{@{}c@{}}CNN + Memory \\ attention pooling\end{tabular}  & \begin{tabular}[c]{@{}c@{}}DeepQA, \\Temporal aggregation\end{tabular}  & \begin{tabular}[c]{@{}c@{}}Learns spatial-temporal sensitivity similar DeepQA, and \\conducts temporal pooling across frames using an aggregation network.\end{tabular}\\
\cline{2-5} 
& \begin{tabular}[c]{@{}c@{}}Wei Sun  
 \\ et al.~\cite{sun2021deep}\end{tabular} & \begin{tabular}[c]{@{}c@{}}CNN + MLP \& \\ Memory effect pooling\end{tabular}  & \begin{tabular}[c]{@{}c@{}}Similarity Calculation, \\ Quality Regression\end{tabular}& \begin{tabular}[c]{@{}c@{}}Different measurements are calculated within intermediate layers depending on \\whether FR or NR assessment is required, followed by regression and pooling.\end{tabular}\\
\cline{1-5} 
\multirow{8}{*}{\begin{tabular}[c]{@{}c@{}} Temporal NN module \\ based VQA \end{tabular}} & C3DVQA~\cite{xu2020c3dvqa} & \begin{tabular}[c]{@{}c@{}} 2D CNN + 3D CNN \\ \& Mask pooling \end{tabular} & 2D CNN, 3D CNN & \begin{tabular}[c]{@{}c@{}}A CNN with 2D and 3D kernels learns distortion visibility thresholds\\ which mask residual frames to exaggerate noticeable distortions.\end{tabular}\\
\cline{2-5} 
& \begin{tabular}[c]{@{}c@{}} Junming Chen  
 \\ et al.~\cite{chen2021deep}\end{tabular} & 2D CNN + 3D CNN & \begin{tabular}[c]{@{}c@{}}2D CNN,\\Spatiotemporal aggregation\end{tabular}& \begin{tabular}[c]{@{}c@{}}Three  distinct approaches are investigated to aggregate patch-level\\ quality scores along both spatial and temporal axes.\end{tabular}\\
\cline{2-5} 
& DVQM-HT~\cite{feng2022deep} & 3D CNN + MLP  & \begin{tabular}[c]{@{}c@{}}3D CNN,\\ shallow 2D CNN\end{tabular} & \begin{tabular}[c]{@{}c@{}}The first model to employ a 3D CNN to predict video quality on\\ patches, where VMAF is used to enrich the training data.\end{tabular}\\
\cline{2-5} 
& STRA-VQA~\cite{10444627} & \begin{tabular}[c]{@{}c@{}}2D CNN \\+ Transformer\end{tabular} & \begin{tabular}[c]{@{}c@{}}ResNet50, Adaptive \\ Weight Transformer\end{tabular} & \begin{tabular}[c]{@{}c@{}}Spatiotemporal quality is modeled in an adaptive weight Transformer\\ which is sensitive to the downsampling spatial and temporal resolutions.\end{tabular}\\
\bottomrule
\end{tabular}
}
\vspace{-12pt}
\end{table*}

\subsubsection{\textbf{Deep Learning Based FR VQA Methods}}
\label{dp_frvqa}
Deep neural networks for video quality assessment aim to capture both spatial and temporal quality variations, which is a challenging task due to larger data inputs and complex frame interactions. A straightforward approach extends FR IQA methods by generating frame-level scores/features followed by temporal fusion, though simple pooling often fails to model temporal distortions effectively. Hence, subjectively-inspired temporal effects have been leveraged by several models to effectively account for temporal quality. Moreover, temporal-aware neural modules, such as 3D CNNs for spatiotemporal features or Transformers for long-range dependencies, have been adopted. Deep learning-based FR VQA methods are thus categorized into temporal pooling-based and temporal NN module-based approaches. Since this review primarily focuses on deep learning-based FR/NR VQA methods, detailed design overviews are provided as below.


\noindent\underline{\textbf{Type i: Temporal pooling-based VQA.}}
A common challenge is that the simple average pooling method does not capture complex temporal dynamics. 
To address this, models like \textbf{FloLPIPS}~\cite{danier2022flolpips} incorporate optical flow to weight the pooling process, enhancing sensitivity to temporal distortions. 
Additionally, models like \textbf{DeepVQA}~\cite{kim2018deep} use spatiotemporal masking effects to adjust pooling weights, thus better identifying frame-level transitions and quality fluctuations. 
Similarly, \textbf{Sun et al.}~\cite{sun2021deep} adopted subjectively-inspired temporal pooling to reflect human perceptual responses.

\begin{itemize}[leftmargin=0em, itemindent=1em, itemsep=0pt, parsep=0pt]
\item  \textbf{FloLPIPS}~\cite{danier2022flolpips} adapts LPIPS\cite{zhang2018unreasonable} for video frame interpolation by improving temporal consistency. Optical flow between consecutive reference and distorted frames is computed using PWC-Net to generate weights for spatial pooling in LPIPS. AlexNet extracts LPIPS features, and the weights are normalized based on flow map differences.

\item \textbf{C3DVQA}~\cite{xu2020c3dvqa} models temporal masking effects using both 2D and 3D CNNs(thus it can be categorized to Type ii, too). The 2D CNN processes distorted and residual frames to extract spatial features, which are concatenated to form spatiotemporal features. The 3D CNN learns distortion thresholds from these features, identifying perceptually significant artifacts by masking residuals with thresholds. Fully connected layers map the artifacts to quality scores.
%
%
\item \textbf{DeepVQA}~\cite{kim2018deep} models spatiotemporal masking and temporal memory effects in a two-step process. First, CNNs calculate a spatiotemporal sensitivity map from distorted frames, spatial/temporal error maps, and frame differences, producing frame-level scores via perceptual error mapping. Second, a memory attention mechanism assigns significance weights to frames, enabling weighted aggregation of frame-level scores into global quality scores.

%
\item \textbf{Sun et al.}\cite{sun2021deep} propose a framework for both FR and NR quality assessment of compressed UGC videos. FR mode calculates texture and structure similarities across CNN layers, while NR mode computes global statistics from hierarchical feature maps. Frame-level scores are derived via fully connected layers and aggregated using subjectively-inspired temporal pooling from VSFA\cite{li2019quality}.
\end{itemize}

\noindent\underline{\textbf{Type ii: Temporal NN module-based VQA.}}
In temporal NN module-based VQA models, the primary approach involves using neural networks to directly model temporal dependencies and quality variations across frames. 
Unlike simpler pooling methods, these models leverage advanced architectures like 3D CNNs and transformers to more comprehensively learn spatiotemporal interactions. 
For instance, \textbf{C3DVQA}~\cite{xu2020c3dvqa} and \textbf{Chen et al.}~\cite{chen2021deep} employ 3D CNNs to capture spatiotemporal quality variations, while models like \textbf{STRA-VQA}~\cite{10444627} incorporate transformer modules to adapt to changes in spatial and temporal resolutions.
A key insight is that these models often use attention mechanisms or adaptive weighting strategies, as seen in \textbf{Chen et al.}~\cite{chen2021deep} and \textbf{STRA-VQA}~\cite{10444627}, to focus on perceptually significant distortions. 
This helps to enhance a model’s ability to account for complex temporal dynamics and quality fluctuations, providing more accurate video quality predictions as compared to simpler pooling methods.
\begin{itemize}[leftmargin=0em, itemindent=1em, itemsep=0pt, parsep=0pt]
%
\item \textbf{Chen et al.}\cite{chen2021deep} propose a unified framework combining spatiotemporal quality prediction and aggregation. A variant of C3DVQA~\cite{xu2020c3dvqa} with a residual attention mechanism highlights noticeable distortions, while an adaptive spatiotemporal aggregation network assigns weights to aggregate quality scores across spatial and temporal dimensions.
%
\item \textbf{DVQM-HT}~\cite{feng2022deep} employs hybrid training by using VMAF as a proxy for subjective scores. A 3D CNN learns patch-wise quality measurements, aggregated into frame- and sequence-level scores using a temporal aggregation network. Training data for 3D CNN includes a database of 614,400 patch pairs with VMAF values, and the temporal aggregation network is trained on the VMAFplus\cite{li2016toward} database.
%
\item \textbf{STRA-VQA}~\cite{10444627} addresses spatiotemporal resolution adaptive coding scenarios. Features from source and reconstructed videos are processed through an adaptive weight transformer, which accounts for spatial and temporal downsampling, followed by an FC layer to predict quality scores.
\end{itemize}

\subsection{No-reference Video Quality Assessment}
\label{nr_vqa}
High-quality reference signals are unavailable in many practical scenarios, and therefore, the development of no-reference video quality assessment models (NR-VQA) is quite important, but remains quite challenging.
NR-VQA algorithms need to be highly sensitive to distortions, while also accounting for aspects of visual content that reduce (mask) distortions, or that are more sensitive to distortion.
\textbf{Knowledge-driven} models often rely on perceptually relevant statistical properties of pictures and videos, similar to those underlying successful FR models like VIF and VMAF. 
\textbf{Deep learning-based} methods learn to predict quality degradations from large databases of distorted visual data.
Similar to the approach taken in Section~\ref{fr_vqa}, we begin by briefly reviewing learning no-reference or blind IQA (BIQA) models and blind VQA (BVQA) methods, then study data-driven deep learning-based blind models.

\subsubsection{\textbf{Knowledge-driven BVQA Methods}}
\label{traditional_bvqa}
Traditional VQA methods typically follow a knowledge-driven approach, whereby handcrafted features are manually extracted to assess video quality. 
For low-level visual feature-based VQA methods, these features are carefully designed based on prior domain knowledge, aiming to capture key aspects of video quality such as sharpness, contrast, or motion artifacts. 
The quality evaluation process relies heavily on predefined rules, which, while effective to some extent, may struggle to generalize across diverse video contents and distortion types. 
These have been largely replaced by more powerful general-purpose BIQA/BVQA models that are based on measurements of distortion-induced statistical deviations of bandpass processed images/videos from perceptually relevant models of natural scene statistics (NSS)~\cite{ruderman1994statistics}, as discussed earlier.

\noindent\underline{\textbf{Type i: Low-level visual feature-based VQA.}}
There are a few works performing quality measurement based on distortion-specific or low-level visual features, such as local spatial features, \textit{e.g.}, edge strength, contour shape, and motion intensity.
The earliest (BIQA or) BVQA  algorithms were designed to analyze and quantify single distortion types, such as blockiness, blur, ringing, sharpness, and space compression~\cite{5246972, 6476013, wang2002no, wang2000blind}, based on the measurement of a small number of image or frame level features, 
For example, \textbf{CPBDM}~\cite{5246972} and \textbf{LPCM}~\cite{6476013} were developed for blur evaluation, while \textbf{NJQA}~\cite{6691937} and \textbf{JPEG-NR}~\cite{wang2002no} aimed to assess noise and JPEG compression, respectively. 
\textbf{TLVQM}~\cite{korhonen2019two} computes diverse handcrafted quality-aware features, including spatial attributes, motion-induced statistics, and aesthetics features, then train a shallow regressor to predict video quality scores. 
\textbf{CORNIA}~\cite{6247789} and \textbf{HOSA}~\cite{7501619} employ unsupervised dictionary learning techniques to learn distortion codebooks from local features extracted from image patches, through which global quality-aware image representations are obtained.

\noindent\underline{\textbf{Type ii: Neurostatistics-based VQA.}}
NSS-based methods are effective due to the fact that high-quality natural images and videos consistently follow specific statistical patterns, which are systematically disrupted by distortions~\cite{mittal2015completely, mittal2012no}.
For example, \textbf{BRISQUE}~\cite{mittal2012no} computes a small set of spatial bandpass and locally normalized features, then uses a support vector regressor (SVR) to learn mapping from them to human opinion scores. 
Similarly, \textbf{GM-LOG}~\cite{xue2014blind} extracts statistical features from smoothed gradient and Laplacian-of-Gaussian bandpass spaces, while \textbf{HIGRADE}~\cite{kundu2017no} computes gradient features in the LAB and gradient of LAB color spaces. 
A variety of color spaces and perception-driven transforms were utilized to extract a larger number of perceptually relevant NSS features in the "bag of features" model called \textbf{FRIQUEE}~\cite{FRIQUEE}.

Building on top of IQA models, a variety of blind VQA methods have been proposed to analyze the perceptual quality based on spatio-temporal scene statistics. 
\textbf{V-BLIINDS}\cite{6705673} utilizes a spatio-temporal model of the statistics of DCT coefficients of local frame differences to predict perceptual video quality scores. 
\textbf{Li et al.}~\cite{li2016spatiotemporal} employ spatiotemporal natural video statistics in the 3D-DCT domain to predict perceptual quality. 
\textbf{VIDEVAL}~\cite{9405420} implements an ensemble model that fuses features selected from several top-performing BVQA models, using a supervised feature selection procedure on a large UGC superset.
\textbf{STS-QA}~\cite{9897565} extracts spatiotemporal statistical features along different orientations of video space-time slices (STS) that capture directional global motion, then regresses the feature vector into video quality predictions using a shallow learner.
\textbf{FAVER}~\cite{zheng2022faver} is the first NR-VQA model able to account for frame rate variations, using extended models of the natural statistics of space-time wavelet-decomposed video signals.

The majority of existing BIQA or BVQA algorithms belong to the ``opinion-aware'' category, wherein a learned regression model, either deep or shallow, is trained on databases of distorted videos that have been human-labeled in the form of mean opinion scores (MOS). 
However, these models can suffer from limited generalization capability on real-world images/videos, where multiple, unknown, commingled distortions may arise that are not present in the training set. 
Therefore, ``opinion-unaware'' (OU) or ``completely blind'' models which do not rely on training on human-labeled videos, have aroused considerable research insterest~\cite{6353522,7094273,mittal2015completely,ghadiyaram2017no,kancharla2021completely,liu2020blind,liu2019unsupervised}.
Among these, \textbf{NIQE}~\cite{6353522} and \textbf{IL-NIQE}~\cite{7094273} were designed using perceptually-inspired neuro-statistical features in the spatial domain that capture deviations from perceptually relevant NSS models. 
Following the design framework of NIQE~\cite{6353522}, \textbf{NPQI}~\cite{liu2020blind} explores NSS features from a local binary map and locally normalized coefficients of images, while \textbf{SNP-NIQE}~\cite{liu2019unsupervised} measures structural variations as well as naturalness deviations.
\textbf{VIIDEO}~\cite{mittal2015completely} models the temporal regularities of natural videos, using them to assess video quality.
A more recent completely blind BVQA model targeting UGC videos, called \textbf{STEM}~\cite{kancharla2021completely}, quantifies losses of ``perceptual straightness''~\cite{henaff2019perceptual} to measure temporal quality.
Similarly, \textbf{TPQI}~\cite{10.1145/3503161.3547849} measures the perceptual straightness and compactness of trajectories of video representations.
Another type of OU quality predictor is based on pseudo-references, such as \textbf{SLEEQ}~\cite{ghadiyaram2017no} and \textbf{BPRI}~\cite{8241862}.
%
The recent \textbf{VIQE}~\cite{9921340} model fuses a variety of patch- and frame-wise video statistics, to capture distortion-predictive space-time statistical irregularities of videos.

\subsubsection{\textbf{Deep Learning Based BIQA Methods}}
\label{dpl biqa}
Following Section~\ref{sssec:dl_fr_iqa}, we review deep learning-based blind IQA and VQA methods separately. 
Such a focused review will enable a deeper analysis of how these no-reference models handle temporal dynamics, motion artifacts, and frame consistency without relying on reference information, providing a more comprehensive understanding of how deep learning techniques are specifically adapted for video-related challenges. Similarly, we first discuss existing deep learning-based BIQA methods.

\noindent\underline{\textbf{Type i: CNN feature extraction and FC fusion.}}
This category of blind IQA models uses CNNs for feature extraction, followed by fully connected (FC) layers for quality score prediction. 
Some models~\cite{ying2020patches} adopt pre-trained networks, benefiting from the generalization ability gained from large datasets. 
This allows for faster convergence and robust performance. 
Others~\cite{kang2014convolutional,bosse2016deep,ma2017end} use end-to-end training with custom-built CNNs, which are tailored to specific IQA tasks, offering greater flexibility and potential for capturing unique distortion patterns. 
After feature extraction, FC layers fuse the features to produce final quality scores.
\textbf{Kang et al.}~\cite{kang2014convolutional} introduce an early BIQA framework using a five-layer CNN to predict patch-level quality, averaged to produce image-level scores.
\textbf{Bosse et al.}~\cite{bosse2016deep} utilize a deep CNN for patch-level quality predictions, aggregated using either average or weighted-average pooling.
\textbf{MEON}~\cite{ma2017end} combines distortion identification and quality prediction via two sub-networks sharing learned features, with outputs fused to estimate overall quality.
\textbf{NIMA}~\cite{talebi2018nima} and \textbf{PQR}~\cite{zeng2017probabilistic} predict probability distributions of quality scores instead of scalar values.
\textbf{DB-CNN}~\cite{zhang2018blind} assesses perceptual quality of synthetic and authentic distortions with two specialized networks, combined using a deep bilinear model.
\textbf{PaQ-2-PiQ}~\cite{ying2020patches} introduces models leveraging ResNet-18\cite{he2016deep} as the backbone and RoIPool~\cite{7410526} for aggregating patch and whole-image quality. The proposed P2P Feedback model, which concatenates patch and image scores for final prediction, achieves the best performance.
\textbf{Sun et al.}~\cite{10109108} propose a staircase neural network to capture multi-level features and use an iterative training strategy for mixed database training.

\noindent\underline{\textbf{Type ii: CNN feature extraction and special fusion.}}
This class of blind IQA methods builds on CNN-based feature extraction but incorporates specialized fusion techniques to enhance quality prediction. Unlike traditional FC fusion approaches, these models apply innovative strategies such as graph-based learning~\cite{9720138}, fuzzy logic~\cite{10.1145/3503161.3547872}, or hallucination~\cite{Lin_2018_CVPR,9894272} techniques to better capture the complex relationships between image distortions and perceived quality.
\textbf{QCN}~\cite{Shin_2024_CVPR} estimates quality by iteratively updating image embeddings and ``score pivots'' through geometric order learning, with the final score derived from the nearest pivot.
\textbf{GraphIQA}~\cite{9720138} uses distortion graph representations for perceptual quality modeling, enhanced by a discrimination network and a fuzzy prediction network.
\textbf{Gao et al.}~\cite{10.1145/3503161.3547872} introduce a fuzzy-based network that maps VGG16 image embeddings to quality score distributions.
\textbf{FPR}~\cite{Lin_2018_CVPR} and \textbf{HIQA}~\cite{9894272} assess quality by referencing hallucinated or pristine versions of distorted images.
\textbf{BIECON}~\cite{kim2016fully} mimics FR-IQA models by predicting local patch scores generated by an FR-IQA algorithm.

%
\noindent\underline{\textbf{Type iii: Transformer-based IQA.}}
Transformers have shown strong potential as a generalist model for broad computer vision tasks~\cite{dosovitskiy2020image,liu2021swin,liu2022video,tu2022maxvit,dong2022cswin}.
Transformer-based IQA models leverage the powerful self-attention mechanism of Transformers to capture both global and local dependencies in images, making them particularly well-suited for image quality assessment tasks.
\textbf{MUSIQ}~\cite{ke2021musiq} and \textbf{TRIQ}~\cite{you2021transformer} use Transformer encoders to process images of varying resolutions, employing embedding methods and learnable classification (CLS) tokens to predict quality scores via fully connected layers.
\textbf{DEIQT}~\cite{Qin_Hu_Liu_Zheng_Liu_Li_Zhang_2023} extends this approach by using CLS tokens from the Transformer encoder as queries in the decoder to learn quality-aware embeddings, which are regressed into quality scores using an MLP.
\textbf{TReS}~\cite{Golestaneh_2022_WACV} combines a CNN and a Transformer encoder to extract local and global features, with specialized loss functions improving prediction accuracy and robustness.
\textbf{MANIQA}~\cite{Yang_2022_CVPR} employs a transpose attention block and Swin Transformer to enhance interactions between global and local features extracted by a Vision Transformer.

\noindent\underline{\textbf{Type iv: Multi-task learning.}}
This category of blind IQA approaches focuses on multi-task learning, where models are designed to address challenges like overfitting, learning efficiency, and the ability to generalize to unseen distortion types, enabling more robust and scalable IQA solutions across multiple domains. They incorporate techniques such as continual learning, meta-learning, and incremental learning to adaptively improve the model’s capacity to assess image quality in diverse scenarios.
\textbf{Zhang et al.}~\cite{zhang2022continual} address continual learning in BIQA by enabling models to learn continually from IQA dataset streams.
\textbf{MetaIQA}~\cite{zhu2020metaiqa} uses meta-learning to create a prior NR-IQA model for distortion-specific tasks, fine-tuning it to predict quality for unknown distortions.
\textbf{Su et al.}~\cite{SU2023109047} propose a two-stream architecture with a shallow network learning distortion manifolds and a deep network refining quality predictions, enhanced by masked distortion labeling and gradual weighting strategies.
\textbf{SLIF}~\cite{10036133} performs multi-task quality assessment using incremental learning with scalable memory units that prune unimportant neurons during training.
\textbf{Li et al.}~\cite{10636288} introduce a channel modulation kernel to capture intra- and inter-domain attention, combined with a multi-dataset learning strategy for mixed datasets.
\textbf{Wang et al.}~\cite{10070789} develop an online mining pipeline where a full and pruned model are trained together, using prediction disagreements to mine hard examples for enhancing the full model.
\textbf{Zhang et al.}~\cite{10462930} propose task-specific normalization parameters for each dataset to produce weighted quality scores for overall predictions.

\noindent\underline{\textbf{Type v: Unsupervised and self-supervised learning.}}
This class of IQA methods focuses on unsupervised and self-supervised learning techniques, aiming to overcome reliance on large labeled datasets for quality prediction. By leveraging unsupervised learning or self-supervised learning, these approaches learn quality-aware representations without requiring explicit quality labels. These methods offer greater flexibility and scalability, particularly in scenarios where labeled data is scarce, while also promoting improved generalization across different datasets and distortion types.
\textbf{CONTRIQUE}~\cite{madhusudana2022image} is an unsupervised contrastive learning-based model that leverages a distortion classifier trained on diverse distortions, achieving SOTA performance across leading IQA databases.
\textbf{Shukla et al.}~\cite{Shukla_2024_WACV} use a variational autoencoder GAN trained on pristine images to assess quality by comparing latent distributions between pristine and distorted images.
\textbf{Re-IQA}~\cite{Saha_2023_CVPR} trains two encoders in an unsupervised setting to capture low-level quality and high-level content features, with a regression model mapping these to quality scores.
\textbf{Babu et al.}~\cite{Babu_2023_WACV} propose a self-supervised contrastive learning framework, pretraining on synthetic data and fine-tuning on authentic data using content separation and positive samples only.
\textbf{ARNIQA}~\cite{Agnolucci_2024_WACV} models the distortion manifold in a self-supervised manner, maximizing similarities between representations of equally distorted images to enhance robustness and generalization across datasets.
\textbf{Zhao et al.}~\cite{Zhao_2023_CVPR}, similar to CONTRIQUE, introduces a quality-aware pretext task and trains an IQA model using diverse augmented distortions constrained by a quality-aware contrastive loss.

\noindent\underline{\textbf{Type vi: Large multimodality model-based IQA.}}
\textbf{CLIP-IQA}~\cite{Wang_Chan_Loy_2023} assesses image quality by leveraging the vision-language prior in CLIP~\cite{pmlr-v139-radford21a}. It uses cosine similarity between image embeddings and text embeddings of antonym prompts (`good photo' and `bad photo') to compute quality scores via softmax. The variant \textbf{CLIP-IQA+} fine-tunes prompts for improved accuracy, supporting zero-shot and fine-tuned quality assessments.
Inspired by CLIP-IQA, \textbf{LIQE}~\cite{Zhangwx_2023_CVPR} uses a multitask learning scheme to optimize CLIP end-to-end, and \textbf{TCDs}~\cite{10235894} employs explicit content descriptions for aesthetics evaluation.

Emerging multimodal large language models (MMLLMs) demonstrate strong visual understanding capabilities~\cite{liu2023visualinstructiontuning,zhu2023minigpt4enhancingvisionlanguageunderstanding,dai2023instructblipgeneralpurposevisionlanguagemodels,li2023ottermultimodalmodelincontext}.
\textbf{Q-Bench}~\cite{wu2023q} explores MMLLMs for image perception, deriving quality scores through softmax pooling of `good' and `poor' tokens. However, its performance is limited by the lack of low-level visual datasets.
\textbf{Q-Instruct}~\cite{Wu_2024_CVPR} addresses this by fine-tuning MMLLMs on a curated low-level instruction dataset, while \textbf{Q-Align}~\cite{wu2023qalign} uses a human-emulating syllabus to improve quality score predictions.
\textbf{Co-Instruct}~\cite{wu2024towards} extends MMLLMs for visual comparisons by fine-tuning on the Co-Instruct-562K dataset with image-text interleaved features, achieving superior performance on multiple tasks.

\begin{figure}[!t]
    \centering
    \includegraphics[width=8.5cm]{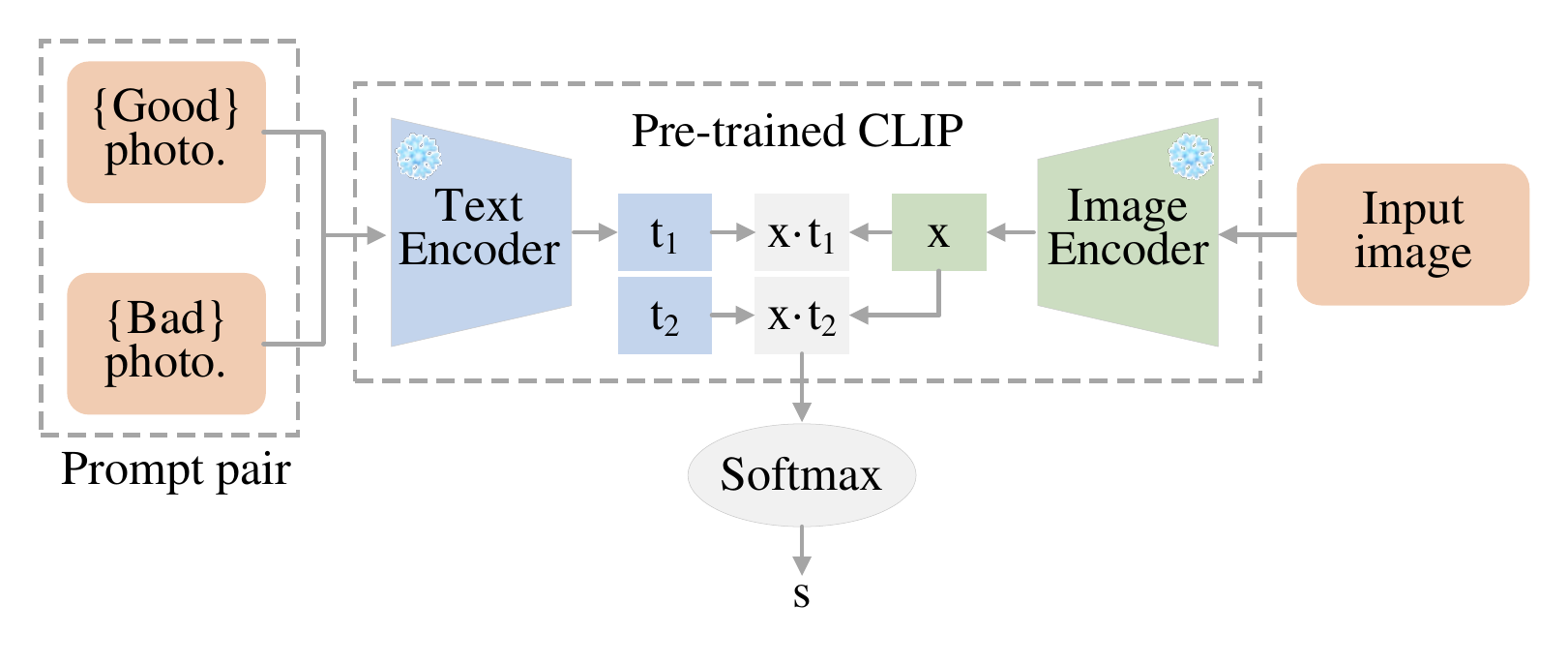}
    \vspace{-8pt}
    \caption{The framework of CLIP-IQA~\cite{Wang_Chan_Loy_2023}.}
    \vspace{-12pt}
    \label{fig:CLIP-IQA}
\end{figure}


\begin{table*}[!t]
\renewcommand{\arraystretch}{0.95}
\centering
\fontsize{6.2pt}{8pt}\selectfont
\caption{Overview of deep no-reference VQA methods.}
\label{table:nr_vqa}
\renewcommand\arraystretch{1.2}
\resizebox{\textwidth}{!}{%
\begin{tabular}{c|c|c|c|c}
\toprule 
\makebox[0.05\textwidth][c]{Type} & \makebox[0.05\textwidth][c]{Method} & \makebox[0.1\textwidth][c]{\begin{tabular}[c]{@{}c@{}} Architecture \\ (Feature extraction \\ + Quality fusion)\end{tabular}} &
\makebox[0.15\textwidth][c]{\begin{tabular}[c]{@{}c@{}} Core block \\ (Pretrained models or \\ crafted modules)\end{tabular}} &
\makebox[0.3\textwidth][c]{Key idea} \\
\hline
\\[-1.em]

\multirow{16}{*}{\begin{tabular}[c]{@{}c@{}} 2D CNNs\\ with simple \\ score/feature \\averaging \end{tabular}} & \begin{tabular}[c]{@{}c@{}}Domonkos Varga 
 \\ et al.~\cite{varga2019noPool}\end{tabular} & CNN + SVR & \begin{tabular}[c]{@{}c@{}}Inception-V3, \\Inception-ResNet-V2\end{tabular} & \begin{tabular}[c]{@{}c@{}}A pre-trained CNN computes frame-level feature \\ vectors after being fine-tuned on an image database.\end{tabular}\\
\cline{2-5}
& NAVE~\cite{martinez2019no} & \begin{tabular}[c]{@{}c@{}}Handcrafted features  \\ + Autoencoders\end{tabular}& \begin{tabular}[c]{@{}c@{}}Autoencoder, \\Classification Layer \end{tabular}& \begin{tabular}[c]{@{}c@{}} Deep autoencoder takes NSS and temporal-spatial features as \\ input and generates more descriptive quality features.\end{tabular}\\
\cline{2-5} 
& CNNTLVQM~\cite{korhonen2020blind} & \begin{tabular}[c]{@{}c@{}}Handcrafted Features \& CNN  \\+ LSTM/SVR\end{tabular} & \begin{tabular}[c]{@{}c@{}} TLVQM, ResNet-50, \\ Sobel Filters, LSTM \end{tabular}& \begin{tabular}[c]{@{}c@{}} VQA model that combines handcrafted motion features \\ and CNN-based spatial features pooled with spatial activity. \end{tabular}  \\
\cline{2-5}
& SIONR~\cite{wu2021semantic} & CNN + MLP& ResNet-50, GAP & \begin{tabular}[c]{@{}c@{}} Temporal variations of semantic features and low-level \\distortions are computed to predict frame-level quality.\end{tabular} \\
\cline{2-5}
& CenseoQoE~\cite{wen2021strong} & CNN + MLP& Shallow CNN, GAP & \begin{tabular}[c]{@{}c@{}} A unified model of both FR and NR quality assessment of \\ images and videos based on a backbone network and FC layers. \end{tabular} \\
\cline{2-5}
& RAPIQUE~\cite{tu2021rapique} & \begin{tabular}[c]{@{}c@{}}Handcrafted features \\ \& CNN + SVR\end{tabular} & ResNet-50 & \begin{tabular}[c]{@{}c@{}} A space-time bandpass statistics model that combines\\ quality-aware NSS features with deep semantic features. \end{tabular} \\
\cline{2-5} 
& \begin{tabular}[c]{@{}c@{}}SWDF-DF-\\VQA~\cite{varga2022no}\end{tabular} & CNN + SVR/GPR & Pre-trained CNN, SWGA & \begin{tabular}[c]{@{}c@{}} 
Parallel pre-trained CNNs extract quality distortions\\ which are then temporally pooled and saliency weighted. \end{tabular} \\
\cline{2-5}
& NR-VMAF~\cite{10564175} & CNN + MLP& \begin{tabular}[c]{@{}c@{}}Pre-trained CNN,\\ Visual saliency extraction  \end{tabular} & \begin{tabular}[c]{@{}c@{}} Deep features are extracted by pre-train CNNs from frame filtered\\ by visual saliency, which are regressed and averaged pooling.\end{tabular}\\
\cline{1-5}
\multirow{22}{*}{\begin{tabular}[c]{@{}c@{}} 2D CNN\\ with temporal \\ aggregation \\networks \end{tabular}} 
& \begin{tabular}[c]{@{}c@{}}MLSP-VQA\\-FF/RN~\cite{Flickrvid-150k} \end{tabular}& \begin{tabular}[c]{@{}c@{}}  CNN + MLP \\  or CNN + LSTM \end{tabular} & \begin{tabular}[c]{@{}c@{}} Inception-ResNet-v2, LSTM, \\ Feed Forward Network\end{tabular} & \begin{tabular}[c]{@{}c@{}} Global average pooling is performed on the activation maps \\ of all kernels in the stem of the pre-trained network\end{tabular} \\
\cline{2-5}
& \begin{tabular}[c]{@{}c@{}}Domonkos Varga 
 \\ et al.~\cite{varga2019noCNN}\end{tabular}
 & CNN + LSTM & Pre-trained CNN, LSTM & \begin{tabular}[c]{@{}c@{}}Frame-level feature vectors are computed by a pre-trained \\ CNN and transferred to an LSTM network. \end{tabular} \\
\cline{2-5}
&VSFA~\cite{li2019quality}  & \begin{tabular}[c]{@{}c@{}}CNN + GRU \&  Memory \\ effect modeling \end{tabular}& ResNet-50, GRU& \begin{tabular}[c]{@{}c@{}} Content-dependency and temporal-memory effects are modeled \\ by classification CNN and GRU networks, respectively.\end{tabular} \\
\cline{2-5} 
& MGQA~\cite{10648736}  & \begin{tabular}[c]{@{}c@{}}CNN + GRU \& Memory \\ effect modeling \end{tabular} & \begin{tabular}[c]{@{}c@{}} Grid mini-patch \\ sampling, VSFA \end{tabular} & \begin{tabular}[c]{@{}c@{}} Generate heavily squeezed videos by spatiotemporal sampling, \\from which a reliable VQA model predict quality scores.  \\  \end{tabular}  \\
\cline{2-5}
& RIRNet~\cite{chen2020rirnet} & \begin{tabular}[c]{@{}c@{}}CNN + RNN \& \\ Motion effect modeling \end{tabular}& ResNet-50, GRU & \begin{tabular}[c]{@{}c@{}} Motion effect of multiple temporal frequencies is characterized \\ by a hierarchical recurrent temporal modeling scheme.\end{tabular} \\
\cline{2-5} 
& STDAM~\cite{xu2021perceptual} & CNN + BiLSTM & \begin{tabular}[c]{@{}c@{}} ResNet-18, Graph \\ Convolution,BiLSTM\end{tabular} & \begin{tabular}[c]{@{}c@{}} A graph convolution and attention module enhances the deep \\frame-level features which are evolved by a BiLSTM network. \end{tabular}\\
\cline{2-5}
& MDTVSFA~\cite{li2021unified} & \begin{tabular}[c]{@{}c@{}}CNN + GRU \&  Memory \\ effect modeling \end{tabular} & VSFA, Nonlinear Mapping & \begin{tabular}[c]{@{}c@{}} VQA model learned using a mixed datasets \\ training strategy to account for various contents and distortions.\end{tabular} \\
\cline{2-5}
& AB-VQA~\cite{20219506420} & \begin{tabular}[c]{@{}c@{}}CNN + GRU \&  Memory \\ effect modeling \end{tabular} & VGG-16, GRU & \begin{tabular}[c]{@{}c@{}} Attention module extracts local distortions while GRU \\ and temporal pooling layers model memory effects. \end{tabular} \\
\cline{2-5}
& \begin{tabular}[c]{@{}c@{}}Junfeng Li \\et al.~\cite{li2022study}\end{tabular} & \begin{tabular}[c]{@{}c@{}}CNN + BiGRU \&  Memory \\ effect modeling \end{tabular}& \begin{tabular}[c]{@{}c@{}} ResNet-50, Inception-V3,\\ BiGRU\end{tabular} & \begin{tabular}[c]{@{}c@{}} Dual network generates frame features which are \\ processed by BiGRU and two temporal pooling modules.\end{tabular} \\
\cline{2-5}
& GSTVQA~\cite{20229452150} & \begin{tabular}[c]{@{}c@{}}CNN + GRU \& \\ Temporal pyramid pooling \end{tabular} & VGG-16, GRU & \begin{tabular}[c]{@{}c@{}} Unified Gaussian distribution constraints are imposed on deep \\features to obtain more generalized quality representations. \end{tabular} \\
\cline{2-5}
& 2BiVQA~\cite{telili20222bivqa} & \begin{tabular}[c]{@{}c@{}}CNN + BiLSTM \end{tabular}& Pre-trained CNN, Bi-LSTM & \begin{tabular}[c]{@{}c@{}} Two Bi-LSTM networks capture both short-range dependencies \\ and long-range dependencies to account for memory effects.\end{tabular} \\
\cline{1-5}
\multirow{20}{*}{\begin{tabular}[c]{@{}c@{}}3D CNNs\\/ Transformation \end{tabular}} & SACONVA~\cite{li2015no} & \begin{tabular}[c]{@{}c@{}}Handcrafted features\\ + CNN \end{tabular}& \begin{tabular}[c]{@{}c@{}}3D Shearlet Transform, \\ 1D CNN, Autoencoder\end{tabular} & \begin{tabular}[c]{@{}c@{}} Spatiotemporal features are extracted by a 3D shearlet transform \\and exaggerated into discriminative features by CNN. \end{tabular} \\
\cline{2-5} 
& V-MEON~\cite{liu2018end} & 3D CNN + MLP& 3D CNN, GDN & \begin{tabular}[c]{@{}c@{}} Multi-task DNN framework not only predicts perceptual \\quality but a probabilistic prediction of codec type. \end{tabular} \\
\cline{2-5}
 & \begin{tabular}[c]{@{}c@{}}Rui Hou \\ et al.~\cite{hou2020no}\end{tabular} & 2D CNN + 3D CNN & VGG-16, 3D CNN & \begin{tabular}[c]{@{}c@{}} Bin-based average pooling is applied on a 3D CNN to \\calculate spatiotemporal information while reducing over-fitting.\end{tabular} \\
\cline{2-5}
&Patch-VQ~\cite{ying2021patch} & \begin{tabular}[c]{@{}c@{}}2D \& 3D CNN + \\ Time series regression\end{tabular}&\begin{tabular}[c]{@{}c@{}} PaQ-2-PiQ, ResNet3D,\\ InceptionTime\end{tabular} & \begin{tabular}[c]{@{}c@{}} Space-time features are extracted by 2D and 3D network streams, \\then a time series network processes the pooled features. \end{tabular} \\
\cline{2-5}
& CoINVQ~\cite{wang2021rich} & 2D \& 3D CNN + FC & EfficientNet-b0, D3D & \begin{tabular}[c]{@{}c@{}} Video quality is analyzed from multiple aspects: \\content, distortion, compression level, and temporal aggregation. \end{tabular} \\
\cline{2-5}
& \begin{tabular}[c]{@{}c@{}}Wei Sun \\et al.~\cite{sun2022deep}\end{tabular} & 2D \& 3D CNN + FC & ResNet-50, SlowFast, MLP & \begin{tabular}[c]{@{}c@{}} Frame-level and chunk-level deep features are calculated,\\ regressed, and pooled to obtain video-level quality scores.\end{tabular} \\
\cline{2-5}
& \begin{tabular}[c]{@{}c@{}}Bowen Li \\et al.~\cite{li2022blindly}\end{tabular} & \begin{tabular}[c]{@{}c@{}} 2D \& 3D CNN + GRU \&  \\ Memory effect modeling \end{tabular} & ResNet-50, SlowFast, GRU & \begin{tabular}[c]{@{}c@{}} Transfers knowledge from IQA datasets of authentic distortions \\ and large-scale action recognition to learn feature extractors.\end{tabular} \\
\cline{2-5} 
& MD-VQA~\cite{Zhang_2023_CVPR}& \begin{tabular}[c]{@{}c@{}}2D \& 3D CNN \& \\ Handcrafted features + MLP\end{tabular}& \begin{tabular}[c]{@{}c@{}}EfficientNetV2, ResNet3D-18, \\Handcrafted features \end{tabular} & \begin{tabular}[c]{@{}c@{}}Pretrained 2D CNN, 3D CNN, and handcrafted distortion descriptors \\ separately extract semantic, distortion, and motion quality-aware features. \end{tabular}\\
\cline{2-5} 
& UCDA~\cite{Chen_2021_ICCV}& 3D CNN + MLP & \begin{tabular}[c]{@{}c@{}} C3D, unsupervised \\domain adaptation \end{tabular}& \begin{tabular}[c]{@{}c@{}} A curriculum-style unsupervised domain adaptation for cross-domain \\generalization by progressively adapting based on prediction confidence.\end{tabular}\\
\cline{2-5} 
& Wenhao Shen et al.~\cite{10375568} & 3D CNN + MLP & \begin{tabular}[c]{@{}c@{}}3D CNN, Spatiotemporal \\ pyramid attention \end{tabular}& \begin{tabular}[c]{@{}c@{}} Hierarchical motion information at different temporal scales is fed into \\spatiotemporal pyramid attention for cross-scale dependency modeling. \end{tabular}\\
\cline{1-5} 
\multirow{6}{*}{\begin{tabular}[c]{@{}c@{}} Transformer-based \\ models\end{tabular}} & StarVQA~\cite{xing2021starvqa} & Transformer + MLP&\begin{tabular}[c]{@{}c@{}} Encoder, Time Attention, \\Space Attention\end{tabular} & \begin{tabular}[c]{@{}c@{}} First work to leverage Transformers on the VQA problem. \\A vectorized loss function is designed to help train the Encoder. \end{tabular}  \\
\cline{2-5} 
 &Junyong You~\cite{you2021long} & \begin{tabular}[c]{@{}c@{}} CNN + Transformer \\ \& MLP \end{tabular}& \begin{tabular}[c]{@{}c@{}}Encoder, Channel Attention,\\ Spatial Attention\end{tabular} & \begin{tabular}[c]{@{}c@{}} A long short-term convolutional Transformer architecture predicts \\perceptual video quality from frame features extracted by a CNN. \end{tabular} \\
\cline{2-5}
& DisCoVQA~\cite{wu2022discovqa} & \begin{tabular}[c]{@{}c@{}} Transformer + \\ Transformer \& MLP \end{tabular}& \begin{tabular}[c]{@{}c@{}}Swin-Transformer, \\ Transformer Encoder, Decoder\end{tabular} & \begin{tabular}[c]{@{}c@{}} Temporal variations and quality attention effects are modeled \\by a distortion extraction module and a content attention module.\end{tabular} \\
\bottomrule
\end{tabular}
}
\end{table*}

\begin{table*}[!t]
\renewcommand{\arraystretch}{0.95}
\centering
\fontsize{6.2pt}{8pt}\selectfont
\caption*{(Continued)}
\renewcommand\arraystretch{1.2}
\resizebox{\textwidth}{!}{%
\begin{tabular}{c|c|c|c|c}
\toprule 
\makebox[0.05\textwidth][c]{Type} & \makebox[0.05\textwidth][c]{Method} & \makebox[0.1\textwidth][c]{\begin{tabular}[c]{@{}c@{}} Architecture \\ (Feature extraction + \\ Quality fusion)\end{tabular}} &
\makebox[0.15\textwidth][c]{\begin{tabular}[c]{@{}c@{}} Core block \\ (Pretrained models or \\ crafted modules)\end{tabular}} &
\makebox[0.3\textwidth][c]{Key idea} \\
\hline
\\[-1.em]

\multirow{8}{*}{\begin{tabular}[c]{@{}c@{}} Transformer \\-based  models\end{tabular}} 
& \begin{tabular}[c]{@{}c@{}}FAST-VQA~\cite{wu2022fast}/\\FasterVQA~\cite{10264158}\end{tabular} & Transformer + MLP& \begin{tabular}[c]{@{}c@{}} Swin-Transformer, \\ Grid Mini-patch Sampling\end{tabular}& \begin{tabular}[c]{@{}c@{}} A grid mini-patch sampling scheme is used to\\ reduce the computational cost of high-resolution VQA.\end{tabular} \\ 
\cline{2-5} 
& SAMA~\cite{Liu_Quan_Xiao_Li_Wu_2024} & Transformer MLP & \begin{tabular}[c]{@{}c@{}} Multi-granularity sampling, \\ Swin Transformer\end{tabular} & \begin{tabular}[c]{@{}c@{}} A novel visual data sampling strategy is proposed based on\\ multi-granularity pyramid sampling and spatiotemporal masking.\end{tabular}\\
\cline{2-5}  
& DOVER~\cite{Wu_2023_ICCV} & \begin{tabular}[c]{@{}c@{}} a) Transformer + MLP \\ b) CNN + Cosine Similarity  \end{tabular}& \begin{tabular}[c]{@{}c@{}} inflated-ConvNext, \\ Video Swin Transformer \end{tabular} & \begin{tabular}[c]{@{}c@{}} Decompose video quality perception into aesthetic and technical perspectives \\and extract features from two pre-trained models.\end{tabular}\\
\cline{2-5} 
& SSL-VQA~\cite{Mitra_Soundararajan_2024} & \begin{tabular}[c]{@{}c@{}} Transformer + 3D CNN \\ \& Distribution distance  \end{tabular} & \begin{tabular}[c]{@{}c@{}} Video Swin Transformer, \\Knowledge transfer  \end{tabular} & \begin{tabular}[c]{@{}c@{}} Integrate a knowledge transfer mechanism within a semi-supervised learning\\ framework to effectively utilize a limited amount of labeled data. \end{tabular}\\
\cline{1-5} 
\multirow{18}{*}{\begin{tabular}[c]{@{}c@{}} Large \\multimodality \\model-based \\methods\end{tabular}} & PTM-VQA~\cite{Yuan_2024_CVPR} & \begin{tabular}[c]{@{}c@{}} 2D \& 3D CNN \&  CLIP \\ visual encoder + MLP \end{tabular}& \begin{tabular}[c]{@{}c@{}} Pretrained model set\\ including CLIP  \end{tabular} & \begin{tabular}[c]{@{}c@{}} Features extracted from pretrained models are integrated to learn \\ final quality representations and scores supervised by diverse loss. \end{tabular} \\
\cline{2-5} 
 & COVER~\cite{He_2024_CVPR} & \begin{tabular}[c]{@{}c@{}}2D CNN \&  CLIP \\visual encoder + MLP \end{tabular}& \begin{tabular}[c]{@{}c@{}} CLIP, ConvNet, \\ Swin Transformer  \end{tabular}& \begin{tabular}[c]{@{}c@{}} Model video quality in semantic, aesthetic, and technical branches with \\ features fused by cross-gating blocks and regressed into quality score. \end{tabular}\\
\cline{2-5} 
 & KSVQE~\cite{Lu_2024_CVPR} &\begin{tabular}[c]{@{}c@{}} 3D Transformer +\\  Attention \& MLP \end{tabular}& \begin{tabular}[c]{@{}c@{}} 3D-Swin Transformer, CLIP, \\ Cross-/self-attention \end{tabular}& \begin{tabular}[c]{@{}c@{}} A CLIP-based content understanding module select quality-aware patches to\\ fed into Transformer-based quality regression module for quality prediction.   \end{tabular}\\
\cline{2-5} 
 & Wen Wen~\cite{Wen_2024_CVPR}  & \begin{tabular}[c]{@{}c@{}} CLIP visual encoder \\ + MLP \end{tabular}& \begin{tabular}[c]{@{}c@{}} CLIP, ResNet-18, \\ SlowFast \end{tabular}& \begin{tabular}[c]{@{}c@{}} A base quality predictor responds to basic visual distortion, with \\ spatial and temporal rectifiers capturing resolution and framerate changes. \end{tabular}\\
\cline{2-5} 
 & BUONA-VISTA~\cite{wu2023exploringopinionunawarevideoquality} & \begin{tabular}[c]{@{}c@{}} a) CLIP + Cosine similarity \\ b) Crafted scores + Pooling \\ c) Crafted scores + Normalization\end{tabular} & CLIP, NIQE, TPQI & \begin{tabular}[c]{@{}c@{}} CLIP with multi-pair antonym prompts forms the semantic quality index\\  while spatial and temporal quality is compensated by NIQE and TPQI. \end{tabular}\\
\cline{2-5} 
 & MaxVQA~\cite{wu2023towards} & \begin{tabular}[c]{@{}c@{}}CLIP \& Transformer\\ + Cosine similarity  \end{tabular}& CLIP, FAST-VQA &  \begin{tabular}[c]{@{}c@{}} Cosine similarity is computed between CLIP textual features \\  of learnable contrastive prompts and fused features from CLIP and FAST-VQA.    \end{tabular}\\
\cline{2-5} 
 & ZE-FESG~\cite{10448422} & \begin{tabular}[c]{@{}c@{}} CLIP + \\ Cosine similarity \end{tabular} & CLIP & \begin{tabular}[c]{@{}c@{}}  52 descriptions accounting for technical and abstract aspects as well as \\ the video frames are fed into CLIP to obtain features of multiple dimensions.  \end{tabular}\\
\cline{2-5} 
 & Q-Align~\cite{wu2023qalign} &\begin{tabular}[c]{@{}c@{}} LMM + rating \\ level conversion \end{tabular}& mPLUG-Owl-2 (LMM) & \begin{tabular}[c]{@{}c@{}}  LMMs are firstly fine-tuned to rate like a human with rating levels and \\ then fed with discrete video frames to rate video quality level.  \end{tabular}\\
\cline{2-5} 
 & LMM-VQA~\cite{ge2024lmm} & \begin{tabular}[c]{@{}c@{}} CLIP \& 3D CNN + LLM \end{tabular}& \begin{tabular}[c]{@{}c@{}} Llama-3-8b-Instruct (LLM), \\ CLIP, SlowFast  \end{tabular} & \begin{tabular}[c]{@{}c@{}} Spatial features from CLIP and temporal feature from SlowFast are projected into \\ text-guided tokens, and are fed into LLM along with prompt tokens.\end{tabular} \\
\bottomrule
\end{tabular}
}
\vspace{-12pt}
\end{table*}

\subsubsection{\textbf{Deep Learning Based BVQA Methods}}
Unlike IQA, BVQA must address the temporal dimension, making temporal distortion modeling a critical challenge, especially without reference information. This requires inferring quality solely from distorted content. Temporal aggregation methods range from simple score averaging to advanced techniques like 3D CNNs and attention mechanisms, enabling models to capture temporal dependencies and assess quality effectively.
Recent deep learning-based BVQA models, listed in Table~\ref{table:nr_vqa}, are categorized into five classes based on their architectures and temporal aggregation strategies. Following~\ref{dp_frvqa}, our discussion focuses on their design approaches.

\noindent\underline{\textbf{Type i: 2D CNNs with simple score/feature averaging.}}
A straightforward approach for video quality prediction involves averaging frame-level scores or features from spatial-only neural networks. While computationally efficient and leveraging established image-based models, these methods often overlook temporal dynamics.
Some approaches, like \textbf{SIONR}~\cite{wu2021semantic}, combine spatial and temporal features or calculate frame-wise variations, but still prioritize spatial analysis. Simple pooling methods, such as \textbf{GAP}~\cite{varga2022no,wen2021strong,wu2021semantic}, offer efficiency but may miss critical spatial and temporal details. Advanced techniques, like saliency-weighted pooling in \textbf{SWDF-DF-VQA}~\cite{varga2022no}, address this by focusing on regions of interest. These methods balance simplicity and efficiency while limiting temporal complexity.
\begin{itemize}[leftmargin=0em, itemindent=1em, itemsep=0pt, parsep=0pt]
    \item \textbf{Varga et al.}~\cite{varga2019noPool} use Inception-V3\cite{szegedy2016rethinking} and Inception-ResNet-V2~\cite{szegedy2017inception} for frame-level feature extraction, applying various pooling methods (\textit{e.g.}, average, median) to obtain video-level features. SVRs map these features to quality scores.
    \item \textbf{NAVE}~\cite{martinez2019no} employs deep autoencoders to process a global feature matrix of extracted NSS and spatial-temporal features. Frame-level quality scores are obtained via a DeepNet and averaged to estimate video-level quality.
    \item \textbf{CNN-TLVQM}~\cite{korhonen2020blind} combines TLVQM temporal features with ResNet-50 spatial features which are weighted by a spatial activity map. Temporal quality variations are modeled either using LSTM or pooling with an SVR.
    %
    \item \textbf{SIONR}~\cite{wu2021semantic} fuses temporal variations in high-level semantic features (ResNet-50) and low-level statistical features, predicting frame-level scores that are temporally pooled for video-level quality.
    %
\item \textbf{CenseoQoE}~\cite{wen2021strong} applies a lightweight ConvNet to extract frame-level features, with scores averaged through GAP to produce video-level quality.
    %
    \item \textbf{RAPIQUE}~\cite{tu2021rapique} combines handcrafted neurostatistical spatial-temporal features with ResNet-50 deep features to produce frame-level vectors, pooled and mapped via SVR for video quality scores.
    %
    \item \textbf{SWDF-DF-VQA}~\cite{varga2022no} uses multiple pre-trained CNNs for frame-level feature extraction, applying saliency-weighted GAP to emphasize important regions. Frame-level features are pooled and regressed to predict video quality.
    \item \textbf{NR-VMAF}~\cite{10564175} targets compression and scaling artifacts by extracting patch-level features using CNN/DNN, aggregating frame scores into video-level quality predictions.
    
\end{itemize}

\noindent\underline{\textbf{Type ii: 2D CNN with temporal aggregation networks.}}
Temporal quality modeling is enhanced in models that capture frame-level interactions and temporal memory effects. Models like \textbf{MLSP-VQA}~\cite{Flickrvid-150k}, \textbf{Varga et al.}~\cite{varga2019noCNN}, and \textbf{VSFA}~\cite{li2019quality} use recurrent networks (LSTMs or GRUs) to aggregate frame-level features, addressing motion and quality fluctuations overlooked by earlier frame-based approaches.
\textbf{VSFA}~\cite{li2019quality} and \textbf{AB-VQA}~\cite{20219506420} emphasize perceptual hysteresis effects, simulating non-linear human responses to quality drops and recoveries for more accurate predictions.
Models like \textbf{RIRNet}~\cite{chen2020rirnet} and \textbf{STDAM}~\cite{xu2021perceptual} incorporate spatial pyramid pooling and multi-scale attention mechanisms to capture both local and global temporal features, handling complex video content such as rapid motion or subtle changes.
Attention mechanisms in \textbf{STDAM}~\cite{xu2021perceptual} and \textbf{GSTVQA}~\cite{20229452150} enhance quality predictions by focusing on salient regions and key frames, balancing spatial and temporal distortions for videos with uneven quality.
\begin{itemize}[leftmargin=0em, itemindent=1em, itemsep=0pt, parsep=0pt]
\item \textbf{MLSP-VQA}~\cite{Flickrvid-150k} uses Inception-ResNet-v2~\cite{szegedy2017inception} to extract spatial features from video frames, which are pooled via GAP to form frame-level feature vectors. Temporal aggregation is achieved using either an LSTM or a temporal CNN, enabling video-level quality predictions.

\item \textbf{Varga et al.}~\cite{varga2019noCNN} employs AlexNet~\cite{krizhevsky2014one}, Inception-V3~\cite{szegedy2016rethinking}, and Inception-ResNet-V2~\cite{szegedy2017inception} pre-trained on KoNViD-10K for transfer learning. Frame-level features are extracted and processed through a two-layer LSTM with a fully connected layer to predict video quality.

\item \textbf{VSFA}~\cite{li2019quality} combines ResNet-50~\cite{he2016deep}-based feature extraction with GRUs to capture temporal dependencies. It models perceptual hysteresis effects by combining memory quality and current quality into frame-level scores, which are pooled into overall video quality predictions.
\begin{figure}[!t]
    \centering
    \includegraphics[width=8.5cm]{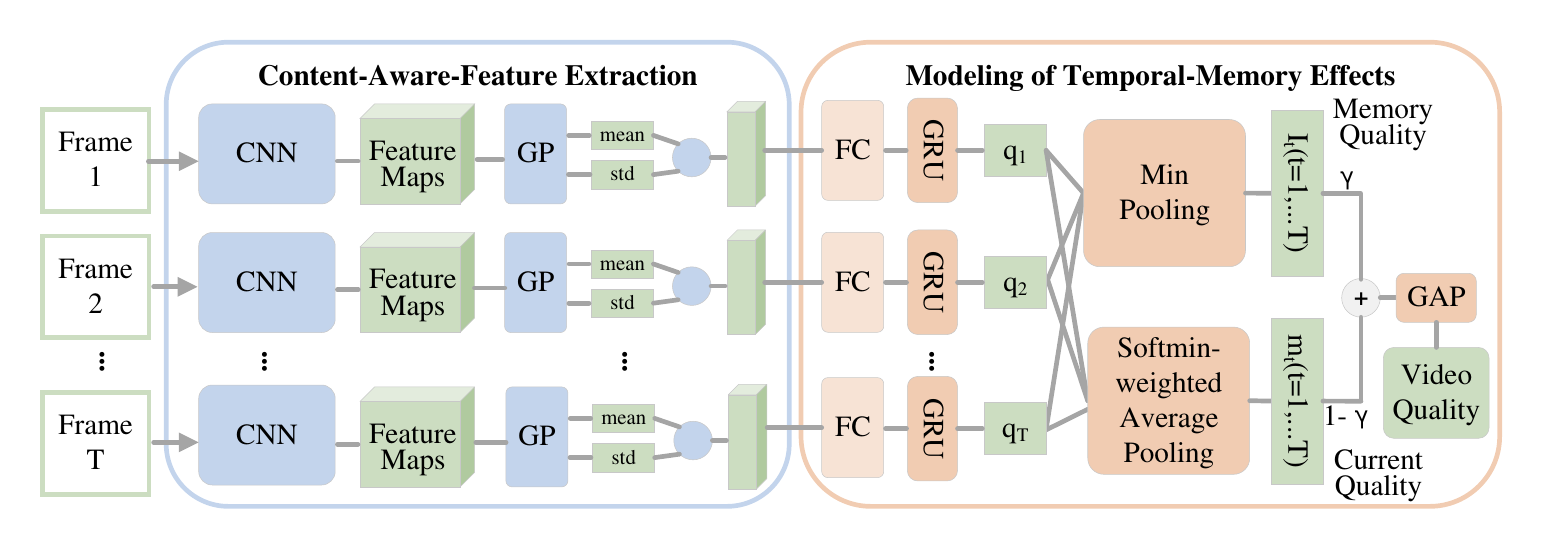}
    \vspace{-5pt}
    \caption{The framework of VSFA~\cite{li2019quality}.}
    \vspace{-15pt}
    \label{fig:vsfa}
\end{figure}
\item \textbf{MGQA}~\cite{10648736} builds on VSFA by incorporating spatial and temporal grid sampling. It evaluates classical temporal sampling methods from video understanding tasks and spatial Grid Mini-patch Sampling (GMS), and uses them to generate squeezed video representations for efficient quality assessment.

\item \textbf{RIRNet}~\cite{chen2020rirnet} uses ResNet-50~\cite{he2016deep} with spatial pyramid pooling~\cite{he2015spatial} to extract motion-aware features. A recurrent temporal dimension (RTD) module models motion over scales, while a recurrent temporal resolution (RTR) module processes nested temporal features, producing final quality predictions via a linear layer.

\item \textbf{STDAM}~\cite{xu2021perceptual} combines graph convolutions, attention mechanisms, and optical flow modules. Graph convolutions extract multi-scale features, while channel and spatial attention enhance salient regions. Motion features are captured using optical flow, and bi-LSTMs aggregate frame-level features for final predictions.

\item \textbf{MDTVSFA}~\cite{li2021unified} decomposes VQA into relative, perceptual, and subjective quality tasks. Relative quality is modeled with VSFA as the backbone, while perceptual alignment uses a 4-parameter nonlinear mapping and dataset-specific adjustments via an FC layer account for subjective quality. Mixed data training enhances generalization.

\item \textbf{AB-VQA}~\cite{20219506420} focuses on UGC videos with complex, uneven distortions. VGG16~\cite{simonyan2014very} extracts frame-level features, attention modules capture long-range dependencies, and HVS-inspired temporal pooling in VSFA models perceptual hysteresis effects. Final quality scores are pooled via minimum and softmin-weighted averaging methods.

\item \textbf{Li et al.}~\cite{li2022study} combines ResNet-50~\cite{he2016deep} and Inception-V3~\cite{szegedy2016rethinking} for feature extraction, with global pooling and convolution fusion. A bi-GRU models temporal dependencies, while regression blocks, including a temporal memory and Gaussian regression block, refine predictions. Final quality scores are obtained via a weighted sum of components.

\item \textbf{GSTVQA}~\cite{20229452150} extracts frame-level features using VGG16~\cite{simonyan2014very}, which are processed with channel attention to create multi-scale features. A Gaussian normalization layer reduces domain gaps, and pyramid pooling aggregates short- and long-term quality features for robust generalization.

\item \textbf{2BiVQA}~\cite{telili20222bivqa} extracts patch-level features from video frames using a pre-trained CNN. Bi-LSTM modules handle spatial pooling of patches and temporal pooling of frames, with an FC layer producing overall quality predictions.
\end{itemize}

\noindent\underline{\textbf{Type iii: 3D CNNs / Transformation.}} 
3D CNN-based models process consecutive frames as 3D space-time volumes, enabling simultaneous spatial and temporal feature extraction. This approach, used in models like \textbf{V-MEON}~\cite{liu2018end}, \textbf{You et al.}~\cite{you2019deep}, and \textbf{Hou et al.}~\cite{hou2020no}, naturally integrates motion dynamics and simplifies architectures compared to 2D CNNs with additional temporal pooling
mechanisms. Advanced methods like \textbf{Li et al.}~\cite{li2022blindly} and \textbf{Sun et al.}~\cite{sun2022deep} further refine predictions by incorporating HVS-inspired hysteresis pooling, aligning better with human perception. Multi-scale approaches, such as \textbf{Patch-VQ}~\cite{ying2021patch} and \textbf{Shen et al.}~\cite{10375568}, leverage pyramidal structures and attention mechanisms to capture fine-grained and global spatiotemporal features. Hybrid models like \textbf{CoINVQ}~\cite{wang2021rich} and \textbf{MD-VQA}~\cite{Zhang_2023_CVPR} combine spatial, motion, and semantic features to handle diverse distortions and artifacts. \textbf{SACONVA}~\cite{li2015no} utilizes 3D shearlet transforms to extract spatiotemporal features before feeding them into a CNN.
\begin{itemize}[leftmargin=0em, itemindent=1em, itemsep=0pt, parsep=0pt]
\item \textbf{SACONVA}~\cite{li2015no} uses 3D shearlet transforms for spatiotemporal feature extraction, and features are then processed by a 1D CNN initialized with autoencoders. Logistic regression predicts quality, while a softmax classifier identifies distortion types.

\item \textbf{V-MEON}~\cite{liu2018end} combines 2D and 3D CNN layers to extract features for both quality prediction and codec classification. Final scores are derived as the inner product of probability and quality vectors.

\item \textbf{You et al.}~\cite{you2019deep} employs a 3D CNN for clip-level scores, aggregated via an LSTM and fully connected layers to produce video-level quality predictions.

\item \textbf{Hou et al.}~\cite{hou2020no} extracts spatial features from eight-frame blocks using VGG-Net~\cite{simonyan2014very}, with a 3D CNN and bin-based pooling to model temporal and spatial dynamics for quality prediction.

\item \textbf{Patch-VQ}~\cite{ying2021patch} combines 2D and 3D features with ROIPool and SOIPool layers. An InceptionTime network~\cite{ismail2020inceptiontime} predicts local and global quality, leveraging labels from the proposed LSVQ database.

\item \textbf{CoINVQ}~\cite{wang2021rich} uses ContentNet, DistortionNet, and CompressionNet to capture content type, distortion type, and compression levels, aggregated through temporal pooling for final predictions.

\item \textbf{Sun et al.}~\cite{sun2022deep} combines ResNet-50 for spatial features and SlowFast R50 for motion features. Multi-scale fusion weights quality scores computed at three resolutions for improved accuracy.

\item \textbf{Li et al.}~\cite{li2022blindly} integrates ResNet-50~\cite{he2016deep} spatial features and SlowFast~\cite{Feichtenhofer_2019_ICCV} motion features. Temporal pooling via HVS-inspired hysteresis yields final video quality predictions.

\item \textbf{MD-VQA}~\cite{Zhang_2023_CVPR} extracts hybrid features using EfficientNetV2 for semantics, ResNet3D-18 for motion, and handcrafted features for UGC live video quality predictions.

\item \textbf{UCDA}~\cite{Chen_2021_ICCV} uses a pre-trained C3D~\cite{Tran_2015_ICCV} backbone for feature extraction, applying unsupervised domain adaptation between labeled source videos and unlabeled target videos.

\item \textbf{Shen et al.}~\cite{10375568} captures multi-scale motion through a spatiotemporal feature pyramid, with pyramid attention modules for channel and spatial selective sensitivity.
\end{itemize}

\noindent\underline{\textbf{Type iv: Transformer-based models.}} 
Transformers, built on self-attention mechanisms, excel at modeling long-range and contextual dependencies, making them highly effective for video quality assessment, where spatial and temporal features must be captured across extended sequences.
Attention modules in Transformer-based models selectively focus on important spatial-temporal regions. For example, \textbf{DisCoVQA}~\cite{wu2022discovqa} uses distortion-aware tokens to identify degraded regions, while \textbf{StarVQA}~\cite{xing2021starvqa} applies attention to non-overlapping space-time patches, enhancing robustness across diverse content types and uneven distortions.
Efficient handling of large video data is a key focus. Models like \textbf{FAST-VQA}~\cite{wu2022fast} and \textbf{FasterVQA}~\cite{10264158} reduce computational costs using techniques like Grid Mini-patch Sampling (GMS), preserving sensitivity to quality features while improving efficiency.
Multi-level feature extraction and hierarchical attention are also prominent trends. Models like \textbf{PHIQNet}~\cite{you2021long} and \textbf{DisCoVQA}~\cite{wu2022discovqa} extract features at spatial, temporal, and clip levels, then apply attention mechanisms to fuse these into unified quality predictions.
This streamlined overview highlights Transformers' adaptability and efficiency in video quality assessment.
\begin{itemize}[leftmargin=0em, itemindent=1em, itemsep=0pt, parsep=0pt]
\item \textbf{StarVQA}~\cite{xing2021starvqa} divides video frames into non-overlapping space-time patches, which are encoded into spatiotemporal vectors with a learnable label token. Encoding blocks apply temporal and spatial attention, and outputs are processed via LayerNorm and a fully connected (FC) network. A vectorized regression loss converts outputs into quality scores using softmax and linear SNR mapping.
    \begin{figure}[!t]
    \centering
    \includegraphics[width=8.5cm]{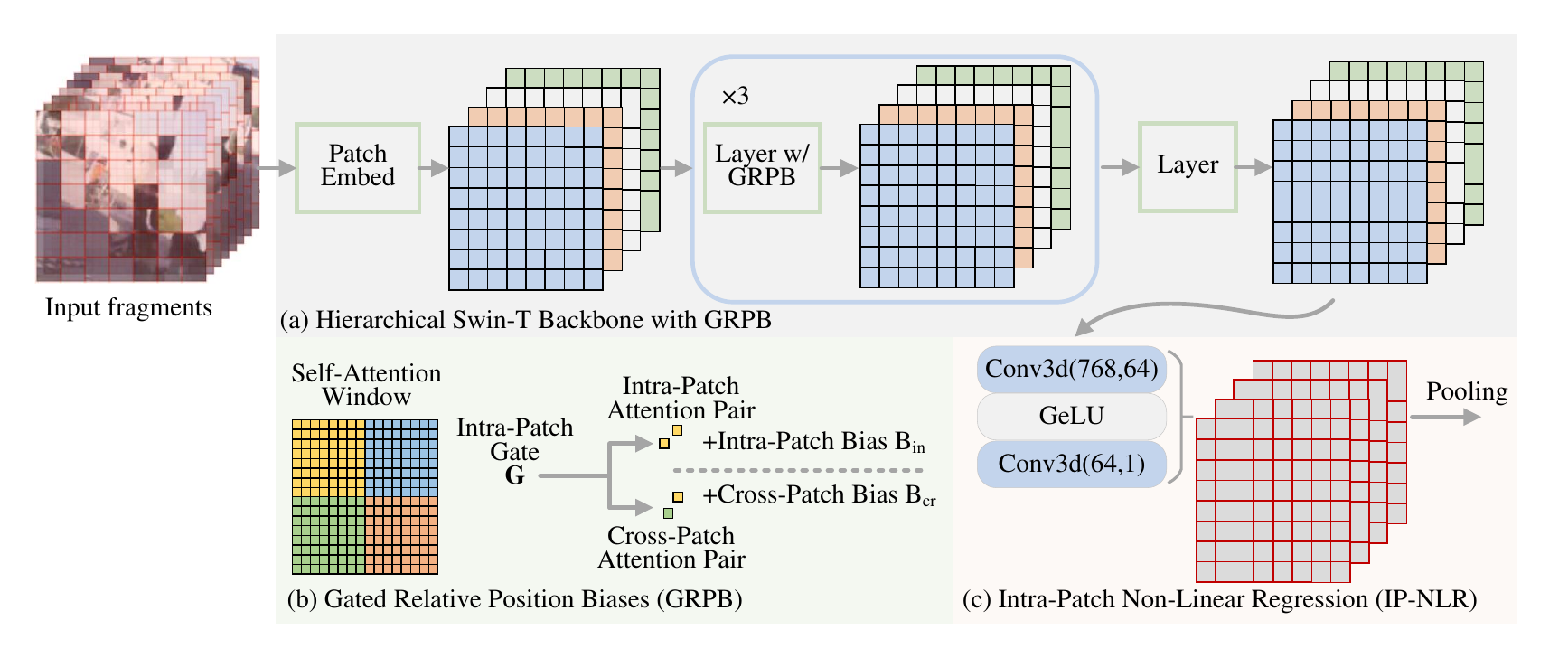}
    \vspace{-5pt}
    \caption{The framework of FAST-VQA~\cite{wu2022fast}.}
    \vspace{-15pt}
    \label{fig:fastvqa}
\end{figure}
\item \textbf{PHIQNet}~\cite{you2021long} combines ResNet-50~\cite{he2016deep} pyramid feature maps with channel and spatial attention blocks to produce global feature vectors. A long short-term convolutional Transformer (LSCT) fuses clip-level features (processed by a 1D CNN) into overall quality predictions, capturing both short- and long-term dependencies.

\item \textbf{DisCoVQA}~\cite{wu2022discovqa} utilizes a Spatial-Temporal Distortion Extraction (STDE) module with a Swin-T backbone to extract distortion-aware tokens and scores. A Temporal Content Transformer (TCT) computes attention weights on these scores, yielding final video quality predictions.

\item \textbf{FAST-VQA}~\cite{wu2022fast} reduces computational cost by 97.6\% through Grid Mini-patch Sampling (GMS), extracting spatially aligned mini-patches (fragments). These fragments are processed by a Fragment Attention Network (FANet), which uses Gated Relative Position Biases (GRPB) and intra-patch non-linear regression (IP-NLR) to predict quality. \textbf{FasterVQA}~\cite{10264158} introduces a temporal GMS module for 4x greater efficiency than FAST-VQA.
\item \textbf{SAMA}~\cite{Liu_Quan_Xiao_Li_Wu_2024} implements a scalable and masking-based video sampling method to preprocess input data into multi-scale representations, improving compatibility with base IQA/VQA models.

\item \textbf{DOVER}~\cite{Wu_2023_ICCV} splits quality evaluation into two branches: a technical branch using Video Swin Transformer~\cite{Liu_2022_CVPR_vst} for objective metrics and an aesthetic branch using inflated-ConvNext~\cite{9879745} for subjective preferences. The two branches supervise separate quality components and combine for overall predictions.

\item \textbf{SSL-VQA}~\cite{Mitra_Soundararajan_2024} leverages semi-supervised learning with a Video Swin Transformer trained via contrastive loss for quality-aware representations. A regressor and a distance model leverage labeled and unlabeled data to generate final quality scores by averaging predictions.

\end{itemize}


\noindent\underline{\textbf{Type v: Large multimodality model-based VQA methods.}}
The visual encoder of CLIP~\cite{pmlr-v139-radford21a}, trained on the WebImageText database, effectively models semantic-level quality of video frames. Models like \textbf{PTM-VQA}~\cite{Yuan_2024_CVPR}, \textbf{COVER}~\cite{He_2024_CVPR}, \textbf{KSVQE}~\cite{Lu_2024_CVPR}, and \textbf{Wen et al.}~\cite{Wen_2024_CVPR} utilize CLIP’s pre-trained encoder for semantic feature extraction to enhance quality prediction.
CLIP’s natural language supervision further enables it to generate specific quality descriptions by matching visual and textual features. Models like \textbf{BUONA-VISTA}~\cite{wu2023exploringopinionunawarevideoquality}, \textbf{MaxVQA}~\cite{wu2023towards}, and \textbf{ZE-FESG}~\cite{10448422} use antonym prompts and frame embeddings to calculate quality scores based on feature affinities.
Large multimodal models (LMMs) extend this capability by combining robust visual and textual representations. For instance, \textbf{Q-Align}~\cite{wu2023qalign} predicts human-aligned quality ratings, while \textbf{LMM-VQA}~\cite{ge2024lmm} integrates spatial and temporal tokens to derive text-based quality scores. These LMM-based approaches showcase the potential to model complex video quality patterns across diverse contexts.

\begin{itemize}[leftmargin=0em, itemindent=1em, itemsep=0pt, parsep=0pt]
\item \textbf{PTM-VQA}~\cite{Yuan_2024_CVPR} uses pre-trained models, including CLIP's visual encoder, to extract quality-related features integrated into a unified latent space. Intra-Consistency and Inter-Divisibility (ICID) losses ensure feature consistency and separability for effective clustering, achieving state-of-the-art performance on multiple VQA datasets.

\item \textbf{COVER}~\cite{He_2024_CVPR} evaluates video quality across technical, aesthetic, and semantic dimensions. A Swin Transformer assesses technical quality, a ConvNet analyzes aesthetics, and CLIP captures semantic features. A cross-gating block fuses these branches to deliver holistic quality scores, excelling on UGC datasets.

\item \textbf{KSVQE}~\cite{Lu_2024_CVPR} is designed for short-form video quality, combining a 3D Swin Transformer backbone with modules for quality-aware region selection (QRS), content-adaptive modulation (CaM), and distortion-aware modulation (DaM). CLIP aids in extracting semantic information, enabling superior handling of quality-relevant regions and complex distortions.

\item \textbf{Wen et al.}~\cite{Wen_2024_CVPR} combines a base quality predictor (using sparse frames and CLIP for semantic features) with spatial and temporal rectifiers to adjust scores based on resolution and frame rate. This approach ensures adaptability to diverse video attributes.

\item \textbf{BUONA-VISTA}~\cite{wu2023exploringopinionunawarevideoquality} is an opinion-unaware model integrating high-level semantic metrics from CLIP (via prompts like "high quality" vs. "low quality") with spatial and temporal naturalness indices. Aggregated scores are computed through Gaussian normalization and sigmoid rescaling, effectively capturing technical and aesthetic quality.

\item \textbf{MaxVQA}~\cite{wu2023towards} combines CLIP’s semantic features with FAST-VQA’s texture and temporal distortion features. Quality predictions are derived by calculating cosine similarity between learnable prompts and fused features, achieving interpretable, state-of-the-art results.

\item \textbf{ZE-FESG}~\cite{10448422} leverages CLIP to extract 52-dimensional feature vectors from video frames using 35 technical and 17 abstract descriptions as semantic guidance. Features are aggregated to produce quality scores, offering a zero-shot NR-VQA solution.

\item \textbf{Q-Align}~\cite{wu2023qalign} fine-tunes large multimodal models (LMMs) to align predictions with human ratings by categorizing MOS into discrete levels (\textit{e.g.}, "excellent," "good"). During inference for video quality assessment, the model processes sparsely sampled frames to output weighted scores.

\item \textbf{LMM-VQA}~\cite{ge2024lmm} models temporal information using SlowFast~\cite{Feichtenhofer_2019_ICCV} motion features processed into temporal tokens, and spatial features via ViT modules in CLIP. These tokens, combined with text tokens, are decoded into sequences indicating video quality scores or levels.
\end{itemize}

\subsection{Loss Function}
\label{loss function}

Next, we introduce loss functions that are commonly used as objectives when training deep video quality models.

\subsubsection{\textbf{$L_1/L_2$ loss}}
$L_p$ ($p=1,2$) norms have been used to learn most visual tasks due to their simplicity and analytical properties.
The $L_1$ loss is also known as the mean absolute error (MAE) between a batch of predicted quality scores and MOS:
\begin{equation}
    L_{1} = \frac{1}{N}\sum_{i=1}^{N}\abs{q_i-\hat{q_i}},
\end{equation}
where $N$ is the batch size, and $q_i$ and $\hat{q_i}$ are the predicted quality scores and the ground truth quality scores of the i-th video in the batch, respectively. Many deep learning based VQA models~\cite{ying2021patch,wang2021rich,sun2022deep,li2019quality,li2022study,wen2021strong,wu2022discovqa} employ the $L_1$ loss.
Many other deep VQA models~\cite{Flickrvid-150k,varga2019noCNN,xu2021perceptual,wu2021semantic,you2021long,li2015no,chen2020rirnet,zhangyu2018blind,hou2020no,telili20222bivqa,kim2018deep,xu2020c3dvqa,chen2021deep,feng2022deep} use the $L_2$ loss, also known as the mean squared error (MSE) between the predicted and ground-truth quality scores:
\begin{equation}
    L_{2} = \frac{1}{N}\sum_{i=1}^{N}\Vert q_i-\hat{q_i} \Vert_2^2.
\end{equation}

\subsubsection{\textbf{Monotonicity Loss}}
This is a type of ranking loss which distinguishes the relative qualities of videos.
The rank value between an arbitrary pair of videos in a batch can be formulated as:
\begin{equation}
    L_{rank}^{ij} = max(0,\abs{q_i-q_j}-e(q_i,q_j)\cdot (\hat{q_i}-\hat{q_j})),
\end{equation}
where
\begin{equation}
e(q_i,q_j)\!=\! 
\left\{\!
\begin{array}{ll}
\!1\! & q_i\!\geq\!q_j 
\\[2ex]
\!-1\! & q_i\!<\!q_j.
\end{array}
\right.
\end{equation}

The monotonicity loss is then:
\begin{equation}
    L_{monotonicity} = \frac{1}{N^2}\sum_{i=1}^{N}\sum_{i=1}^{N}L^{ij}_{rank}.
\end{equation}

The authors of ~\cite{sun2022deep,wen2021strong} calculate the weighted average of monotonicity loss and $L_1$ loss to jointly optimize the monotonicity and accuracy of their video quality prediction models.
They found that using monotonicity loss can accelerate model convergence.

\subsubsection{\textbf{Cross entropy loss}}
Some authors have cast the VQA problem in a classification setting~\cite{zeng2018blind,tu2021regression}, where different quality levels correspond to different classes.
Then the widely used cross entropy loss between the predicted qualities and the discrete set of labels can be used to optimize model efficiency.
The cross entropy loss is:
\begin{equation}
    L_{cross-entropy} = -\frac{1}{N}[\sum_{i=1}^{N}\sum_{j=1}^{K}\hat{p}(l^{(i)}=j)\log p(l^{(i)}=j)],
\end{equation}
where $K$ is the number of classes, and $\hat{p}$ and $p$ are K-dimensional probability vectors, where each entry indicates the probability of a quality level of the ground-truth and prediction, respectively.
For example, V-MEON~\cite{liu2018end} predicts the types of codecs used to compress the input videos, ContentNet in~\cite{wang2021rich} outputs the content labels along with quality, and DistortionNet ~\cite{wang2021rich} and the Softmax classification module in ~\cite{li2015no} both predict the distortion types. 

\subsubsection{\textbf{Cosine Similarity Loss}}
Vectorized regression ~\cite{xing2021starvqa} generates two probabilistic quality vectors corresponding to ground-truth $\bm{q}$ and prediction $\bm{y}$.
In this case, one may deploy the cosine similarity loss between them:
\begin{equation}
    L_{cosine} = 1 - \cfrac{\langle \bm{q} \cdot \bm{y} \rangle}{\Vert \bm{q} \Vert \cdot \Vert \bm{y} \Vert},
\end{equation}
where $\langle \cdot \rangle$ is the inner product, $\Vert \cdot \Vert$ is the $L_2$-norm.

\subsubsection{\textbf{SROCC$/$PLCC Loss}}
The Spearman Rank Order Correlation Coefficient (SROCC) and Pearson Linear Correlation Coefficient (PLCC) are commonly used as metrics to evaluate objective VQA model predictive performance against subjective labels, and hence it is natural to employ them as loss functions~\cite{liu2018end,li2022blindly,wu2022fast}.
The differentiable SROCC~\cite{blondel2020fast} is expressed as:
\begin{equation}
    SROCC = \cfrac{\sum_{i=1}^N(q_{pr}^i-\overline{q_{pr}})(q_r^i-\overline{q_r})}{\sqrt{\sum_{i=1}^N(q_{pr}^i-\overline{q_{pr}})^2\sum_{i=1}^N(q_r^i-\overline{q_r})^2}},
\end{equation}
where $\{q_{pr}^i\}_{i=1}^N$ and $\{q_{r}^i\}_{i=1}^N$ are the ranks of the predicted quality scores $\{q_{p}^i\}_{i=1}^N$ and ground-truth opinions $\{q^i\}_{i=1}^N$, respectively, and $\overline{q_{pr}}$ and $\overline{q_{r}}$ denote the mean values of $\{q_{pr}^i\}_{i=1}^N$ and $\{q_{r}^i\}_{i=1}^N$.
The SROCC loss is then defined as:
\begin{equation}
    L_{SROCC} = 1-SROCC.
\end{equation}

Similarly, the differentiable PLCC is calculated as:
\begin{equation}
    PLCC = \cfrac{\sum_{i=1}^N(q_{m}^i-\overline{q_{m}})(q^i-\overline{q})}{\sqrt{\sum_{i=1}^N(q_{m}^i-\overline{q_{m}})^2\sum_{i=1}^N(q^i-\overline{q})^2}},
\end{equation}
where the $\{q_m^i\}_i^N$ are the fitted predictions from a 4-parameter nonlinear mapping on $\{q_{p}^i\}_{i=1}^N$, following the recommendation of the Video Quality Experts Group~\cite{video2000final}, and $\overline{q_m}$ is the mean value of $\{q_m^i\}_i^N$. Then the PLCC loss can be defined as:
\begin{equation}
    L_{PLCC} = \cfrac{1-PLCC}{2}.
\end{equation}

\subsubsection{\textbf{Other Loss Functions}}
A variety of other loss functions have been used to design deep VQA models. A mixed dataset training strategy is adopted in~\cite{li2021unified} to better generalize VQA models, wherein monotonicity loss, PLCC loss and normalized $L_1$ loss are averaged to obtain a compound loss on each individual dataset, and where the overall loss used to train the unified VQA model across the multiple datasets is a softmax weighted average of the losses on all the datasets.
NIMA~\cite{talebi2018nima} applies the squared Earth Mover's Distance (EMD)~\cite{hou2016squared} to measure the distances between the cumulative distributions of the predicted quality ratings and the ground-truth human ratings.
CNN-TLVQM~\cite{korhonen2020blind} leverages Huber loss to train a CNN model to conduct local feature extraction, and to regress segment-level feature vectors into final video quality scores produced by an LSTM network.
CoINVQ~\cite{wang2021rich} uses pairwise loss and contrastive loss to evaluate compression level differences in CompressNet, while calculating the pairwise hinge loss in DistortionNet.

%% file: sections/5_Performance_benchmark.tex
\section{Performance Benchmark}
\label{performance_bench}
In this section, we delve deeper into the comparison of both FR and NR IQA/VQA algorithms across several open-source video quality assessment datasets, each curated to challenge algorithms with diverse distortion types that are increasingly relevant in contemporary video streaming and broadcast scenarios.

\begin{table*}[!t]
\renewcommand{\arraystretch}{1.2}
\centering
\fontsize{7pt}{7.5pt}\selectfont
\setlength{\tabcolsep}{5pt}
\caption{Performance comparison of full reference IQA/VQA models on FR video quality assessment databases. The image and video quality assessment models are indicated by \textit{italic} and orthographic font, respectively. Metrics are SROCC/PLCC, and the results of the best performing model are boldfaced.}
\vspace{-5pt}
\label{table:performance_fr}
\resizebox{\textwidth}{!}{%
\begin{tabular}{l|l|cccccc}
\toprule
                 &              & \multicolumn{2}{c}{General distortion}   & & \multicolumn{3}{c}{Bespoke for temporal distortion}                                       \\ \cline{3-4} \cline{6-8} \\[-1.em] 
Type& Model & LIVE-VQA~\cite{LIVE-VQA}  & MCL-V~\cite{MCL-V} & & LIVE-YT-HFR~\cite{madhusudana2021subjective} & BVI-HFR~\cite{mackin2018study} &  BVI-VFI~\cite{10304617}  \\
                \midrule
\multirow{5}{*}{\begin{tabular}[c]{@{}l@{}}Knowledge-driven \\ (IQA/VQA) \end{tabular}} 
& \emph{PSNR} &0.6958/0.7499 & 0.5410/0.5430 &  & 0.7802/0.7481& 0.2552/0.3155&0.5200/0.4710\\
& \emph{SSIM}~\cite{wang2004image} &  0.7211/0.7883 & 0.7100/0.7090&  & 0.5566/0.5418& 0.1958/0.3532&0.6000/0.5400\\
& \emph{VIF}~\cite{sheikh2006image} &0.6861/0.7601 & 0.7430/0.7470 &  & 0.6810/0.7020& 0.2500/0.2640&0.5350/0.4890\\
& VMAF~\cite{arjovsky2017wasserstein}  & 0.8163/0.8115 &0.8280/0.8300 &  &0.7782/0.7419 &0.1888/0.3703 &0.5950/0.5640\\
& ST-GREED~\cite{madhusudana2021st} &0.6869/0.7049 & 0.7817/0.7996
&  &\textbf{0.8822}/\textbf{0.8869} &\textbf{0.8042}/\textbf{0.8312} &0.1120/0.2140\\
\midrule
\multirow{4}{*}{\begin{tabular}[c]{@{}l@{}}Deep learning \\ (IQA) \end{tabular}} 
& \emph{DeepQA}~\cite{kim2017deepsen} & 0.8678/0.8692&0.6395/0.6384 &  &0.0815/0.3162 &0.0222/0.3061 &0.4438
/0.4671\\
& \emph{LPIPS}~\cite{zhang2018unreasonable} & 0.5243/0.5658 & 0.7610/0.7520&  &0.6920/0.7050 & 0.2388/0.3753
&0.5990/0.5970\\
& \emph{DISTS}~\cite{ding2020image} &0.4582/0.4868  &0.7960/0.7820  &  & 0.7210/0.7290&0.4046/0.6789
 &0.5000/0.5500\\
& \emph{TOPIQ}~\cite{10478301} & 0.7095/0.7345 &0.7309/0.7262 &  & 0.1216/0.3080 & 0.0016/0.2823&0.2342/0.3639\\
\midrule
\multirow{4}{*}{\begin{tabular}[c]{@{}l@{}}Temporal pooling \\ / NN module \\(VQA) \end{tabular}}  
& FloLPIPS~\cite{danier2022flolpips} & 0.5485/0.5653 & 0.5447/0.5833 &  & 0.0716/0.1603&0.1041/0.2456 &\textbf{0.6830}/\textbf{0.7060}\\
& DeepVQA~\cite{kim2018deep}  & 0.9152/0.8952  & -/-&  & 0.4331/0.3996& 0.1469/0.2013&-/-\\  
& C3DVQA~\cite{xu2020c3dvqa}& \textbf{0.9261}/\textbf{0.9122}&0.7850/0.7920 &  & 0.7300/0.7410& -/-&0.5080/0.3510\\
& STRA-VQA~\cite{10444627} & -/-&\textbf{0.8570}/\textbf{0.8640} &  &0.7990/0.8060 & -/-&-/-\\ \bottomrule 
\end{tabular}
}
\end{table*}

\begin{table*}[!t]
\renewcommand{\arraystretch}{1.2}
\centering
\fontsize{7pt}{7.5pt}\selectfont
\setlength{\tabcolsep}{5pt}
\caption{Performance comparison of no-reference IQA/VQA models on NR video quality assessment databases. The image and video quality assessment models are indicated by \textit{italic} and orthographic font, respectively. Metrics are SROCC/PLCC, and the results of the best performing model are boldfaced.}
\vspace{-5pt}
\label{table:performance_nr}
\resizebox{\textwidth}{!}{%
\begin{tabular}{l|l|cccccc}
\toprule
                 &              & \multicolumn{3}{c}{User-generated Content}   & & \multicolumn{2}{c}{AI-generated Content}                                       \\ \cline{3-5} \cline{7-8} \\[-1.em] 
Type& Model & LIVE-VQC~\cite{LIVE-VQC} & KoNViD-1k~\cite{Konvid-1k} & YouTube-UGC~\cite{Youtube-UGC} & & T2VQA-DB~\cite{kou2024subjectivealigneddatasetmetrictexttovideo}  &  GAIA~\cite{chen2024gaiarethinkingactionquality}  \\
                \midrule
\multirow{5}{*}{Knowledge-driven (IQA/VQA)} 
& \emph{BRISQUE}~\cite{mittal2012no} & 0.5948/0.6267&0.6567/0.6576 & 0.3820/0.3952 & &0.1880/0.2554 &0.0967/0.2120
\\
& \emph{NIQE}~\cite{6353522} & 0.5896/0.6205&0.5417/0.5530 & 0.2379/0.2776&  & 0.0047/0.2045& 0.0615/0.1904\\
& TLVQM~\cite{korhonen2019two}  & 0.7964/0.7975 &  0.7729/0.7688  & 0.6693/0.6590 &  & 0.4891/0.4960 &   0.4655/0.4783   \\   
&  VIDEVAL~\cite{9405420} & 0.7148/0.7258 &  0.7832/0.7803  & 0.7787/0.7733  &  & 0.5246/0.5435 &   0.4684/0.4801    \\
& FAVER~\cite{zheng2022faver} & 0.7888/0.7982& 0.7906/0.7912& 0.7367/0.7395&  & 0.5105/0.5351& 0.2004/0.2691\\
\midrule
\multirow{4}{*}{Deep learning (IQA)} 
& \emph{PaQ-2-PiQ}~\cite{ying2020patches} & 0.6436/0.6683& 0.6130/0.6014& 0.2658/0.2935&  & 0.1518/0.1409& 0.2261/0.2348\\
& \emph{DB-CNN}~\cite{zhang2018blind} & 0.6391/0.7125& 0.7187/0.7300& 0.4793/0.5224&  & 0.0129/0.0552&0.1727/0.1797 \\
& \emph{MUSIQ}~\cite{ke2021musiq} &0.6252/0.7101 &0.7266/0.7468 & 0.5299/0.5622&  &0.0712/0.0698 &0.1572/0.1609\\
& \emph{CLIP-IQA+}~\cite{Wang_Chan_Loy_2023} & 0.7276/0.7789&0.7813/0.7817 &  0.5374/0.5801&  & 0.0759/0.1351 &0.1538/0.1677 \\

       \midrule
\multirow{2}{*}{\begin{tabular}[c]{@{}l@{}} 2D CNNs with simple \\ score/feature averaging (VQA) \end{tabular}} 
&  RAPIQUE~\cite{tu2021rapique} & 0.7287/0.7594 &  0.8031/0.8175  & 0.7591/0.7684  &  &  0.3130/0.4510& 0.2728/0.3246
    \\
& SIONR~\cite{wu2021semantic} & 0.7361/0.7821& 0.8109/0.8180& 0.3621/0.3949&   & 0.2434/0.2554&0.1263/0.1520\\
\midrule    
\multirow{2}{*}{\begin{tabular}[c]{@{}l@{}} 2D CNN with temporal \\ aggregation networks (VQA)\end{tabular}}   
&  VSFA~\cite{li2019quality} & 0.6978/0.7426 &  0.7728/0.7754  &  0.7240/ 0.7430 &  & 0.1011/0.1193  & 0.5085/0.5215  \\  
& 2BiVQA~\cite{telili20222bivqa} &0.7610/0.8320 &0.8150/0.8350 &0.7710/0.7900 &  & -/- & -/-\\

\midrule    
\multirow{2}{*}{\begin{tabular}[c]{@{}l@{}}3D CNNs/ Transformation \\ (VQA)\end{tabular}}  
& BVQA~\cite{li2022blindly} & 0.8340/0.8420 &   0.8340/0.8360  &  0.8180/0.8260 &  & 0.7390/0.7486 & 0.5201/0.5289\\   
& Shen et al.~\cite{10375568} & 0.7620/0.7660&0.7920/0.7880 &0.7740/0.7660 &  & 0.2736/0.3147 &0.0811/0.1438\\
\midrule    
\multirow{3}{*}{Transformer (VQA)}  
& FAST-VQA~\cite{wu2022fast} & 0.8211/0.8359 &  0.8543/0.8508  & 0.8617/0.8669 &  & 0.7173/0.7295 &  0.5276/0.5475  \\  
& DOVER~\cite{Wu_2023_ICCV} &  0.7989/0.8348 & 0.8752/0.8816   &  0.8761/0.8753 &  & 0.7609/0.7693 & \textbf{0.5335}/\textbf{0.5502}  \\  
& SAMA~\cite{Liu_Quan_Xiao_Li_Wu_2024} & \textbf{0.8600}/\textbf{0.8780}&0.8920/0.8920 & 0.8810/0.8800&  & 0.0136/0.0447&0.2361/0.2432\\
\midrule    
\multirow{4}{*}{\begin{tabular}[c]{@{}l@{}} Large multimodality models \\ (LMMs) (VQA)\end{tabular}}  
& COVER~\cite{He_2024_CVPR} & 0.8093/0.8478 & 0.8933/0.8947   & \textbf{0.9143}/\textbf{0.9165} &  & 0.1276/0.2463 & 0.2254/0.2318  \\   
& MaxVQA~\cite{wu2023towards} & 0.8540/0.8730&  \textbf{0.8940}/\textbf{0.8950}&  0.8940/0.8900&  &0.1941/0.2232 & 0.2528/0.2509\\
& Q-align~\cite{wu2023qalign} & 0.7730/0.8300  &   0.8650/0.8770  &0.8340/0.8480  &  & \textbf{0.7601}/\textbf{0.7768} & -/-  \\ 
& LMM-VQA~\cite{ge2024lmm} & 0.8310/0.8630&0.8750/0.8760 & 0.8580/0.8770&  & -/-& -/-\\
\bottomrule 
\end{tabular}
}
\vspace{-8pt}
\end{table*}

\subsection{Evaluation Criteria}
Generally, an objective model's performance can be evaluated by examining its accuracy, monotonicity, and consistency with human ratings. There are three commonly employed performance metrics: SROCC, PLCC, and the root mean squared error (RMSE). Specifically, SROCC evaluates the monotonicity between the predicted and ground truth quality scores, PLCC evaluates the model’s prediction linearity, and RMSE measures the prediction accuracy. Higher SROCC, PLCC, and lower RMSE scores denote better performances. Note that PLCC and RMSE are computed after performing a nonlinear four-parametric logistic regression to linearize the objective predictions to be on the same scale as MOS \cite{5404314}:

\begin{equation}
\label{eq:logistic}
f(x)=\beta_2+\frac{\beta_1-\beta_2}{1+\exp{(-x+\beta_3/|\beta_4|})}.
\end{equation}

\subsection{Experimental Setup}
For the FR models, we conducted the evaluation using the LIVE-VQA~\cite{LIVE-VQA} and MCL-V~\cite{MCL-V} datasets known for their inclusion of conventional distortions such as compression artifacts, transmission, and scaling, and three additional datasets, LIVE-YT-HFR~\cite{madhusudana2021subjective}, BVI-HFR~\cite{mackin2018study}, and BVI-VFI~\cite{10304617}, specifically chosen for their representation of temporal distortions, including variable frame rates and frame interpolation. For the NR models, the test beds include three UGC VQA datasets, LIVE-VQC~\cite{LIVE-VQC}, KoNViD-1k~\cite{Konvid-1k}, and YouTube-UGC~\cite{Youtube-UGC}, and two cutting-edge AIGC VQA datasets, T2VQA-DB~\cite{kou2024subjectivealigneddatasetmetrictexttovideo} and GAIA~\cite{chen2024gaiarethinkingactionquality}. Details describing these datasets can be found in Section~\ref{subjective-vqa}.

Representative models from each category of full-reference (FR) and no-reference (NR) video quality assessment algorithms, as previously introduced, are selected for performance comparison on datasets having diverse distortion types. The performance comparison results of FR models and NR models are summarized in Table~\ref{table:performance_fr} and Table~\ref{table:performance_nr}, respectively. If available, performance data was taken from the original papers, otherwise we conducted the evaluation.

This analysis aims to provide insights into the accuracy and adaptability of these models, presenting a comprehensive overview of current capabilities and emerging trends in video quality assessment. Next, we separately analyzed the performance results of FR and NR models by type and provide insights on the effective employment of neural network modules in VQA models.

\subsection{FR VQA Models Evaluation}

As shown in Table~\ref{table:performance_fr}, among general-purpose knowledge-driven models, VMAF achieved the highest SROCC/PLCC scores on the LIVE-VQA and MCL-V databases, indicating superior effectiveness in handling traditional quality degradations. However, these general-purpose models are less effective in scenarios with frame rate variability or frame interpolation distortions. ST-GREED, bespoke for perceiving framerate variation, yields high correlations with human ratings on two HFR databases, yet shows limitations on measuring quality degradations induced by video frame interpolation. 

Deep learning-based IQA models perform reasonably well on general distortion datasets. DeepQA, in particular, shows a notable advantage in terms of SROCC/PLCC scores, suggesting that these models can capture complex spatial artifacts introduced by compression, transmission, and scaling distortions. However, they struggle when applied to HFR and VFI datasets due to limited temporal modeling capabilities.

By incorporating effective temporal pooling strategies or temporally aware NN modules, deep learning-based VQA models show improved performance over knowledge-driven and deep learning-based IQA models. For instance, FloLPIPS, a recently proposed bespoke metric for VFI that combines distortions in optical flow with LPIPS, delivers the highest SROCC/PLCC scores on the BVI-VFI database. C3DVQA performs notably well on the LIVE-VQA dataset due to its robust handling of spatiotemporal interactions, underscoring the importance of temporal awareness in VQA tasks. Moreover, STRA-VQA, which uses a Transformer-based architecture to model spatiotemporal quality, yields remarkable performance on MCL-V and LIVE-YT-HFR databases.

We offer the following design insights for deep learning-based FR VQA models:
\begin{itemize}
\item Spatial quality modeling. Reference videos enable precise spatial quality comparisons. Structural similarity metrics, such as SSIM in traditional models and DISTS in deep learning, effectively model spatial quality. Visual sensitivity maps in models like DeepQA also incorporate human perception, enhancing spatial distortion assessment.
    \item Temporal feature extraction. Incorporating optical flow (\textit{e.g.}, FloLPIPS) or spatiotemporal 3D block features (\textit{e.g.}, C3DVQA) improves performance on datasets with significant temporal distortions.
    \item Temporal effect modeling. Attention mechanisms, memory architectures, and memory-aware pooling strategies enhance temporal quality prediction. Transformer-based models like STRA-VQA excel at capturing long-range temporal dependencies, particularly in scenarios like variable frame rates and frame interpolation, by dynamically prioritizing critical temporal features.
\end{itemize}

\subsection{NR VQA Models Evaluation}
As shown in Table~\ref{table:performance_nr}, knowledge-driven IQA models like BRISQUE and NIQE do not perform well in dynamic video environments. By incorporating deep features and end-to-end training, deep learning-based IQA models deliver better performance on UGC databases. However, knowledge-driven VQA models, such as TLVQM, VIDEVAL, and FAVER, can outperform both handcrafted and learning-based IQA models on both UGC and AIGC video datasets, highlighting the effectiveness of handcrafted spatiotemporal features. It can be seen that, in no reference scenarios, traditional IQA models, even deep learning-based, struggle when applied directly to video datasets, particularly those involving complex motions or AI-generated contents. Their reliance on frame-level features without considering temporal dependencies limits applicability in dynamic video environments. Nevertheless, CLIP-IQA+, outperforms handcrafted and learning-based IQA models on UGC databases, benefiting from vision-language prior in CLIP.

Models like RAPIQUE and SIONR incorporate simple temporal pooling strategies, allowing better performance in UGC datasets. VSFA and 2BiVQA use temporal aggregation networks to model temporal dependencies better than the handcrafted methods, improving performance in user-generated content (UGC). However, when tested on AI-generated content, their inability to model complex visual-text alignment or temporal coherence becomes evident. BVQA leverages 3D CNNs for better temporal feature extraction, performing significantly better on UGC datasets and starting to bridge the performance gap on AI-generated content.

Both Transformer-based and LMM-based models demonstrate the most excellent performance as well as a strong generalization to UGC content, and notably, show promise for AI-generated content. These models highlight the potential of Transformers and multimodal perception in VQA, but further optimization is also needed for broader deployment in AI-generated content, which often has lower frame rates and may require more refined temporal modeling.

We supply some insights for model design in deep learning-based NR VQA models as follows.
\begin{itemize}
\item Temporal quality modeling. Capturing temporal dynamics is a key challenge. While 2D CNNs excel at spatial features, they lack temporal distortion modeling. The use of 3D CNNs and advanced Transformers has significantly improved temporal quality prediction by effectively capturing long-range dependencies.
    \item Multimodality Integration. Models like COVER and LMM-VQA leverage large multimodal approaches, combining text-video alignment and video fidelity assessment, and are particularly essential for UGC and AIGC evaluation. Such models are increasingly important given the rise of generative AI, highlighting how multimodal-based approaches can bridge the gap between human perception and objective quality models.
\end{itemize}

%% file: sections/6_application.tex
\begin{figure*}[!t]
\centering
\includegraphics[width=0.85\linewidth]{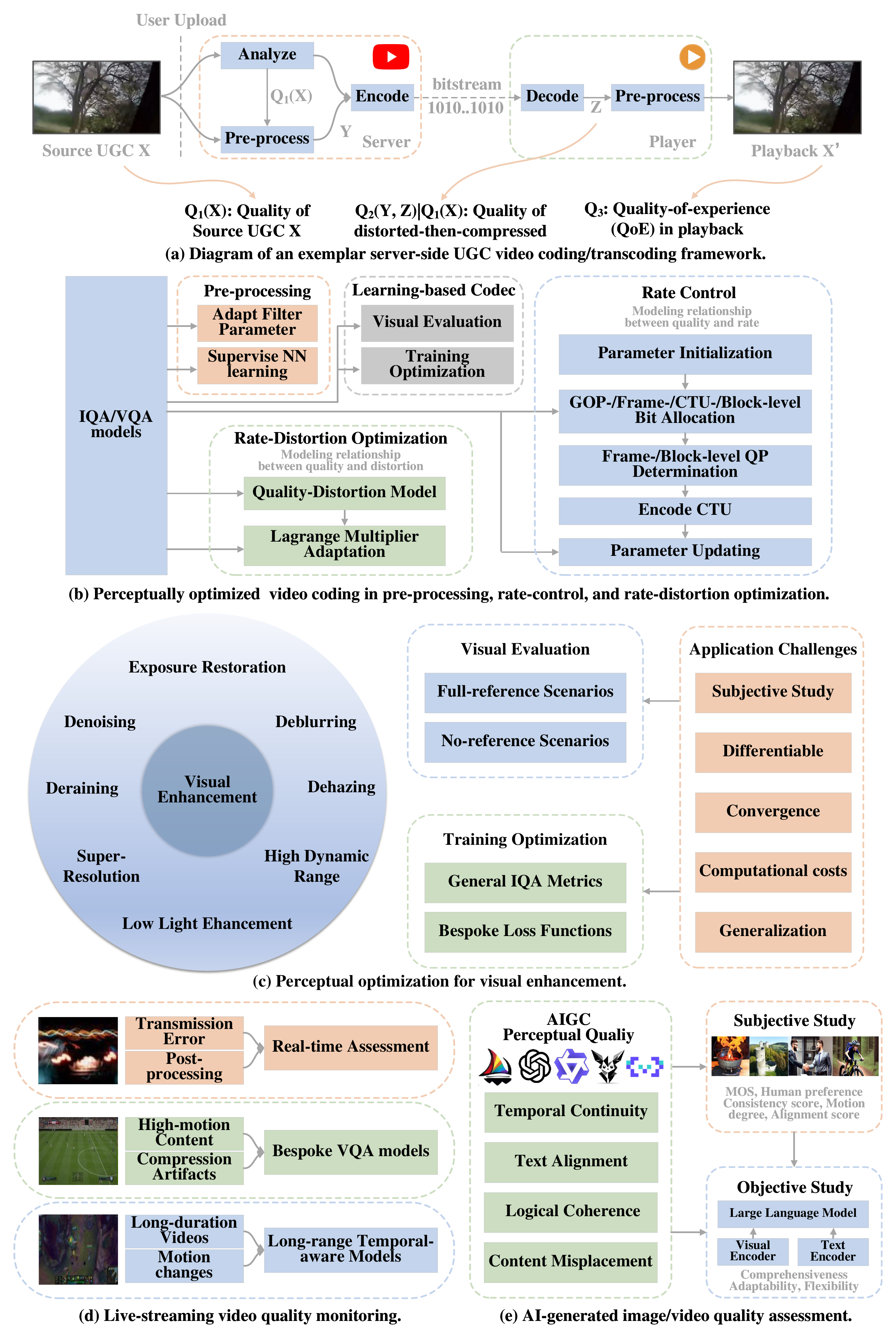}
\caption{Application overview of video quality assessment.}
\label{fig:ugc-transcode-fw}
\end{figure*}

\section{Applications and Challenges}
\label{sec:application_challenge}

\subsection{Server-Side UGC Video Coding and Transcoding}
Streaming platforms like YouTube, TikTok, Facebook, and Bilibili use video transcoding to deliver high-quality, low-bitrate videos. Modern codecs like H.26x, VPx, AV1, and AVSx employ a hybrid framework (prediction, transformation, quantization, entropy coding) guided by rate-distortion optimization (RDO) to minimize distortion under bitrate constraints. However, defining distortion in this context can be challenging. Traditional pixel-based metrics like SAD, MSE, and PSNR poorly correlate with human perception. Advanced full-reference VQA models like SSIM and VMAF provide more accurate predictions of perceptual quality and are widely used for compressing professionally-generated content (PGC). However, user-generated content (UGC) often lacks high-quality reference videos and suffers from diverse distortion types. Compressing UGC using distorted originals as references can lead to inaccurate quality predictions.

Towards resolving this issue, consider the general task of predicting the quality of distorted-then-compressed videos~\cite{9525062,smirnov2024aim,conde2024ais}.
If the existing distortion is also compression, then this may be viewed as a type of video transcoding problem (albeit, likely without decoding the original compression).
Fig.~\ref{fig:ugc-transcode-fw} (a) illustrates the typical server-side processing flow of a UGC coding and communication service.
In this pipeline, if we disregard other pre-processing and post-processing stages, a source UGC video $X$, previously distorted and/or compressed by unknown processes, can be quality-evaluated using a blind VQA model to generate a no-reference quality estimate, $Q_1(X)$.
The overall quality of the subsequent distorted-then-compressed UGC video can then be modeled using a Bayesian approach~\cite{bovik2020assessing}, allowing for prediction of the final quality of the compressed UGC video, based on the estimated compressed video quality $Q_2(Y,Z)$ conditioned on the input quality $Q_1(X)$.
Being able to more accurately predict and monitor the compressed quality of UGC encodes or transcodes can enable more efficient compression rate, extending the limits of traditional UGC coding units.

\subsection{Perceptually Optimized Video Coding}
Significant progress in perceptual image/video quality assessment models has advanced their integration into video codec design, optimization, and testing, enhancing coding efficiency. We refer the readers to~\cite{zhao2013use} for systematic discussions on the rationale and methodology of utilizing perceptual VQA algorithms, like SSIM, in video coding. Below, we review key areas: preprocessing, rate control, rate-distortion optimization, and end-to-end codec evaluation and optimization, as shown in Fig.~\ref{fig:ugc-transcode-fw} (b).
%

Preprocessing input video frames before encoding can enhance the perceptual quality of the reconstructed/decoded videos. For instance, sharpening filters~\cite{deng2020vmaf} counteract encoding-induced blur, with parameters tuned using VMAF. Deep preprocessors~\cite{chadha2021deep} employ perceptual loss functions like MS-SSIM and NIMA, while JND thresholds~\cite{ding2015video} guide adaptive Gaussian filtering for perceptual improvements.

Rate control in video encoding allocates bit budgets by constant bit-rate (CBR), average bit-rate (ABR), or variable bit-rate (VBR) control.
A deep learning-based framework proposed in~\cite{cai2022quality,xing2019predicting} predicts rate factors (RF) for video segments, ensuring constant high quality measured by VMAF. Similarly, Netflix's Dynamic Optimizer~\cite{katsavounidis2018dynamic} generates R-D optimal convex hulls on a per-shot basis, optimizing bit allocation while maintaining VMAF-based distortion metrics. Deng et al.\cite{deng2020vmaf} demonstrated that higher motion content allows higher distortion thresholds for the same VMAF score, enabling QP adjustments to balance quality and bitrate. SSIM-based perceptual rate control\cite{wang2020ssim} models QP-perceptual quality relationships, optimizing QP settings per frame, while~\cite{ou2011ssim} employs SSIM in a rate-distortion model to meet bit budget targets.

The main objective of video compression is to achieve optimal trade-offs between rate and distortion.
A commonly known strategy to achieve this balance is via rate-distortion optimization (RDO), which can be transformed into an unconstrained optimization problem using a Lagrange multiplier approach~\cite{733497}. Thus find:
\begin{equation}
    \min\{J\} \quad \text{where} \quad J = D + \lambda \cdot R,
\end{equation}
where $J$, $D$, $\lambda$, and $R$ denote rate-distortion cost, distortion, Lagrange multiplier, and rate, respectively.
In previous work focusing on RDO based on perceptual distortion, the distortion component $D$ has generally been replaced by VMAF~\cite{deng2020vmaf} or SSIM~\cite{wang2011rate,wang2011ssim,ou2011ssim}.
%

%


Deep learning-based end-to-end image and video compression frameworks have garnered attention for efficiently encoding visual data with high perceptual quality. IQA/VQA metrics are essential for evaluating and optimizing these frameworks, employing full-reference metrics like PSNR, SSIM, MS-SSIM, and LPIPS, and no-reference metrics like NIQE and CLIP-IQA~\cite{NEURIPS2020_8a50bae2, NEURIPS2023_ccf6d8b4,He_2022_CVPR,Jiang_Tan_Tan_Yan_Shen_2023}. These metrics also function as loss functions during training, guiding perceptually optimized compression~\cite{NEURIPS2023_ccf6d8b4,9259253,NEURIPS2020_8a50bae2}. For example, ProxIQA~\cite{9259253} approximates IQA metrics like VMAF and SSIM, serving as a perceptual loss layer. Integrating these metrics advances end-to-end compression, improving both coding efficiency and perceptual quality.

\subsection{Perceptual Optimization for Visual Enhancement}
Visual enhancement technologies aim to restore and improve degraded content, significantly enhancing perceptual quality in industries from entertainment to real-time applications. In mobile photography, for instance, advancements in hardware and computational algorithms address inherent limitations, such as small sensors and constrained optics. Key techniques, including denoising, HDR, super-resolution, color correction, and white balance, play a pivotal role in improving visual quality. Integrating human visual perception into these methods ensures that the results are not only computationally efficient but also visually appealing. IQA/VQA metrics serve as critical tools, functioning as both evaluation benchmarks and loss functions to refine visual enhancement algorithms effectively, as illustrated in Fig.~\ref{fig:ugc-transcode-fw} (c).


IQA/VQA metrics are critical for assessing how well enhancement algorithms improve perceptual quality. Full-reference metrics like PSNR and SSIM are widely used for tasks such as denoising~\cite{Brooks_2019_CVPR, Zamir_2020_CVPR}, deblurring~\cite{Cho_2021_ICCV, Mao_Liu_Liu_Li_Shen_Wang_2023}, dehazing~\cite{Tu_2022_CVPR, 10447111}, and super-resolution~\cite{Liang_2021_ICCV, Zhang_2021_ICCV}. Advanced metrics like MS-SSIM and LPIPS are applied to HDR and multiple restoration tasks~\cite{Chen_2023_ICCV, Zhang_2024_CVPR}. In scenarios lacking reference data, such as autonomous driving under day-night transitions, no-reference metrics like NIQE, MUSIQ, and NIMA evaluate low-light enhancement and HDR ~\cite{li2024light, Zhu_2024_CVPR}, highlighting the necessity of no-reference metrics in challenging scenarios. 

IQA/VQA-based perceptual loss functions are pivotal in training deep learning models for visually natural results. For example, SSIM preserves structural details~\cite{Shyam_2024_WACV}, NLPD optimizes HDR tone mapping~\cite{Laparra:17,cao2022perceptually}, and PTQE finetune video frame interpolation models~\cite{10667010}. Custom loss functions like 1D-Wasserstein distances between CNN activations aid multi-task learning for tasks such as denoising and JPEG artifact removal~\cite{9466271}. Specialized loss functions, including those mimicking human eye response, improve exposure restoration~\cite{9693338}, showcasing the versatility of IQA metrics in perceptual optimization.

Despite their utility, IQA/VQA metrics face several challenges: 1) Subjective gaps. Existing metrics may not align with perceptual quality when measuring restoration results~\cite{10.1007/978-3-030-58621-8_37,10.1145/3457905}, requiring large-scale subjective studies to benchmark and calibrate metrics. 2) Optimization issues. Non-differentiability and non-smooth gradients complicate model training. 3) Computational Costs. Deep learning-based metrics are resource-intensive, limiting real-time applications. 4) Generalization. Metrics tailored for specific tasks, like super-resolution, often underperform for others, such as video enhancement, where temporal consistency is crucial. Developing adaptable metrics remains a significant challenge. 

\subsection{Live-Streaming Video Quality Monitoring}

Live-streaming platforms like Twitch, TikTok, Facebook Live, and YouTube Live have surged in popularity, offering diverse content from gaming and sports to concerts and personal vlogs. Ensuring consistent perceptual video quality is critical but challenging~\cite{shang2021study} due to several inherent factors, as depicted in Fig.~\ref{fig:ugc-transcode-fw} (d).

Firstly, real-time transmission over User Datagram Protocol (UDP) can result in quality degradation, such as frame drops and flickering. Addressing these issues requires robust error recovery mechanisms and real-time VQA metrics to monitor user experiences. Secondly, unlike video-on-demand (VoD), live-streaming has limited time for pre- and post-processing. Videos must be encoded, transmitted, and decoded almost instantly, leaving little room for quality measurement and optimization. Additionally, high-motion live-streams, such as sports or eSports, are prone to distortions like stutter, motion blur, and deinterlacing motion mismatches under constrained bandwidth. Complex temporal variations further challenge traditional VQA models, which may struggle with such dynamic content. Given these challenges, VoD-optimized VQA techniques often fall short for live-streaming due to real-time and long-duration requirements. Effective VQA models for live-streaming must efficiently handle long-range temporal distortions and sudden motion changes.

\subsection{AI-Generated Content Picture/Video Quality Assessment}

The rapid advancements in machine learning (ML) have revolutionized AI-Generated Content (AIGC)\cite{cao2023comprehensive,li20244k4dgen}, encompassing images, videos, and other media created by models like DALLE-2, Imagen, Imagen-Video, and Stable Diffusion. These platforms enable efficient, customized, and diverse content production but also introduce unique challenges, as seen in Fig.~\ref{fig:ugc-transcode-fw} (e).

Unlike traditional content, AIGC faces novel distortions such as Uncanny Valley~\cite{DINATALE2023100288, 10143584},
unrealistic object placement, and a lack of temporal coherence in videos, where frame sequences may lack logical continuity. These issues, absent in human-generated content, significantly impact perceptual quality and fall outside the scope of existing IQA/VQA algorithms, which requires new benchmarks and algorithms tailored to these unique challenges. Recent datasets, such as FETV~\cite{NEURIPS2023_c481049f}, VBench~\cite{Huang_2024_CVPR}, GAIA~\cite{chen2024gaiarethinkingactionquality}, and LGVQ~\cite{zhang2024benchmarkingaigcvideoquality}, provide a foundation by focusing on spatiotemporal consistency, text alignment, and motion quality.  As AIGC evolves, new distortions may emerge, necessitating flexible and adaptable IQA/VQA models. Addressing these challenges will enhance user acceptance and ensure high perceptual quality, fostering growth in this transformative domain, whereby robust IQA/VQA algorithms and benchmarks will be key to unlocking the full potential of AIGC.

%% file: sections/8_conclusion.tex
\section{Conclusion and Future Directions}
\label{conclusion}
We have offered a comprehensive survey of deep learning-based video quality assessment studies, focusing on both subjective and objective quality assessment methods.
The general workflows of subjective quality assessment data gathering and the existing panoply of popular databases were summarized.
We then examined objective full-/no-reference algorithms from the past two decades, with a heavy focus on more recent deep learning-based VQA models.
We also conducted a comprehensive comparison of the effectiveness and adaptability of existing FR and NR algorithms on databases of emerging content, providing insights on the employment of effective neural network modules in VQA models.
We discussed current limitations and challenges in practical applications of deep learning VQA research, and elaborated on significant directions for future research.

Deep learning-based IQA/VQA models face several challenges in datasets, model architectures, and training strategies, with ample opportunities for future improvement. The limitations of existing video quality assessment datasets hinder the potential of data-driven learning. Constructing large-scale, unbiased datasets for VQA is labor-intensive, requiring accurate data collection methodologies and substantial human involvement in psychometric studies. Future datasets are required to match the ever-expanding demands of streaming videos, encompassing advanced formats like high framerate (HFR), high dynamic range (HDR), and virtual/extended reality (VR/XR). Future datasets should also represent varied content types, including screen content, UGC, PGC, AIGC, telepresence data, and point cloud data, ensuring relevance to emerging video technologies and consumer expectations.

In terms of model architectures, video quality assessment presents unique challenges distinct from high-level computer vision tasks. While pre-trained neural networks excel at extracting spatial features, bespoke architectures are needed to address low-level degradation in video content. Future architectures should effectively integrate psychovisual principles such as memory effects and perceptual straightening to align predictions with human visual experiences. Additionally, leveraging large models, including prompt-driven and feature-based methods, offers promising avenues. However, challenges such as aligning textual outputs with human-like quality judgments and optimizing feature extraction for dynamic content must be addressed. Furthermore, designing models that balance prediction accuracy with computational efficiency remains a critical goal, as demonstrated by efficient frameworks like RAPIQUE, FAST-VQA, and Faster-VQA.

Training strategies also warrant further exploration. The scarcity of adequately annotated datasets remains a bottleneck, emphasizing the importance of patch-/clip- level training, leveraging proxy scores, and unsupervised learning. The selection of suitable loss functions is equally critical, with approaches like Huber loss or compound loss functions potentially mitigating the challenges posed by unbounded loss functions in challenging scenarios. These innovations can drive the development of robust, accurate VQA models capable of handling diverse content types and distortion phenomena.

By addressing existing challenges and exploring new avenues, future research can unlock the full potential of VQA in applications ranging from media production to telecommunications, ultimately enhancing the quality of visual experiences for end-users worldwide. We hope this survey stimulates further research within the visual analysis community, encouraging interdisciplinary collaborations to advance the state of video quality assessment.